\documentclass[manuscript,nonacm]{acmart}

\AtBeginDocument{%
  }

\setcopyright{none}

\usepackage{amsmath}
\usepackage[inline]{enumitem}
\usepackage{algorithmic}
\usepackage{alltt}
\usepackage{array}
\usepackage{booktabs}
\usepackage{boxedminipage}
\usepackage[capitalise, nameinlink, noabbrev]{cleveref}
\usepackage{caption}
\usepackage{colortbl}
\usepackage{color}
\usepackage{enumitem}
\usepackage{fp}
\usepackage{graphicx}
\usepackage{hyperref}
\usepackage{lineno}
\usepackage{listings}
\usepackage{xltabular}
\usepackage{MnSymbol}
\usepackage{multicol}
\usepackage{multirow}
\usepackage{orcidlink}
\usepackage{pifont}
\usepackage{relsize}
\usepackage{subcaption}
\usepackage{syntax}
\usepackage{tabularx}
\usepackage[textsize=tiny]{todonotes}
\usepackage{tcolorbox}
\usepackage{textcomp}
\usepackage{tikz}
\usepackage{url}
\usepackage{wasysym}
\usepackage{wrapfig}
\usepackage{xspace}

\usetikzlibrary{tikzmark}
\usetikzlibrary{arrows.meta}

\def\ACCURACY{\emph{Accuracy}\xspace}

\def\ALHAZEN{{\scshape{}alhazen}\xspace}
\def\AMPLE{{\smaller{}AMPLE}\xspace}

\def\AVICENNA{{\scshape{}Avicenna}\xspace}
\def\BUG{{\scshape{}bug}\xspace}
\def\BUGSINPY{{\scshape{}BugsInPy}\xspace}

\def\C{{\scshape{}C}\xspace}

\def\DSTAR{{\smaller{}D*}\xspace}

\def\DEFECTS4J{{\scshape{}Defects4J}\xspace}
\def\EFDD{{\scshape{}EFDD}\xspace}
\def\ENTBUG{{\scshape{}entbug}\xspace}
\def\EXAM{{\smaller{}EXAM}\xspace}

\def\FONESCORE{\emph{F1 Score}\xspace}

\def\GP{{\smaller{}GP}\xspace}
\def\GPOT{{\smaller{}GP13}\xspace}

\def\JACCARD{{\scshape{}jaccard}\xspace}
\def\JAVA{{\scshape{}Java}\xspace}

\def\MACRO{{macro}\xspace}

\def\NAISHT{{\smaller{}NAISH2}\xspace}
\def\NOBUG{{\scshape{}no-bug}\xspace}
\def\OCHIAI{{\smaller{}OCHIAI}\xspace}
\def\PRECISION{\emph{Precision}\xspace}

\def\PYTHON{{\scshape{}Python}\xspace}
\def\RECALL{\emph{Recall}\xspace}
\def\REFACTORY{{\scshape{}refactory}\xspace}

\def\SCIKITLEARN{{\scshape{}scikit-learn}\xspace}

\def\SFLKIT{{\scshape{}SFLKit}\xspace}

\def\TARANTULA{{\smaller{}TARANTULA}\xspace}
\def\TESTS4PY{{\scshape{}Tests4Py}\xspace}

\def\WORD2VEC{{\smaller{}WORD2VEC}\xspace}

\definecolor{rltred}{rgb}{0.5,0,0}
\definecolor{rltgreen}{rgb}{0,0.5,0}
\definecolor{rltblue}{rgb}{0,0,0.5}
\hypersetup{
colorlinks=true,
urlcolor=rltblue,
linkcolor=rltred,
citecolor=rltgreen,
}

\newcommand{\mathid}[1]{\textit{#1}}
\newcommand{\codeid}[1]{\texttt{#1}}
\def\|#1|{\mathid{#1}}
\def\<#1>{\codeid{#1}}

\definecolor{deepblue}{rgb}{0,0,0.5}
\definecolor{deepred}{rgb}{0.6,0,0}
\definecolor{deepgreen}{rgb}{0,0.5,0}
\definecolor{deeporange}{rgb}{0.8,0.45,0}
\definecolor{Grey}{rgb}{0.5,0.5,0.5}
\definecolor{LightGrey}{rgb}{0.9,0.9,0.9}
\definecolor{Green}{rgb}{0.0,0.6,0.0}
\definecolor{Red}{rgb}{0.6,0.0,0.0}
\definecolor{Blue}{rgb}{0.0,0.0,0.6}

\definecolor{DarkBlue}{rgb}{0.0859, 0.308, 0.523}
\definecolor{DarkOrange}{rgb}{0.8, 0.4, 0.0}
\definecolor{DarkGreen}{rgb}{0.00,0.40,0.00}
\definecolor{ScarletRed}{rgb}{0.60,0.00,0.00}
\definecolor{AlmostWhite}{rgb}{0.80,0.80,0.80}
\definecolor{Gray}{gray}{0.85}

\def\rowstrut{\rule{0pt}{1.2em}}
\definecolor{row}{rgb}{0.8627, 0.9098, 0.9804}

\definecolor{Cornsilk}{rgb}{0.98, 0.94, 0.9}

\newcommand{\PASS}{\text{\color{Green}\ding{52}}\xspace}
\newcommand{\FAIL}{\text{\color{Red}\ding{56}}\xspace}

\newcolumntype{a}{>{\columncolor{TableHeader}}l}

\newcounter{confusions}
\let\backslashpipe=\|
\let\backslashstar=\*  % in case we ever need \*
\let\backslashangle=\<
\def\*#1*{\codeid{#1}}  % Code

\newcommand{\maketodo}[2]{\expandafter\newcommand\csname #1\endcsname[1]{\todo[bordercolor=#2!80!black,color=#2]{\textbf{\MakeUppercase #1:} ##1}\xspace}}

\presetkeys{todonotes}{fancyline}{}
\definecolor{todored}{rgb}{1, 0.6, 0.6}
\definecolor{todoorange}{rgb}{1, 0.8, 0.4}
\definecolor{todoblue}{rgb}{0.4, 0.8, 1.0}
\definecolor{todogreen}{rgb}{0.8, 1.0, 0.4}
\definecolor{todopurple}{RGB}{255, 100, 127}

\maketodo{NOTE}{todogreen}
\maketodo{TODO}{todored}
\maketodo{CHECK}{todoblue}
\newcommand{\DONE}[1]{} % Older arguments, now addressed
\newcommand{\WONTFIX}[1]{} % Other arguments, won't be addressed
\newcommand{\ALSO}[1]{} % Additional arguments

\newenvironment{result}%
{\medskip
\noindent
\let\emph=\textbf
\begin{boxedminipage}{\linewidth}\begin{center}\em}%
{\end{center}\end{boxedminipage}%
\medskip
}

\newlist{questions}{enumerate}{1}
\setlist[questions,1]{label=\bfseries RQ\arabic*:,ref=RQ\arabic*,leftmargin=3\parindent}
\crefname{question}{}{}
\Crefname{question}{}{}

\title{How Execution Features Relate to Failures}
\subtitle{An~Empirical~Study and Diagnosis~Approach}

\author{Marius Smytzek}
\orcid{0000-0002-4899-9031}
\affiliation{%
 \institution{CISPA Helmholtz Center for Information Security}
 \city{Saarbr{\"u}cken}
 \country{Germany}
}
\email{marius.smytzek@cispa.de}

\author{Martin Eberlein}
\orcid{0000-0003-4268-7632}
\affiliation{%
 \institution{Humboldt-Universit{\"a}t zu Berlin}
 \city{Berlin}
 \country{Germany}
}
\email{martin.eberlein@hu-berlin.de}

\author{Lars Grunske}
\orcid{0000-0002-8747-3745}
\affiliation{%
 \institution{Humboldt-Universit{\"a}t zu Berlin}
 \city{Berlin}
 \country{Germany}
}
\email{grunske@hu-berlin.de}

\author{Andreas Zeller}
\orcid{0000-0003-4719-8803}
\affiliation{%
 \institution{CISPA Helmholtz Center for Information Security}
 \city{Saarbr{\"u}cken}
 \country{Germany}
}
\email{zeller@cispa.de}



\begin{abstract}
    Fault localization is a fundamental aspect of debugging, aiming to identify code regions likely responsible for failures. Traditional techniques primarily correlate statement execution with failures, yet program behavior is influenced by diverse execution features—such as variable values, branch conditions, and definition-use pairs—that can provide richer diagnostic insights.

    In an empirical study of 310~bugs across 20~projects, we analyzed 17 execution features and assessed their correlation with failure outcomes. Our findings suggest that fault localization benefits from a broader range of execution features:
       \begin{enumerate*}[label=(\arabic*)]
           \item \emph{Scalar pairs} exhibit the strongest correlation with failures;
           \item Beyond line executions, \emph{def-use pairs} and \emph{functions executed} are key indicators for fault localization; and
           \item Combining \emph{multiple features} enhances effectiveness compared to relying solely on individual features.
       \end{enumerate*}
   
    Building on these insights, we introduce a debugging approach to diagnose failure circumstances. 
    The approach extracts fine-grained execution features and trains a decision tree to differentiate passing and failing runs.
    From this model, we derive a diagnosis that pinpoints faulty locations and explains the underlying causes of the failure. 
   
    Our evaluation demonstrates that the generated diagnoses achieve high predictive accuracy, reinforcing their reliability. 
    These interpretable diagnoses empower developers to efficiently debug software by providing deeper insights into failure causes.
\end{abstract}

\begin{document}

\maketitle

\section{Introduction}%
\label{sec:introduction}

Debugging is the process of identifying and fixing faults in a program.
Be it automated or manual, a central task in debugging is \emph{fault localization}---that is, identifying the locations in the code that are likely to contain the root cause of a failure.
Fault localization techniques typically rely on the analysis of program executions to identify the locations that are most relevant for the failure.

\begin{wrapfigure}{l}{0.5\textwidth}
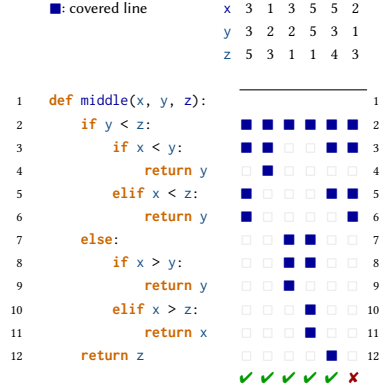

	\def\*{{\color{Blue}$\blacksquare$}}
	\def\+{{\color{Blue}$\blacksquare$}}
	\def\-{{\color{LightGrey}$\Box$}}
	\def\ind{\qquad}
	\scriptsize
    \centering
	\begin{tabular}{@{}>{\tiny}r>{\tt}l@{\quad}l@{\ \ }r@{\ \ }r@{\ \ }r@{\ \  }r@{\ \ }r@{\ \ }r@{\ \ }>{\tiny}r}
	\\
	& \textsf{\color{black}\*: covered line} 
	     & \color{Blue}\texttt{x}      & 3   & 1   & 3   & 5   & 5   & 2  \\
	   & & \color{DarkBlue}\texttt{y}  & 3   & 2   & 2   & 5   & 3   & 1  \\
	   & &  \color{DarkBlue}\texttt{z} & 5   & 3   & 1   & 1   & 4   & 3  \\
	   \\ \cline{4-9}
	1  & \textbf{\color{DarkOrange}def} {\color{Blue}middle}({\color{DarkBlue}x}, {\color{DarkBlue}y}, {\color{Blue}z}):  & & & & & & & & 1 \\
	2  & \ind  \textbf{\color{DarkOrange}if} {\color{DarkBlue}y} < {\color{DarkBlue}z}:         
	& & \*  & \*  & \*  & \*  & \*  & \*  & 2 \\ %\cline{3-8}
	3  & \ind \ind  \textbf{\color{DarkOrange}if} {\color{DarkBlue}x} < {\color{DarkBlue}y}:       
	& & \*  & \*  & \-  & \-  & \*  & \*  & 3 \\ %\cline{3-8}
	4  & \ind \ind \ind  \textbf{\color{DarkOrange}return} {\color{DarkBlue}y}      
	& & \-  & \*  & \-  & \-  & \-  & \-  & 4 \\ %\cline{3-8}
	5  & \ind \ind  \textbf{\color{DarkOrange}elif} {\color{DarkBlue}x} < {\color{DarkBlue}z}:  
	& & \*  & \-  & \-  & \-  & \*  & \*  & 5 \\ %\cline{3-8}
	6  & \ind \ind \ind  \textbf{\color{DarkOrange}return} {\color{DarkBlue}y}      
	& & \*  & \-  & \-  & \-  & \-  & \*  & 6 \\ %\cline{3-8}
	7  & \ind  \textbf{\color{DarkOrange}else}:            
	& & \-  & \-  & \*  & \*  & \-  & \-  & 7 \\ %\cline{3-8}
	8 & \ind \ind  \textbf{\color{DarkOrange}if} {\color{DarkBlue}x} > {\color{DarkBlue}y}:       
	& & \-  & \-  & \*  & \*  & \-  & \-  & 8 \\ %\cline{3-8}
	9 & \ind \ind \ind  \textbf{\color{DarkOrange}return} {\color{DarkBlue}y}      
	& & \-  & \-  & \*  & \-  & \-  & \-  & 9 \\ %\cline{3-8}
	10 & \ind \ind  \textbf{\color{DarkOrange}elif} {\color{DarkBlue}x} > {\color{DarkBlue}z}:  
	& & \-  & \-  & \-  & \*  & \-  & \-  & 10 \\ %\cline{3-8}
	11 & \ind \ind \ind  \textbf{\color{DarkOrange}return} {\color{DarkBlue}x}      
	& & \-  & \-  & \-  & \*  & \-  & \-  & 11 \\ %\cline{3-8}
	12 & \ind  \textbf{\color{DarkOrange}return} {\color{DarkBlue}z}             
	& & \-  & \-  & \-  & \-  & \*  & \-  & 12 \\ %\cline{3-8}
	   &                           
	& & \PASS{} & \PASS{} & \PASS{} & \PASS{} & \PASS{} & \FAIL{}\\
	\end{tabular}
	\caption{Statistical fault localization~\cite{jones2002visualization}. The \texttt{middle()} function takes three values and returns the one that is neither the single smallest nor the single largest one; execution of Line~6 correlates most with the failure.}
	\label{fig:middle-example}
\end{wrapfigure}

As a classic example, consider \Cref{fig:middle-example}, used by its authors to demonstrate the \TARANTULA{} localization~\cite{jones2002tarantula}.
The \texttt{middle()} function returns neither the minimum nor the maximum of its arguments \texttt{x}, \texttt{y}, and~\texttt{z}.
However, \texttt{middle()} is faulty, as \texttt{middle(2, 1, 3)} returns \texttt{1} rather than \texttt{2}.

\TARANTULA{} runs a suite of tests on the program, correlating the failures with the execution of code lines.
Given the four test cases from~\cite{jones2002tarantula}, Line~6 is executed only in the one failing test, which strongly correlates with the failure.
(Line~6 also \emph{is} the faulty line.)
However, the effectiveness of line-based fault localization depends heavily on the test cases:
Adding a test \texttt{middle(1, 1, 3)} breaks the strong correlation---Line~6 is executed, but the test returns \texttt{1}, passing.

The advantage of line coverage is that it is universal—pretty much any programming language provides a means to measure it.
The coverage of a line, however, is just one of many features of a program execution.
Besides line coverage, other features could also correlate with failures---such as variable values, definition-use pairs (i.e., pairs of locations in which values are defined and later used), branch conditions, and more.

Moreover, debugging the locations alone may not be sufficient to understand the root cause of a failure from a developer's perspective.
A developer relies on questions that can not be answered by a single line of code, such as:
\begin{enumerate*}[label=(\arabic*)]
    \item Why did the failure occur? 
    \item What led to the failure? 
    \item What contributed to the failure?
\end{enumerate*}

Recent toolkits allow retrieving all of these features from a program's execution, thus providing a much more ideal choice of execution features to use for fault localization.
The \SFLKIT{}~\cite{smytzek2022sflkit} toolkit for Python programs, among many other execution features, allows extracting \emph{scalar pairs} that occurred during execution---that is, relationships between two variables.
Aided by \SFLKIT{} on the \texttt{middle()} example, we could extract a feature that the test fails if $\texttt{y} < \texttt{x}$ hold.
However, this feature is only one of many that could be extracted from the execution.

Do various execution features work better than lines to characterize failures or locate faults? Moreover, which execution features would these be?
To answer these questions, we conduct an empirical study to determine \emph{which execution features are most relevant to failures.}
During program execution, we collect 17~different execution features inspired by test coverage criteria---from branches covered to definition-use pairs and many more---and assess how effective these features are in detecting and localizing faults.
Our study encompasses 310~faults from 20~open-source projects from the \TESTS4PY{} dataset~\cite{smytzek2024tests4py}, which is based on the faults in \BUGSINPY{}~\cite{widyasari2020bugsinpy}.

To the best of our knowledge, this is the first study to analyze and compare the interplay of a large set of execution features with software failures.
Our findings suggest that fault localization \emph{should rely on a larger and more diverse range of execution features:}
\begin{enumerate}[topsep=5pt]
    \item Features that capture data \emph{and} control flow, such as scalar pairs, showed the highest correlation with failures;
    \item Besides lines executed, \emph{def-use pairs} and \emph{functions executed} are the best features for localizing faults; and
    \item Considering \emph{several features} yields better fault localization than executed lines alone.
\end{enumerate}
Insights such as these can
\begin{enumerate*}[label=(\arabic*)]
 \item help in identifying root causes of program failures more accurately,
 \item provide better hints for automated code repairs, and
 \item open up new possibilities for debugging techniques and research.
\end{enumerate*}

Building on these insights, we introduce a novel debugging approach that learns relevant execution features and generates diagnoses to explain the underlying causes of failures.
We leverage the investigated execution features to train a \emph{classifier} for predicting failures based on the collected features.
In the case of \texttt{middle()} and the executions in \Cref{fig:middle-example}, for instance, the model determined by our execution-feature-driven debugging (or short \EFDD{}) approach predicts a failure if two properties hold:
\begin{enumerate*}
    \item The \texttt{return y} in Line~6 is executed, and
    \item $y$ is less than $x$.
\end{enumerate*}
These two conditions form an accurate \emph{diagnosis}---the failure occurs precisely under these conditions that a developer can understand and act upon, i.e., the fix needs to be in Line~6, which returns $y$, but $y$ is less than $x$, and from the previous Line~5 we know that $x$ is less than $z$, hence $x$ should be returned.

Note that our debugging approach does not need a specification of correct behavior (which, even for a seemingly simple function like \texttt{middle()}, is not straightforward).
Nor does it assume a \emph{generic} failure such as a crash (for which the execution feature that most correlates with failure is obvious.\footnote{It is the crash.}).
Moreover, our approach is not limited to specific kinds of tests, such as unit tests, but works with any test suite that provides labeled outcomes, such as inputs for which we know correct and incorrect behavior.
To the best of our knowledge, this is the first approach that learns relevant features based on the execution and generates fault diagnoses that pinpoint faulty locations and explain the underlying causes of failures.

Overall, we make the following contributions:
\begin{description}[leftmargin=5pt,noitemsep,partopsep=0pt,topsep=0pt,parsep=0pt]
 \item[An \emph{empirical study} assessing how execution features relate to failures.] We conduct an \emph{empirical study} to investigate the effectiveness of the various execution features in detecting and localizing faults, which guides future research in this area and can inspire novel debugging ideas.
 \item[An investigation on \emph{combinations of features}.] We investigate how collecting and leveraging \emph{several features} from program executions can improve fault localization.
 This way, we can identify the \emph{features that are most relevant to failures} instead of relying on one feature (say, line coverage) only.
 \item[A comprehensive \emph{dataset}.] We provide a dataset of faults and executions, i.e., all events of a corresponding execution, to enable further research.
 \item[A debugging approach] We present an execution-feature-driven-debugging approach to produce failure diagnoses that relate failures to \emph{fine-grained} execution features such as individual lines and variable values, which can \emph{precisely pinpoint failure conditions}.
 The diagnoses produced by our debugging approach come as \emph{explicit conditions} that \emph{explain} the failure and can be assessed by humans.
\end{description}

The remainder of this paper is organized as follows.
\Cref{sec:execution-features} describes the execution features analyzed in our study.
Additionally, this section introduces the concept of using multiple execution features for fault localization.
\Cref{sec:study} details our study's design for evaluating these features' effectiveness in fault detection and localization.
This section concludes with a presentation of our study's results and implications.
After introducing execution features and their value for debugging, we present a debugging approach that leverages these features to generate fault diagnoses in \Cref{sec:debugging}.
\Cref{sec:evaluation} evaluates the effectiveness of our debugging approach in generating fault diagnoses.
We address potential threats to validity in \Cref{sec:threats-to-validity} and review related work in \Cref{sec:related-work}.
Finally, \Cref{sec:conclusion} concludes the paper and suggests avenues for future research.
All data and tools developed for this study are available as open source (\Cref{sec:tool}).

\section{Execution Features}%
\label{sec:execution-features}

Our study analyzes program executions to identify features most pertinent to failures, aiming to capture behavioral nuances of the program during execution.
We collect various data points—such as executed lines, variable states, method calls, and condition outcomes—recorded as \emph{events} across different execution instances.
These events encapsulate control and data flow, providing insight into the program's operational context.

From these events, we derive \emph{features}, representing whether specific behaviors occurred during execution or if conditions relevant to failure were met.
By statistically analyzing these features, we identify those most strongly associated with failure, extending beyond the typical scope of statement coverage used in traditional fault localization.

\subsection{Statistical Fault Localization}%
\label{sub:statistical-fault-localization}

Statistical fault localization techniques aim to pinpoint code locations most associated with failures by correlating program execution data with the occurrence of failures.
Many of these techniques rely on \emph{spectra}-based information, typically as the line or block coverage, to identify suspect areas.
For instance, the \TARANTULA{} tool~\cite{jones2002tarantula} calculates the \emph{suspiciousness} of each line based on its coverage across failing and passing test cases, with lines more frequently executed by failing tests flagged as potential fault sites.
Beyond line coverage, some techniques expand to other execution details. Liblit et al.~\cite{liblit2005sd} focus on \emph{branches}, identifying branches that appear disproportionately in failing test cases.
Similarly, Santelices et al.~\cite{santelices2009defuse} introduce \emph{definition-use pairs} as a data flow feature for fault localization, highlighting variable definitions and uses that correlate with failure.
Although these methods improve diagnostic granularity, they typically lack direct mappings to specific fault locations in the code, limiting their practical diagnostic utility.
Traditional methods also rely on specific metrics, such as \TARANTULA{}~\cite{jones2002tarantula}, \OCHIAI{}~\cite{abreu2006ochiai}, and \JACCARD{}~\cite{chen2002jaccard}, which calculate line suspiciousness by comparing the presence of features across failing and passing executions.
While these metrics have shown utility, they primarily operate on executed lines or specific statements, potentially overlooking richer data sources within the program execution.

In our study, we expand statistical fault localization by applying suspiciousness metrics to a diverse set of \emph{execution features} beyond line and branch coverage alone.
By integrating features like data flow events, variable values, and condition outcomes, we aim to capture a more comprehensive view of program behavior that could more accurately reflect fault locations.
These broader investigation targets of our study enable us to assess whether a feature-centric perspective, rather than a line-centric one, can yield improved fault localization outcomes.

\subsection{Events}%
\label{sub:events}

During execution, we capture events that reflect the program's behavior.
These events are categorized using standard test coverage metrics and further enriched to support broader fault analysis.
This collection process involves instrumenting a faulty program, allowing us to obtain a detailed execution trace during any program run.
The primary types of captured events are outlined below:

\begin{description}[leftmargin=5pt,noitemsep,partopsep=0pt,topsep=0pt,parsep=0pt]
    \item[Lines.]
 We monitor the execution of each program line, recording every executed line as an event in the trace.
 This fine-grained detail helps pinpoint fault locations and understand execution flow.
    
    \item[Branches.]
 Whenever a conditional branch is executed, it is logged in the trace.
 This event provides insights into the program's decision-making process and helps detect logical errors.
    
    \item[Functions.]
 All function calls, including their names and passed parameters, are logged.
 These calls offer a detailed view of the program's modular structure and function interactions.
    
    \item[Variable Definitions.]
 An event is generated whenever a variable is defined, capturing its name and value. 
This event helps to understand the program's state evolution and identify issues related to variable initialization and scope.
    
    \item[Variable Uses.]
 Events are recorded when variables are referenced or manipulated, allowing for better analysis of variable utilization and detection of incorrect variable usage.
    
    \item[Return Values.]
 Function return values are logged to enhance the trace, enabling a deeper understanding of function outcomes and their influence on execution.
 This tracking is crucial for diagnosing faults related to incorrect returns.
    
    \item[Loops.]
 We track the frequency of loop executions, capturing the number of iterations and controlling conditions.
 This frequency helps identify infinite loops and incorrect loop conditions.
    
    \item[Conditions.]
 The evaluation results of branch conditions, including nested and compound conditions, are recorded.
 This event aids in analyzing complex decision-making processes and detecting logical errors.
\end{description}

This comprehensive instrumentation ensures that each program execution—whether from a unit test or an arbitrary input—generates a detailed trace of these defined events.
After instrumentation, we iteratively execute the provided test cases to assemble an event trace for each.
Given that test cases include labels indicating their ``pass'' or ``fail'' status, we can differentiate between passing and failing traces.

By capturing a wide range of events, we gain a holistic view of the program's execution, which enhances fault localization and diagnosis.
This detailed event trace forms the foundation of our subsequent analysis, particularly in our debugging approach, where we leverage machine learning to identify patterns and correlations indicative of faults.
Understanding the exact sequence of events leading to a fault enables developers to pinpoint root causes and implement appropriate fixes efficiently.

\subsection{Features}%
\label{sub:features}

We derive \emph{features} from the event traces that capture meaningful execution behaviors.
These features indicate the program's behavior under both normal and failing conditions.
Inspired by statistical fault localization research~\cite{jones2002tarantula,jones2005tarantula,liblit2005sd,santelices2009defuse}, we construct a total of 17~features, which can be categorized into four key groups: coverage information, value properties, condition outcomes, and triggered exceptions.

\begin{description}[leftmargin=5pt,noitemsep,partopsep=0pt,topsep=0pt,parsep=0pt]
 \item[Coverage.]
 We focus on line execution, as suggested in spectrum-based fault localization~\cite{jones2002tarantula}, and also consider specific branches taken during execution~\cite{liblit2005sd}.
 Additionally, we capture definition-use pairs as proposed by Santelices et al.~\cite{santelices2009defuse}.
 Loop coverage is assessed by tracking whether a loop was skipped, entered once, or executed multiple times.

 \item[Values.]
 We extract features based on logged \emph{values} in events, such as variable assignments and function return values.
 Examples include detecting whether a number is zero, a variable is \emph{NULL}, or a return value contains non-ASCII characters.
 We also compare values to other variables to create \emph{scalar pairs}, an approach influenced by Liblit et al.~\cite{liblit2005sd}.
 Furthermore, for variables with lengths, such as strings or lists, we classify their length as zero, one, or more than one.

 \item[Conditions.]
 Since \emph{branch conditions} are already recorded in our events, they are directly incorporated as features.

 \item[Exceptions.]
 \emph{Exceptions} play a critical role in program control flow and can hint at fault occurrences.
 Hence, we include a feature that records whether a function exited normally or with an exception.
\end{description}

A feature is considered \emph{satisfied} for a specific execution if the related event was triggered and the feature's defined condition was met during execution.
For instance, a feature monitoring the execution of a particular line is satisfied if that line was executed within the program run.
Similarly, a feature based on a variable's value is satisfied if the variable was defined during execution and met a specified condition, such as an integer value less than zero.
Once derived from events, these features are analyzed using statistical fault localization techniques to assess their relevance to failures.
We identify features most strongly associated with faults by correlating feature satisfaction across individual test cases with failure occurrences.
This correlation helps pinpoint execution characteristics that indicate likely fault locations.

Our feature selection is inspired by the extensive set of events, spectra, and predicates available in \SFLKIT{}~\cite{smytzek2022sflkit}, providing a robust foundation for constructing and evaluating feature conditions for fault localization.

As an example, consider the \texttt{middle()} function from \Cref{fig:middle-example}.
\Cref{fig:features-example} illustrates the features derived from executing this function.
To maintain conciseness, we collect executed lines, branches, definition-use pairs, scalar pairs, and conditions instead of all possible features.
For instance, a scalar pair feature is collected when a variable is defined and constructed by comparing the variable to another variable, e.g., \texttt{y < x}, for all compatible types.
This feature holds if the comparison evaluates to \texttt{True} and does not hold if it evaluates to \texttt{False}.
If a feature is not recorded for an execution, we assign a value indicating its absence.
Similarly, a definition-use pair feature is collected when a variable is used, incorporating its definition and usage, e.g., \texttt{y} defined in line~1 and used in line~3.
More straightforward features, such as line execution, branch outcomes, or condition evaluations, are collected whenever the corresponding event occurs, e.g., execution of line~2 or evaluation of \texttt{y < z}.
Again, absent features are assigned values indicating their non-occurrence.

Later in our debugging approach, we transform these features into binary and tertiary representations.
A binary feature either holds ($1$) or does not hold ($0$) during a run, while a tertiary feature can hold ($1$), not hold ($0$), or be unobserved ($-1$).
For example, line execution is a binary feature, as it can only be executed or not executed. In contrast, condition evaluation is tertiary, as it can evaluate to \texttt{true}, \texttt{false}, or remain unobserved.

Since not all runs include every feature, we assign default values during feature vector construction to normalize the dataset.
The default value for binary features is \texttt{false}, while for features with three states, it is \texttt{not observed}.
This standardization ensures that feature vectors maintain consistent size and structure across all runs.

\begin{figure}
    \def\ind{\qquad}
    \centering
    \resizebox{\columnwidth}{!}{%
    \begin{tabular}{@{}>{\scriptsize}r>{\tt\color{Blue}}l@{\quad}l@{\ \ }>{\raggedright}p{.3\columnwidth}@{\ \ }>{\raggedright}p{.3\columnwidth}@{\ \ }>{\raggedright}p{.3\columnwidth}@{\ \ }>{\scriptsize}r}
      & &\color{Blue}\texttt{x}& \multicolumn{1}{c}{1} &  \multicolumn{1}{c}{3} &  \multicolumn{1}{c}{2} & \\
      & &\color{Blue}\texttt{y}& \multicolumn{1}{c}{2} &  \multicolumn{1}{c}{1} &  \multicolumn{1}{c}{1} & \\
      & &\color{Blue}\texttt{z}& \multicolumn{1}{c}{3} &  \multicolumn{1}{c}{2} &  \multicolumn{1}{c}{3} & \\
    1 & \textbf{\color{deeporange}def} middle(x, y, z):     & 
    & \multicolumn{3}{@{}>{\raggedright}p{\columnwidth}}{\scriptsize ScalarPair(y<x), ScalarPair(y>x), ScalarPair(y==x), ScalarPair(y<=x), ScalarPair(y>=x), ScalarPair(y!=x), 
    ScalarPair(z<x), ScalarPair(z>x), ScalarPair(z==x), ScalarPair(z<=x), ScalarPair(z>=x), ScalarPair(z!=x),
    ScalarPair(z<y), ScalarPair(z>y), ScalarPair(z==y), ScalarPair(z<=y), ScalarPair(z>=y), ScalarPair(z!=y)} 
    & 1 \\ %\cline{4-8}
    2 & \ind  \textbf{\color{deeporange}if} y < z:          & 
    & \scriptsize Line(2), DefUse(x,1,2), DefUse(z,1,2), Condition(y<z)
    & \scriptsize Line(2), DefUse(x,1,2), DefUse(z,1,2), Condition(y<z)
    & \scriptsize Line(2), DefUse(x,1,2), DefUse(z,1,2), Condition(y<z)
    & 2 \\ %\cline{3-8}
    3 & \ind \ind  \textbf{\color{deeporange}if} x < y:     & 
    & \scriptsize Line(3), Branch(1), DefUse(x,1,3), DefUse(y,1,3), Condition(x<y)
    & \scriptsize Line(3), Branch(1), DefUse(x,1,3), DefUse(y,1,3), Condition(x<y)
    & \scriptsize Line(3), Branch(1), DefUse(x,1,3), DefUse(y,1,3), Condition(x<y)
    & 3 \\ %\cline{3-8}
    4 & \ind \ind \ind  \textbf{\color{deeporange}return} y & 
    & \scriptsize Line(4), Branch(3), DefUse(y,1,4) 
    & \scriptsize 
    & \scriptsize 
    & 4 \\ %\cline{3-8}
    5 & \ind \ind  \textbf{\color{deeporange}elif} x < z:   & 
    & \scriptsize 
    & \scriptsize Line(5), Branch(4), DefUse(x,1,5), DefUse(y,1,5), Condition(x<z)
    & \scriptsize Line(5), Branch(4), DefUse(x,1,5), DefUse(y,1,5), Condition(x<z)
    & 5 \\ %\cline{3-8}
    6 & \ind \ind \ind  \textbf{\color{deeporange}return} y & 
    & \scriptsize 
    & \scriptsize 
    & \scriptsize Line(6), Branch(5), DefUse(y,1,6) 
    & 6 \\ %\cline{3-8}
    7 & \ind  \textbf{\color{deeporange}elif} x > y:        & & \scriptsize & \scriptsize & \scriptsize & 7 \\ %\cline{3-8}
    8 & \ind \ind  \textbf{\color{deeporange}return} y      & & \scriptsize & \scriptsize & \scriptsize & 8 \\ %\cline{3-8}
    9 & \ind  \textbf{\color{deeporange}elif} x > z:        & & \scriptsize & \scriptsize & \scriptsize & 9 \\ %\cline{3-8}
    10 & \ind \ind  \textbf{\color{deeporange}return} x     & & \scriptsize & \scriptsize & \scriptsize & 10 \\ %\cline{3-8}
    11 & \ind  \textbf{\color{deeporange}return} z          & 
    & \scriptsize 
    & \scriptsize Line(11), Branch(6), DefUse(z,1,11) 
    & \scriptsize 
    & 11 \\ %\cline{3-8}
    & & & \multicolumn{1}{c}{\PASS} & \multicolumn{1}{c}{\PASS} & \multicolumn{1}{c}{\FAIL} & \\
    \end{tabular}%
    }%
    \caption{Collecting of Features. 
    For the \texttt{middle()} function, we derive features from the execution. For example, we show the collection of executed lines, branches, definition-use pairs, scalar pairs, and conditions.
    We show when a feature is collected by assigning it to the corresponding line in the code.
    The values in parentheses indicate the parameters of a feature; for lines, the line number; for branches, the branch ID; for definition-use pairs, the variable and the lines for the definition and use; for scalar pairs, the two variables, and the applied comparison; for conditions, the actual condition.
    }%
    \label{fig:features-example}
\end{figure}

\subsection{Soundness and Completeness of Features}%
\label{sub:soundness-completeness}

\subsubsection*{Soundness}
Our feature selection is designed to be sufficiently detailed to represent executions accurately.
By ensuring that each set of features is unique to its corresponding execution or closely similar, we avoid incorrect outcomes, thereby affirming the soundness of our features in distinguishing between passing and failing runs.
Soundness is crucial because it guarantees that the features we select are reliable indicators of the execution's behavior.
This reliability is essential for debugging and optimizing the system. 
It ensures that the features accurately represent the underlying processes and reduces false positives and negatives, leading to more accurate insights.

\vspace{-.5\baselineskip}
\subsubsection*{Completeness.}
While we strive for completeness by covering a broad range of general scenarios, there are instances where our features may fall short.
For example, consider a fault arising from a function that should have been executed.
Our current features might not detect such faults because they focus on observable effects rather than potential omissions.
Although we may pick up indirect signs of this fault, these indicators are not always definitive but may be sufficient to identify the fault.
However, we cannot guarantee the latter scenario.
Enhancing our feature set to capture such scenarios is possible, but some unique situations will inevitably still challenge our feature coverage's completeness.
Completeness is vital because it ensures that all relevant aspects of the execution are considered, providing a holistic view of the system's behavior.
This comprehensive coverage is necessary to identify and address all potential issues, improving the system's robustness and reliability.
To achieve greater completeness, we can incorporate additional features that capture a more comprehensive range of execution behaviors, including those that are less common or more subtle.
However, it is essential to balance the need for completeness with the practical constraints of feature selection, such as computational overhead and data availability.

\subsection{Why Using a Single Feature?}%
\label{sub:multi-feature-technique}

While each feature could be leveraged to localize faults, combining these features that highlight their benefits while mitigating their drawbacks could provide a more comprehensive fault localization.
Hence, we propose to leverage \emph{multi-feature techniques}, similar to Santelices et al.~\cite{santelices2009defuse}, that leverages these derived features and combines the strengths of various features considering the different information sources and their relation to failures to localize faults more accurately and precisely.
These techniques aim to verify whether combining multiple features can provide a more comprehensive fault localization than relying on a single feature alone.

We have implemented such an approach that evaluates all features simultaneously, calculating the correlation of each based on a provided coefficient, e.g., \TARANTULA{}, which considers the number of passed and failed test cases that satisfy the feature.
We then order the features based on their suspiciousness scores, identifying the most relevant features for the failure to occur.
Each feature corresponds to a specific code location, such as a single line or a block of lines, allowing us to assign a suspiciousness score to each line.
A challenge arises when multiple features correspond to the same line, necessitating a method for combining their suspiciousness scores.
We provide an interface in our technique to utilize various metrics to integrate the suspiciousness of multiple features mapping to the same line.
Intuitively, we can select among metrics such as maximum, minimum, median, or average suspiciousness to determine the final score for each line in the program, where we consider mainly the maximum suspiciousness as the metric of choice.
After calculating the suspiciousness scores for each line, we can rank them based on their scores, identifying the lines most likely to contain the root cause of the failure as traditional fault localization techniques do.

%%%%
%%%%
%%%%

\section{Empirical Study}%
\label{sec:study}

\subsection{Research Questions}%
\label{sub:research-questions}

In our empirical study, we aim to investigate the effectiveness of the presented execution features in detecting and localizing faults.
Hence, we formulate the following research questions to guide our study:

\begin{questions}[topsep=2pt]
 \item\label[question]{rq1} \textbf{Correlation.} To what degree does each execution feature correlate with the presence of failures?
 \item\label[question]{rq2} \textbf{Localization.} How well does each execution feature localize the faults in the program?
 \item\label[question]{rq3} \textbf{Multi-Features.} Can we combine execution features to provide a more comprehensive fault localization?
\end{questions}

In the traditional evaluation of fault localization, the faulty lines are leveraged to assess the effectiveness of a technique, which comes from a fixed version of the program that may not be the only possible fix for the fault, which may lead to a bias in the evaluation.
To address this issue, we introduced \Cref{rq1} that shifts the focus to the correlation of features with faults as an initial assessment of the features' effectiveness in identifying the faults' presence.

\subsection{Subjects}%
\label{sub:subjects}

Our study utilizes open-source projects from the \TESTS4PY{} dataset~\cite{smytzek2024tests4py}, a curated collection of Python projects with documented faults and corresponding test cases.
Each project in this dataset includes a faulty and a fixed version, where the faults have been addressed.
To ensure fault validity, we first executed the provided test cases on the faulty versions, confirming that the expected failures occurred and that each fault had at least one passing test case.
This passing test case is essential for calculating meaningful correlations in fault localization.
We then verified the fixed versions by rerunning the test cases, confirming that all tests passed and the faults were correctly resolved.
Through this process, we successfully reproduced failures and validated test case accuracy for 310~subjects across 20~projects in the dataset.
These 310~subjects were selected for our empirical analysis as they provide a reliable and reproducible basis for evaluating fault localization techniques.

\subsection{Collecting the Data}%
\label{sub:data-collection}

For each project, we collected the trace of events during the execution of the test cases for the faulty version of the program.
\TESTS4PY{} provides a set of test cases for each project relevant to the failure.
This set includes test cases that trigger the failure and those that explore related functionality without triggering the bug, such as executing the same function with different parameters.
We ensured that all test case executions were correctly categorized, confirming that the test cases expected to fail did indeed fail.

\subsection{Analysis}%
\label{sub:analysis}

To address \Cref{rq1}, we conducted a statistical analysis of the collected execution features.
We computed the correlation of each feature with the presence of failures using \TARANTULA{}~\cite{jones2002tarantula}, \OCHIAI{}~\cite{abreu2006ochiai}, \DSTAR{}~\cite{wong2012dstar}, \NAISHT{}~\cite{naish2011sbfl}, and \GPOT{}~\cite{xie2013gp} coefficients across all feature types (e.g., lines, branches).
For each feature type, we calculated four metrics:
\begin{enumerate*}[label=(\arabic*)]
    \item the highest correlation among features,
    \item the mean correlation of all features,
    \item the median correlation of all features, and
    \item the lowest correlation among features.
\end{enumerate*}
These metrics provide a comprehensive view of the suspiciousness of how likely each feature type correlates with the presence of failures.
For instance, we have line coverage features for two subjects.
For the first subject, Lines 1, 2, 3, and 4 have correlations of 0.8, 0.6, 0.4, and 0.2, respectively.
For the second subject, Lines 1, 2, and 3 have correlations of 0.7, 0.5, and 0.3.
To determine the average correlation metrics for the line feature class:
the highest average correlation is $(0.8 + 0.7) / 2 = 0.75$;
the mean correlation across all features is $((0.8 + 0.6 + 0.4 + 0.2) / 4 + (0.7 + 0.5 + 0.3) / 3) / 2 = 0.5$;
the median correlation is $(0.5 + 0.5) / 2 = 0.5$; and the lowest average correlation is $(0.2 + 0.3) / 2 = 0.25$.

Additionally, we calculate the actual correlation between the features and the presence of failures by relating the number of failing runs for which a feature is satisfied to the total number of runs for which the feature is satisfied. This relation will show how well the feature correlates with the presence of failures, i.e., if a feature is primarily satisfied if the run is failing, which we will evaluate using Spearman's rank correlation coefficient.
Similar to the similarity coefficients, we calculate the overall correlation (i.e., considering all subjects at once) for each feature type along with the highest, mean, median, and lowest correlation when considering each subject individually.

We evaluated fault localization for \Cref{rq2} by applying the same coefficients to each feature and ranking suggested lines based on faulty lines extracted from patch files in the \TESTS4PY{} dataset.
While these patch files may not exclusively contain lines that directly address the fault (since fixes may involve code outside the directly faulty lines), we chose this approach to remain consistent with established methods in prior work~\cite{jones2002tarantula,jones2005tarantula,abreu2006ochiai,widyasari2022sfl,pearson2017sfl,wong2007wong,landsberg2015sfl}. 
Moreover, the analysis of \Cref{rq1} mitigates the potential bias introduced using the patch files for fault localization.

Following Widyasari et al.~\cite{widyasari2022sfl} and Pearson et al.~\cite{pearson2017sfl}, we assessed localization performance using the top-1, top-5, top-10, and top-200 lines suggested by each coefficient~\cite{le2015topk}.
We calculated the \EXAM{} score~\cite{wong2008crosstab} and \emph{wasted effort}~\cite{xuan2014test} for each feature type, guided by evaluation methods used in previous studies~\cite{just2018sfl,long2016search}.
The \EXAM{} score for a statement $s$ is defined as:
\begin{equation}
 \text{\EXAM{}}_s = \frac{\text{rank}(s)}{\text{Total number of statements}}
\end{equation}
\begin{wraptable}{l}{.61\textwidth}
    \caption{Suspiciousness and Correlation. 
    The calculated suspiciousness for each metric \TARANTULA{}, \OCHIAI{}, \DSTAR{}, \NAISHT{}, and \GPOT{} and the correlation. Highlighted values represent the best results when compared to the other features. Underlined values for the correlation represent statistical significance ($p < 0.05$).}%
    \label{tab:results-correlation-1}
    \centering
    \setlength\extrarowheight{-3pt}
    \resizebox{.6\textwidth}{!}{%
    \begin{tabular}{llrrrrrrrrr}
    \toprule
    \multicolumn{1}{c}{\multirow{2}*{Feature}} & \multicolumn{1}{c}{\multirow{2}*{Metric}} & \multicolumn{4}{c}{Suspiciousness} & \multicolumn{5}{c}{Correlation} \\\cmidrule(lr){3-6}\cmidrule(lr){7-11}
    & & \multicolumn{1}{c}{Best} & \multicolumn{1}{c}{Mean} & \multicolumn{1}{c}{Median} & \multicolumn{1}{c}{Worst}& \multicolumn{1}{c}{Overall} & \multicolumn{1}{c}{Best} &  \multicolumn{1}{c}{Mean} & \multicolumn{1}{c}{Median} & \multicolumn{1}{c}{Worst} \\\midrule
     & \TARANTULA{}\rowstrut{} & 0.930 & \textbf{\color{deepblue}0.387} & \textbf{\color{deepblue}0.412} & 0.010 &  &  &  &  & \\
     & \OCHIAI{} & 0.809 & \textbf{\color{deepblue}0.246} & \textbf{\color{deepblue}0.261} & 0.013 &  &  &  &  & \\
    Lines & \DSTAR{} & 8.009 & 0.797 & \textbf{\color{deepblue}0.708} & 0.019 & \underline{0.473} & \underline{1.000} & \underline{0.791} & \underline{0.836} & \textbf{\color{deepblue}\underline{0.240}}\\
     & \NAISHT{} & 1.712 & \textbf{\color{deepblue}0.630} & \textbf{\color{deepblue}0.742} & -0.577 &  &  &  &  & \\
     & \GPOT{} & 2.504 & \textbf{\color{deepblue}1.372} & \textbf{\color{deepblue}1.552} & 0.032 &  &  &  &  & \\[.2em]
\rowcolor{row}
     & \TARANTULA{}\rowstrut{} & 0.893 & 0.179 & 0.088 & 0.000 &  &  &  &  & \\
\rowcolor{row}
     & \OCHIAI{} & 0.755 & 0.123 & 0.073 & 0.000 &  &  &  &  & \\
\rowcolor{row}
    Branches & \DSTAR{} & 7.923 & 0.455 & 0.156 & 0.000 & \underline{0.624} & \textbf{\color{deepblue}\underline{1.000}} & \underline{0.673} & \underline{0.757} & \underline{0.064}\\
\rowcolor{row}
     & \NAISHT{} & 1.628 & 0.276 & 0.216 & -0.551 &  &  &  &  & \\
\rowcolor{row}
     & \GPOT{} & 2.399 & 0.603 & 0.383 & 0.000 &  &  &  &  & \\[.2em]
     & \TARANTULA{}\rowstrut{} & 0.863 & 0.300 & 0.305 & \textbf{\color{deepblue}0.025} &  &  &  &  & \\
     & \OCHIAI{} & 0.708 & 0.207 & 0.204 & \textbf{\color{deepblue}0.026} &  &  &  &  & \\
    Functions & \DSTAR{} & 7.786 & \textbf{\color{deepblue}0.805} & 0.268 & \textbf{\color{deepblue}0.034} & \underline{0.504} & \textbf{\color{deepblue}\underline{1.000}} & 0.615 & \underline{0.713} & \underline{0.222}\\
     & \NAISHT{} & 1.596 & 0.467 & 0.451 & -0.496 &  &  &  &  & \\
     & \GPOT{} & 2.319 & 1.023 & 1.039 & \textbf{\color{deepblue}0.084} &  &  &  &  & \\[.2em]
\rowcolor{row}
     & \TARANTULA{}\rowstrut{} & 0.570 & 0.025 & 0.000 & 0.000 &  &  &  &  & \\
\rowcolor{row}
     & \OCHIAI{} & 0.501 & 0.022 & 0.000 & 0.000 &  &  &  &  & \\
\rowcolor{row}
    Function Errors & \DSTAR{} & 3.153 & 0.175 & 0.000 & 0.000 & \underline{0.679} & \textbf{\color{deepblue}\underline{1.000}} & 0.484 & \underline{0.524} & \underline{0.173}\\
\rowcolor{row}
     & \NAISHT{} & 0.959 & 0.051 & -0.000 & -0.147 &  &  &  &  & \\
\rowcolor{row}
     & \GPOT{} & 1.537 & 0.082 & 0.000 & 0.000 &  &  &  &  & \\[.2em]
     & \TARANTULA{}\rowstrut{} & 0.929 & 0.278 & 0.266 & 0.009 &  &  &  &  & \\
     & \OCHIAI{} & 0.823 & 0.197 & 0.184 & 0.011 &  &  &  &  & \\
    Def-Use Pairs & \DSTAR{} & 8.100 & 0.729 & 0.280 & 0.016 & \underline{0.520} & \textbf{\color{deepblue}\underline{1.000}} & \underline{0.617} & \underline{0.679} & \underline{0.096}\\
     & \NAISHT{} & 1.695 & 0.451 & 0.417 & -0.586 &  &  &  &  & \\
     & \GPOT{} & 2.508 & 0.960 & 0.935 & 0.027 &  &  &  &  & \\[.2em]
\rowcolor{row}
     & \TARANTULA{}\rowstrut{} & 0.685 & 0.112 & 0.011 & 0.000 &  &  &  &  & \\
\rowcolor{row}
     & \OCHIAI{} & 0.527 & 0.076 & 0.008 & 0.000 &  &  &  &  & \\
\rowcolor{row}
    Loops & \DSTAR{} & 6.601 & 0.293 & 0.018 & 0.000 & \underline{0.620} & \textbf{\color{deepblue}\underline{1.000}} & 0.569 & \underline{0.696} & \underline{0.116}\\
\rowcolor{row}
     & \NAISHT{} & 1.275 & 0.159 & 0.013 & -0.412 &  &  &  &  & \\
\rowcolor{row}
     & \GPOT{} & 1.880 & 0.356 & 0.028 & 0.000 &  &  &  &  & \\[.2em]
     & \TARANTULA{}\rowstrut{} & 0.888 & 0.175 & 0.078 & 0.000 &  &  &  &  & \\
     & \OCHIAI{} & 0.756 & 0.121 & 0.064 & 0.000 &  &  &  &  & \\
    Conditions & \DSTAR{} & 7.931 & 0.453 & 0.139 & 0.000 & \underline{0.607} & \textbf{\color{deepblue}\underline{1.000}} & \underline{0.662} & \underline{0.749} & \underline{0.047}\\
     & \NAISHT{} & 1.621 & 0.274 & 0.197 & -0.550 &  &  &  &  & \\
     & \GPOT{} & 2.409 & 0.593 & 0.346 & 0.000 &  &  &  &  & \\[.2em]
\rowcolor{row}
     & \TARANTULA{}\rowstrut{} & \textbf{\color{deepblue}0.952} & 0.184 & 0.043 & 0.000 &  &  &  &  & \\
\rowcolor{row}
     & \OCHIAI{} & \textbf{\color{deepblue}0.875} & 0.120 & 0.039 & 0.000 &  &  &  &  & \\
\rowcolor{row}
    Scalar Pairs & \DSTAR{} & \textbf{\color{deepblue}8.222} & 0.473 & 0.044 & 0.000 & \underline{0.845} & \textbf{\color{deepblue}\underline{1.000}} & \underline{0.800} & \underline{0.871} & \underline{0.027}\\
\rowcolor{row}
     & \NAISHT{} & \textbf{\color{deepblue}1.745} & 0.276 & 0.059 & -0.629 &  &  &  &  & \\
\rowcolor{row}
     & \GPOT{} & \textbf{\color{deepblue}2.604} & 0.622 & 0.139 & 0.000 &  &  &  &  & \\[.2em]
\bottomrule
\end{tabular}

    }
\end{wraptable}
where $\text{rank}(s)$ is the rank of the statement $s$ in the list of suggested lines ordered by the suspiciousness of each line.
If multiple lines have the same suspiciousness, we consider the average rank of these lines, calculated as ${\frac{n}{2} + (k - 1)}$~(\cite{widyasari2022sfl,pearson2017sfl,steimann2013threats}), where $n$ is the number of lines with the same suspiciousness as $s$ and $k$ is the position of the first line with the same suspiciousness, i.e., the highest possible rank.
The wasted effort for a line $s$ is defined as the number of lines that need to be inspected to find the target line, starting from the top of the list of suggested lines and including the target line itself and the lines with the same suspiciousness.
We calculated the average over all subjects for each feature type and metric.
Similar to Widyasari et al.~\cite{widyasari2022sfl} and Pearson et al.~\cite{pearson2017sfl} we included three debugging scenarios,
\begin{enumerate*}[label=(\arabic*)]
    \item finding one,
    \item half, and
    \item all faulty lines
\end{enumerate*}
named
\begin{enumerate*}[label=(\arabic*)]
    \item \textit{best-case},
    \item \textit{average-case}, and
    \item \textit{worst-case}
\end{enumerate*}
debugging respectively.

We considered the same similarity coefficients for the localization to answer \Cref{rq3}, which. Still, we calculated the suspiciousness for each line based on the features that map to the line.
We considered two cases for the multi-feature technique: one in which we consider the maximum suspiciousness of a line when considering each feature and one in which we consider the mean suspiciousness of the features that map to the line.

\subsection{Implementation}%
\label{sub:implementation}

We implemented the empirical study using the fault localization tool \SFLKIT{}~\cite{smytzek2022sflkit}, a versatile framework supporting a range of statistical fault localization techniques, including \TARANTULA{}~\cite{jones2002tarantula}, \OCHIAI{}~\cite{abreu2006ochiai}, \DSTAR{}~\cite{wong2012dstar}, \NAISHT{}~\cite{naish2011sbfl}, and \GPOT{}~\cite{xie2013gp}.
\SFLKIT{} provides an interface that allows us to integrate custom execution features into its pipeline, instrumenting the program to capture events during execution and offering the necessary infrastructure to analyze these events and derive features.
We have constructed a workflow that instruments the program, extracts the events, assembles the features, and analyzes their effectiveness for all three research questions while providing intermediate results after each step.
With its intermediate results, this process is ideal for reproducing all our findings and verifying that all steps work as intended.

\begin{wraptable}{r}{.61\textwidth}
    \caption{Suspiciousness and Correlation.
    Continuation of \Cref{tab:results-correlation-1}}%
    \label{tab:results-correlation-2}
    \centering
    \setlength\extrarowheight{-3pt}
    \resizebox{.6\textwidth}{!}{%
    \begin{tabular}{llrrrrrrrrr}
    \toprule
    \multicolumn{1}{c}{\multirow{2}*{Feature}} & \multicolumn{1}{c}{\multirow{2}*{Metric}} & \multicolumn{4}{c}{Suspiciousness} & \multicolumn{5}{c}{Correlation} \\\cmidrule(lr){3-6}\cmidrule(lr){7-11}
    & & \multicolumn{1}{c}{Best} & \multicolumn{1}{c}{Mean} & \multicolumn{1}{c}{Median} & \multicolumn{1}{c}{Worst}& \multicolumn{1}{c}{Overall} & \multicolumn{1}{c}{Best} &  \multicolumn{1}{c}{Mean} & \multicolumn{1}{c}{Median} & \multicolumn{1}{c}{Worst} \\\midrule
     & \TARANTULA{}\rowstrut{} & 0.801 & 0.181 & 0.076 & 0.000 &  &  &  &  & \\
     & \OCHIAI{} & 0.643 & 0.120 & 0.066 & 0.000 &  &  &  &  & \\
    Variable Values & \DSTAR{} & 6.910 & 0.435 & 0.119 & 0.000 & \underline{0.709} & \textbf{\color{deepblue}\underline{1.000}} & \underline{0.782} & \underline{0.835} & \underline{0.158}\\
     & \NAISHT{} & 1.466 & 0.288 & 0.215 & -0.465 &  &  &  &  & \\
     & \GPOT{} & 2.246 & 0.628 & 0.362 & 0.000 &  &  &  &  & \\[.2em]
\rowcolor{row}
     & \TARANTULA{}\rowstrut{} & 0.827 & 0.141 & 0.036 & 0.000 &  &  &  &  & \\
\rowcolor{row}
     & \OCHIAI{} & 0.670 & 0.097 & 0.037 & 0.000 &  &  &  &  & \\
\rowcolor{row}
    Return Values & \DSTAR{} & 6.890 & 0.312 & 0.038 & 0.000 & \underline{0.650} & \textbf{\color{deepblue}\underline{1.000}} & 0.619 & \underline{0.700} & \underline{0.139}\\
\rowcolor{row}
     & \NAISHT{} & 1.460 & 0.153 & 0.043 & -0.545 &  &  &  &  & \\
\rowcolor{row}
     & \GPOT{} & 2.173 & 0.423 & 0.114 & 0.000 &  &  &  &  & \\[.2em]
     & \TARANTULA{}\rowstrut{} & 0.870 & 0.216 & 0.154 & 0.000 &  &  &  &  & \\
     & \OCHIAI{} & 0.714 & 0.139 & 0.108 & 0.000 &  &  &  &  & \\
    Null Values & \DSTAR{} & 7.610 & 0.368 & 0.193 & 0.000 & \underline{0.767} & \textbf{\color{deepblue}\underline{1.000}} & \textbf{\color{deepblue}\underline{0.841}} & \textbf{\color{deepblue}\underline{0.892}} & \underline{0.234}\\
     & \NAISHT{} & 1.614 & 0.360 & 0.312 & -0.543 &  &  &  &  & \\
     & \GPOT{} & 2.348 & 0.772 & 0.615 & 0.000 &  &  &  &  & \\[.2em]
\rowcolor{row}
     & \TARANTULA{}\rowstrut{} & 0.893 & 0.129 & 0.000 & 0.000 &  &  &  &  & \\
\rowcolor{row}
     & \OCHIAI{} & 0.766 & 0.084 & 0.000 & 0.000 &  &  &  &  & \\
\rowcolor{row}
    Lengths & \DSTAR{} & 7.965 & 0.296 & 0.000 & 0.000 & \underline{0.844} & \textbf{\color{deepblue}\underline{1.000}} & \underline{0.795} & \underline{0.852} & \underline{0.058}\\
\rowcolor{row}
     & \NAISHT{} & 1.642 & 0.196 & 0.000 & -0.572 &  &  &  &  & \\
\rowcolor{row}
     & \GPOT{} & 2.419 & 0.440 & 0.000 & 0.000 &  &  &  &  & \\[.2em]
     & \TARANTULA{}\rowstrut{} & 0.433 & 0.005 & 0.000 & 0.000 &  &  &  &  & \\
     & \OCHIAI{} & 0.279 & 0.004 & 0.000 & 0.000 &  &  &  &  & \\
    Empty Strings & \DSTAR{} & 4.835 & 0.030 & 0.000 & 0.000 & \underline{0.692} & \textbf{\color{deepblue}\underline{1.000}} & 0.500 & \underline{0.578} & \underline{0.214}\\
     & \NAISHT{} & 0.769 & 0.011 & 0.000 & -0.224 &  &  &  &  & \\
     & \GPOT{} & 1.295 & 0.020 & 0.000 & 0.000 &  &  &  &  & \\[.2em]
\rowcolor{row}
     & \TARANTULA{}\rowstrut{} & 0.816 & 0.107 & 0.077 & 0.000 &  &  &  &  & \\
\rowcolor{row}
     & \OCHIAI{} & 0.638 & 0.066 & 0.037 & 0.000 &  &  &  &  & \\
\rowcolor{row}
    ASCII Strings & \DSTAR{} & 7.580 & 0.249 & 0.025 & 0.000 & \underline{0.898} & \textbf{\color{deepblue}\underline{1.000}} & \underline{0.808} & \underline{0.883} & \underline{0.081}\\
\rowcolor{row}
     & \NAISHT{} & 1.513 & 0.146 & 0.020 & -0.467 &  &  &  &  & \\
\rowcolor{row}
     & \GPOT{} & 2.249 & 0.349 & 0.171 & 0.000 &  &  &  &  & \\[.2em]
     & \TARANTULA{}\rowstrut{} & 0.689 & 0.032 & 0.000 & 0.000 &  &  &  &  & \\
     & \OCHIAI{} & 0.508 & 0.019 & 0.000 & 0.000 &  &  &  &  & \\
    Digit Strings & \DSTAR{} & 6.467 & 0.074 & 0.000 & 0.000 & \underline{0.862} & \textbf{\color{deepblue}\underline{1.000}} & 0.713 & \underline{0.816} & 0.000\\
     & \NAISHT{} & 1.278 & 0.046 & 0.000 & -0.334 &  &  &  &  & \\
     & \GPOT{} & 2.039 & 0.106 & 0.000 & 0.000 &  &  &  &  & \\[.2em]
\rowcolor{row}
     & \TARANTULA{}\rowstrut{} & 0.782 & 0.086 & 0.034 & 0.000 &  &  &  &  & \\
\rowcolor{row}
     & \OCHIAI{} & 0.609 & 0.052 & 0.021 & 0.000 &  &  &  &  & \\
\rowcolor{row}
    Special Strings & \DSTAR{} & 7.085 & 0.236 & 0.021 & 0.000 & \textbf{\color{deepblue}\underline{0.915}} & \textbf{\color{deepblue}\underline{1.000}} & \underline{0.803} & \underline{0.868} & \underline{0.100}\\
\rowcolor{row}
     & \NAISHT{} & 1.461 & 0.116 & 0.011 & -0.414 &  &  &  &  & \\
\rowcolor{row}
     & \GPOT{} & 2.222 & 0.279 & 0.076 & 0.000 &  &  &  &  & \\[.2em]
     & \TARANTULA{}\rowstrut{} & 0.061 & 0.000 & 0.000 & 0.000 &  &  &  &  & \\
     & \OCHIAI{} & 0.039 & 0.000 & 0.000 & 0.000 &  &  &  &  & \\
    Empty Bytes & \DSTAR{} & 0.030 & 0.000 & 0.000 & 0.000 & \underline{0.502} & \textbf{\color{deepblue}\underline{1.000}} & 0.065 & 0.000 & 0.000\\
     & \NAISHT{} & 0.074 & 0.000 & 0.000 & \textbf{\color{deepblue}-0.047} &  &  &  &  & \\
     & \GPOT{} & 0.136 & 0.001 & 0.000 & 0.000 &  &  &  &  & \\[.2em]
\bottomrule
\end{tabular}

    }
\end{wraptable}

\subsection{Results}%
\label{sub:study-results}

All summarized results of our empirical study are presented in \Cref{tab:results-correlation-1,tab:results-correlation-2,tab:results-localization-1,tab:results-localization-2,tab:results-localization-multiple}.
\Cref{tab:results-correlation-1,tab:results-correlation-2} show the correlation of the execution features with the presence of failures for the metrics \TARANTULA{}, \OCHIAI{}, \DSTAR{}, \NAISHT{}, \GPOT{}, and the corresponding Spearman's $\rho$ values for a feature.
Moreover, \Cref{tab:results-localization-1,tab:results-localization-2} show the localization of the faults for the metrics \TARANTULA{}, \OCHIAI{}, \DSTAR{}, \NAISHT{}, and \GPOT{} according to top-1, top-5, top-10, top-200, \EXAM{} score, and wasted effort for all features.
Finally, \Cref{tab:results-localization-multiple} shows the results when considering a näive multi-feature technique from \Cref{sub:multi-feature-technique}.

\subsubsection{\bfseries\Cref{rq1}: Correlation with Failures}%
\label{sub:eval-correlation}

To address \Cref{rq1}, we conducted a detailed analysis of the correlation between execution features and the presence of failures, using the \TARANTULA{}, \OCHIAI{}, \DSTAR{}, \NAISHT{}, and \GPOT{} coefficients along with Spearman's $\rho$ values.
Generally, a high suspiciousness score indicates a likely causality between a feature and the presence of failures, while high correlation coefficients suggest a strong relationship.
In other words, a feature is satisfied when a failure occurs, and the failure occurs when the feature is present.
If both values are high, the feature is likely to indicate the presence of failures.
In contrast, if the suspiciousness score is low, the feature is not commonly satisfied when a failure occurs. 
If the correlation coefficient is low, the feature is often satisfied when a failure does not occur.

Our findings reveal that features capturing data flow and data relationships, such as def-use pairs, scalar pairs, variable values, and null values, exhibit exceptionally high correlations with failures.
Additionally, specific features that verify the contents of strings show an increased correlation based on the overall correlation coefficient, while they do not show a general high suspiciousness score.
This discrepancy suggests that these features indicate failures when they occur but are not standard for most failures.

Features representing control flow, including branches and conditions, also show strong correlations, though they appear slightly less indicative than data flow features.
Interestingly, loop-related features correlate with failures only when they capture variations in loop frequency, such as the number of iterations. 

These findings suggest that while control flow is essential, the data flow and relationships within the program data may be more critical in understanding failure correlations.

\begin{result}
 Features that capture data and control flow, significantly \emph{scalar pairs}, are highly correlated with failures.
\end{result}

Additionally, \emph{Lines} show a high correlation with the presence of failures, particularly when considering the best feature of each class, but also on the mean and median of all features, compared with the other feature classes.
While \emph{functions} have slightly lower correlations than lines, they still show significant correlations for the mean and median of all features.
However, these features cover almost the entire program, with exceptionally executed functions corresponding to an entire code block. 
Because of this information overload, they might not be specific enough to pinpoint the causes of a fault.
Note that the fault's cause may differ from the faulty location.
This finding is supported by Spearman's $\rho$ values, which show that lines and functions are less correlated with failures.

\begin{result}
    \emph{Lines} and \emph{functions} show a high suspicion for the occurrence of failures but are not specific enough to accurately detect a fault's causes based on Spearman's $\rho$ values.
\end{result}

While \emph{return values}, \emph{variables values}, \emph{null values}, and \emph{lengths} provide a high suspiciousness, other value-related features---for instance, if a string is empty or consists of digits.
However, \emph{return values} do not show a high correlation with the presence of failures, suggesting that they might not be as relevant to failure occurrences as the variables themselves and their properties, e.g., properties corresponding to well-known failure patterns, for instance, null pointers or index-out-of-bounds.
Specifically to highlight are \emph{special strings}, i.e., strings that contain characters other than digits or letters, which show a higher suspiciousness and an extremely high correlation with the presence of failures.

\begin{result}
    \emph{Variables values}, \emph{null values}, \emph{lengths}, and if a \emph{string contains special characters} highly correlate with the occurrence of failures.
\end{result}

Counterintuitively, features that capture if a function terminates with exceptions do \emph{not} highly correlate with failure occurrence, suggesting they might not be as relevant.

\begin{figure}
    \centering
    \begin{subfigure}[h]{.32\textwidth}
        \includegraphics[width=\textwidth]{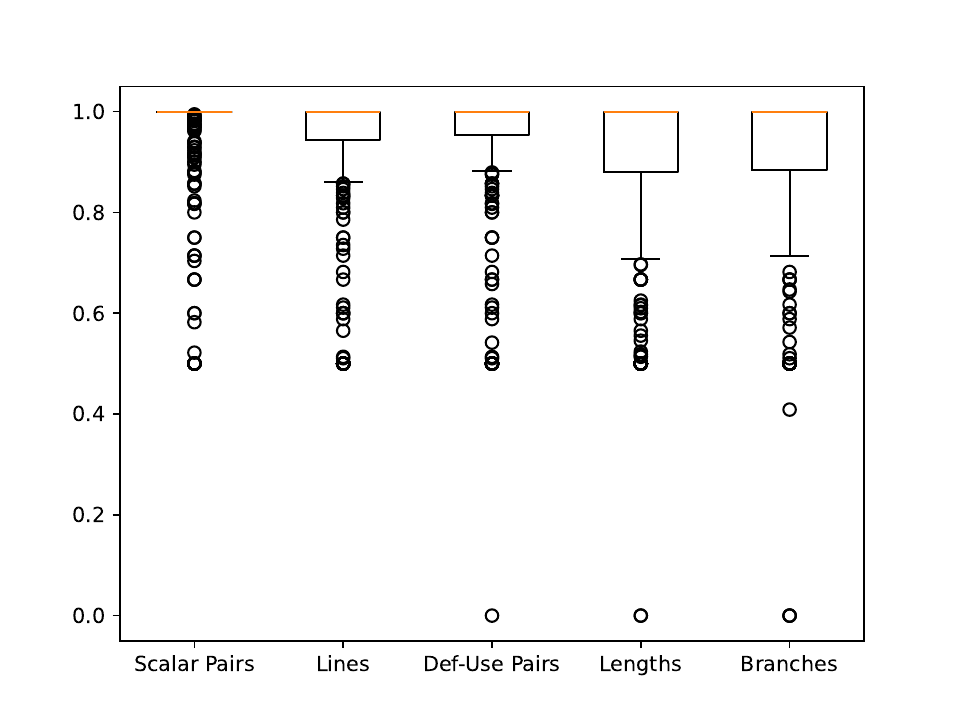}
        \caption{\TARANTULA{}}%
        \label{fig:sus-tarantula}
    \end{subfigure}
    \begin{subfigure}[h]{.32\textwidth}
        \includegraphics[width=\textwidth]{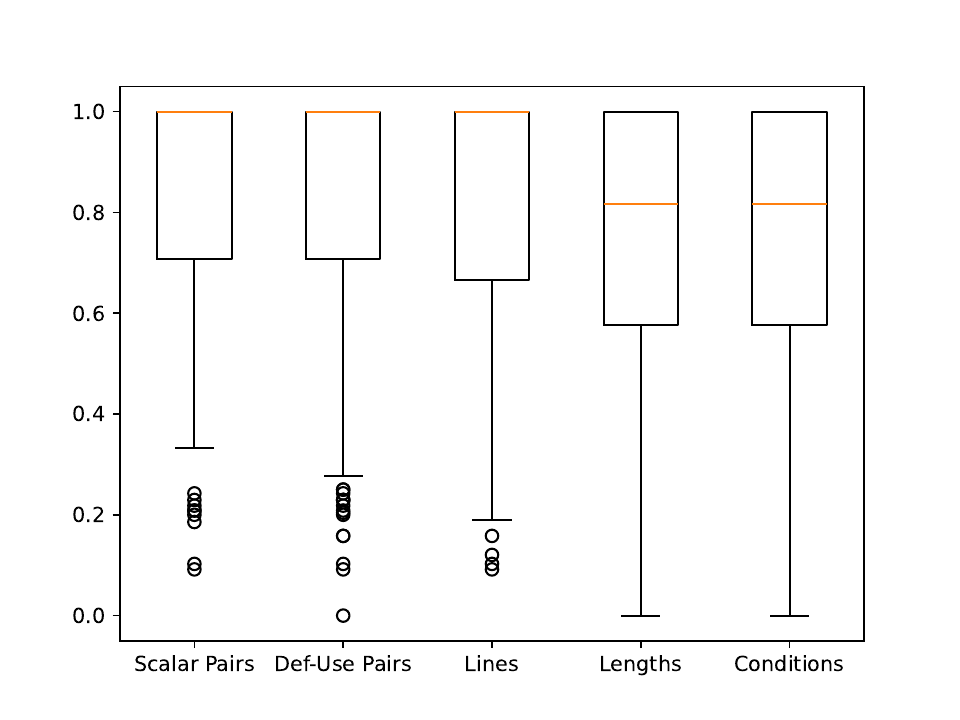}
        \caption{\OCHIAI{}}%
        \label{fig:sus-ochiai}
    \end{subfigure}
    \begin{subfigure}[h]{.32\textwidth}
        \includegraphics[width=\textwidth]{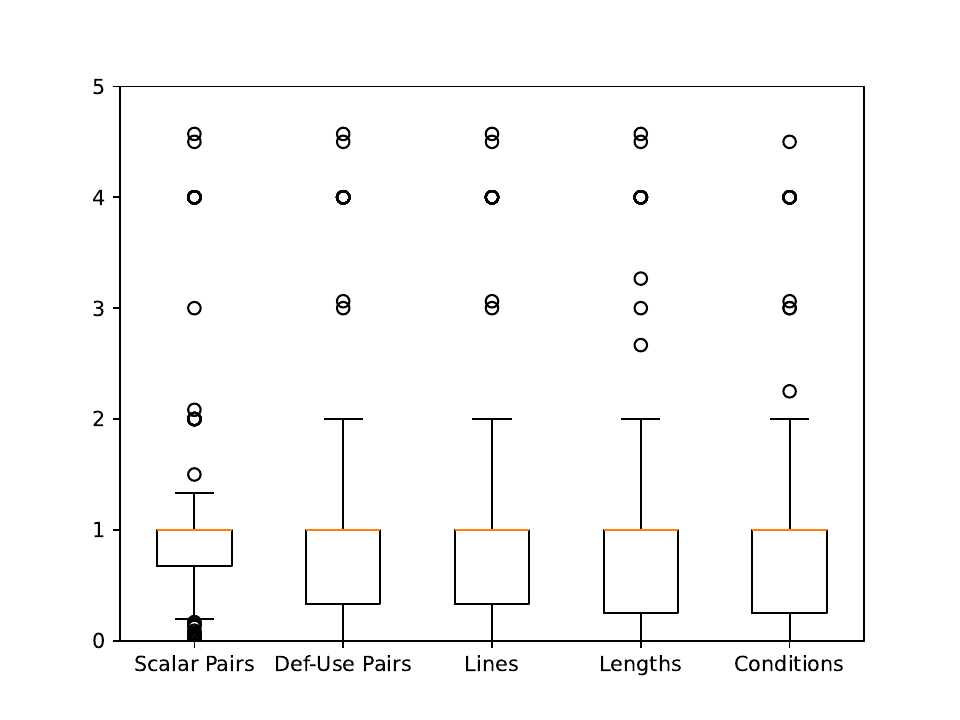}
        \caption{\DSTAR{}}%
        \label{fig:sus-dstar}
    \end{subfigure}
    \begin{subfigure}[h]{.32\textwidth}
        \includegraphics[width=\textwidth]{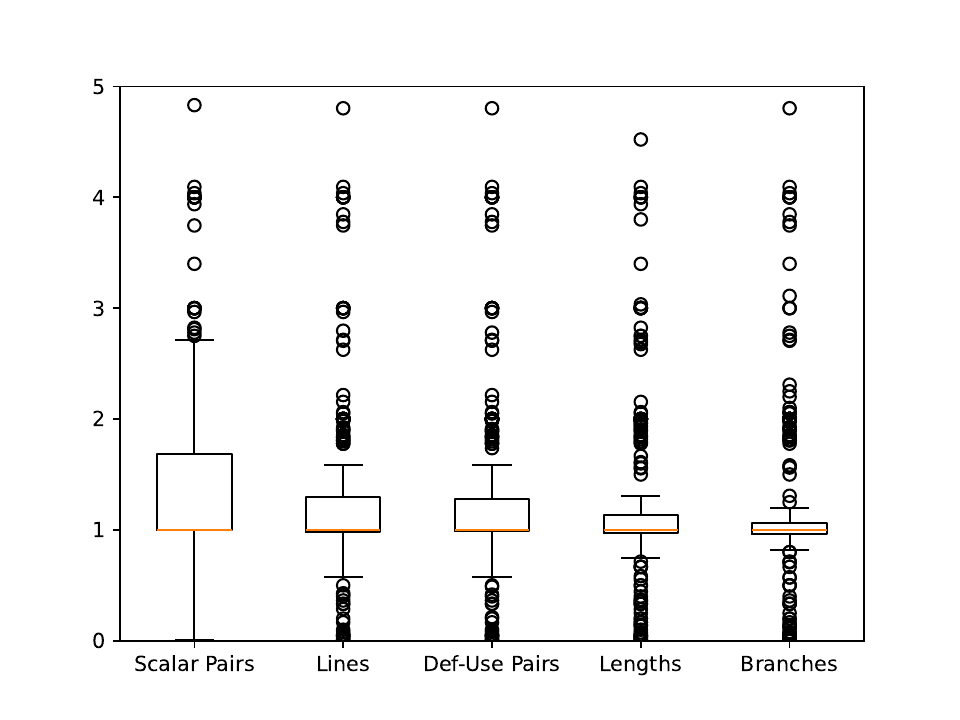}
        \caption{\NAISHT{}}%
        \label{fig:sus-naish}
    \end{subfigure}
    \begin{subfigure}[h]{.32\textwidth}
        \includegraphics[width=\textwidth]{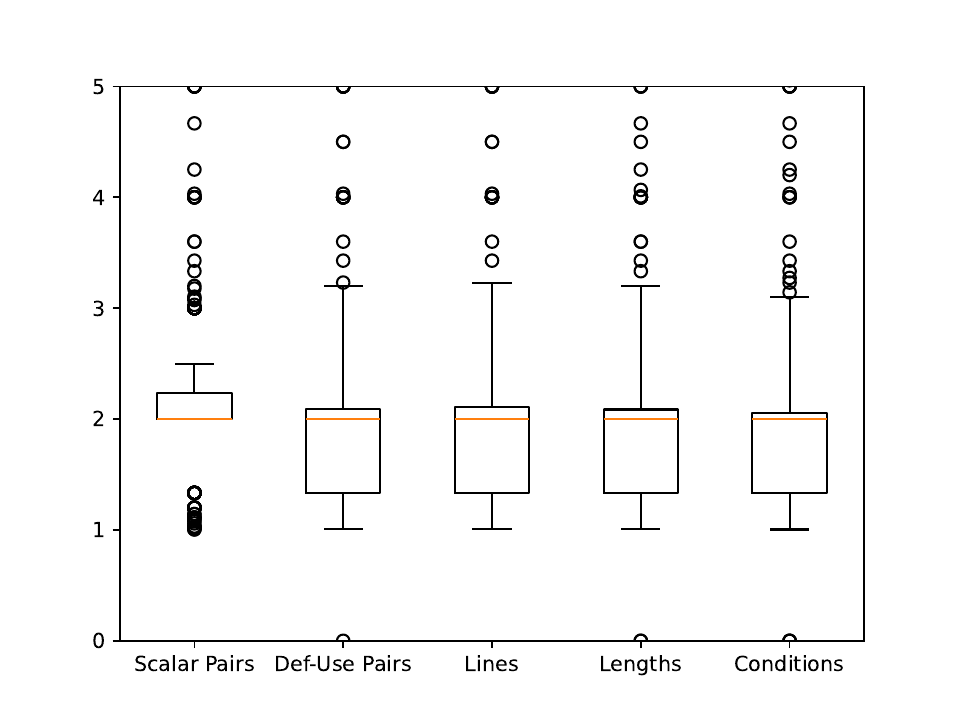}
        \caption{\GPOT{}}%
        \label{fig:sus-gp}
    \end{subfigure}
    \caption{Suspiciousness. Results of the suspiciousness that a feature correlates with the presence of failures for the metrics \TARANTULA{}, \OCHIAI{}, \DSTAR{}, \NAISHT{}, and \GPOT{}. Each feature is evaluated according to the best feature of each class for each subject.}%
    \label{fig:suspiciousness}
\end{figure}

\Cref{fig:suspiciousness} illustrates the suspiciousness results for the five best-performing execution features leveraging \TARANTULA{}, \OCHIAI{}, \DSTAR{}, \NAISHT{}, and \GPOT{} for each subject.
These results support our findings that the features that capture data flow and the relation between a program's data, lines, functions, and return values are more relevant to failures.

\subsubsection{\bfseries\Cref{rq2}: Localizing Faults}%
\label{sub:eval-localization}

\begin{table}
    \caption{Localization. Results for the localization of faults leveraging \TARANTULA{}, \OCHIAI{}, \DSTAR{}, \NAISHT{}, and \GPOT{}. Each feature is evaluated according to Top-1, Top-5, Top-10, Top-200, \EXAM{} score and wasted effort for three debugging scenarios.}%
    \label{tab:results-localization-1}
    \setlength\extrarowheight{-3pt}
    \centering
    \resizebox{\textwidth}{!}{%
    \begin{tabular}{llrrrrrrrrrrrrrrrrrr}
    \toprule
    \multicolumn{1}{c}{\multirow{4}*{Feature}} & \multicolumn{1}{c}{\multirow{4}*{Metric}} & \multicolumn{6}{c}{Best-Case Debugging} & \multicolumn{6}{c}{Average-Case Debugging} & \multicolumn{6}{c}{Worst-Case Debugging} \\\cmidrule(lr){3-8}\cmidrule(lr){9-14}\cmidrule(lr){15-20}
    & & \multicolumn{4}{c}{Top-k} & \multicolumn{1}{c}{\multirow{2}*{\EXAM{}}} & \multicolumn{1}{c}{\multirow{2}*{Effort}}
 & \multicolumn{4}{c}{Top-k} & \multicolumn{1}{c}{\multirow{2}*{\EXAM{}}} & \multicolumn{1}{c}{\multirow{2}*{Effort}}
 & \multicolumn{4}{c}{Top-k} & \multicolumn{1}{c}{\multirow{2}*{\EXAM{}}} & \multicolumn{1}{c}{\multirow{2}*{Effort}}
\\\cmidrule{3-6}\cmidrule{9-12}\cmidrule{15-18}
    & & \multicolumn{1}{c}{1} & \multicolumn{1}{c}{5} & \multicolumn{1}{c}{10} & \multicolumn{1}{c}{200} & &
 & \multicolumn{1}{c}{1} & \multicolumn{1}{c}{5} & \multicolumn{1}{c}{10} & \multicolumn{1}{c}{200} & &
 & \multicolumn{1}{c}{1} & \multicolumn{1}{c}{5} & \multicolumn{1}{c}{10} & \multicolumn{1}{c}{200} & &
\\\midrule
     & \TARANTULA{}\rowstrut{} & \textbf{\color{deepblue}14.2\%} & 25.4\% & 29.5\% & 47.2\% & 0.101 & 6.1k & \textbf{\color{deepblue}9.8\%} & 18.5\% & 21.7\% & 37.4\% & 0.169 & 10.8k & \textbf{\color{deepblue}7.6\%} & 13.9\% & 16.0\% & 28.7\% & 0.316 & 29.3k \\
     & \OCHIAI{} & \textbf{\color{deepblue}14.0\%} & 25.0\% & 28.7\% & 46.2\% & 0.101 & 6.1k & \textbf{\color{deepblue}10.0\%} & 18.0\% & 21.0\% & 36.7\% & 0.169 & 10.8k & \textbf{\color{deepblue}7.7\%} & 13.8\% & 15.6\% & 28.2\% & 0.316 & 29.3k \\
    Lines & \DSTAR{} & \textbf{\color{deepblue}9.2\%} & 14.9\% & 18.6\% & 38.0\% & 0.118 & 6.6k & \textbf{\color{deepblue}6.1\%} & 9.5\% & 13.3\% & 30.1\% & 0.185 & 11.2k & \textbf{\color{deepblue}4.5\%} & 6.9\% & 9.6\% & 23.0\% & 0.329 & 29.7k \\
     & \NAISHT{} & \textbf{\color{deepblue}14.4\%} & 23.7\% & 26.5\% & 45.9\% & 0.102 & 6.1k & \textbf{\color{deepblue}10.1\%} & 17.2\% & 19.8\% & 37.3\% & 0.170 & 10.8k & \textbf{\color{deepblue}7.8\%} & 13.2\% & 15.0\% & 28.6\% & 0.316 & 29.3k \\
     & \GPOT{} & \textbf{\color{deepblue}14.5\%} & 23.6\% & 26.5\% & 45.0\% & 0.102 & 6.1k & \textbf{\color{deepblue}10.1\%} & 17.1\% & 19.9\% & 36.1\% & 0.170 & 10.8k & \textbf{\color{deepblue}7.8\%} & 13.1\% & 15.1\% & 27.7\% & 0.317 & 29.3k \\[.2em]
\rowcolor{row}
     & \TARANTULA{}\rowstrut{} & 8.0\% & 17.6\% & 21.4\% & 39.0\% & 0.244 & 16.3k & 4.8\% & 12.6\% & 16.1\% & 31.1\% & 0.291 & 22.4k & 3.4\% & 9.3\% & 12.0\% & 23.8\% & 0.392 & 37.4k \\
\rowcolor{row}
     & \OCHIAI{} & 7.7\% & 17.1\% & 21.0\% & 38.8\% & 0.244 & 16.3k & 4.7\% & 12.3\% & 15.7\% & 30.8\% & 0.292 & 22.5k & 3.3\% & 9.1\% & 11.7\% & 23.4\% & 0.392 & 37.5k \\
\rowcolor{row}
    Branches & \DSTAR{} & 4.6\% & 11.8\% & 15.3\% & 33.9\% & 0.250 & 16.4k & 2.6\% & 7.8\% & 10.6\% & 26.9\% & 0.297 & 22.5k & 1.8\% & 5.9\% & 7.6\% & 20.5\% & 0.396 & 37.5k \\
\rowcolor{row}
     & \NAISHT{} & 7.2\% & 15.9\% & 19.3\% & 39.2\% & 0.245 & 16.3k & 4.6\% & 11.7\% & 14.6\% & 32.3\% & 0.292 & 22.4k & 3.3\% & 8.9\% & 11.2\% & 24.7\% & 0.392 & 37.4k \\
\rowcolor{row}
     & \GPOT{} & 7.3\% & 15.9\% & 19.1\% & 37.7\% & 0.245 & 16.4k & 4.7\% & 11.7\% & 14.5\% & 30.5\% & 0.292 & 22.5k & 3.3\% & 8.8\% & 11.1\% & 23.2\% & 0.392 & 37.5k \\[.2em]
     & \TARANTULA{}\rowstrut{} & 6.2\% & 20.8\% & 27.0\% & 55.9\% & \textbf{\color{deepblue}0.087} & \textbf{\color{deepblue}5.4k} & 3.9\% & 15.3\% & 20.5\% & \textbf{\color{deepblue}49.0\%} & \textbf{\color{deepblue}0.107} & \textbf{\color{deepblue}6.0k} & 2.7\% & 11.1\% & 14.8\% & \textbf{\color{deepblue}39.4\%} & \textbf{\color{deepblue}0.223} & \textbf{\color{deepblue}20.7k} \\
     & \OCHIAI{} & 6.1\% & 20.2\% & 26.6\% & 55.2\% & \textbf{\color{deepblue}0.087} & \textbf{\color{deepblue}5.4k} & 3.9\% & 15.1\% & 20.4\% & 48.0\% & \textbf{\color{deepblue}0.107} & \textbf{\color{deepblue}6.0k} & 2.8\% & 11.2\% & 15.0\% & \textbf{\color{deepblue}38.9\%} & \textbf{\color{deepblue}0.222} & \textbf{\color{deepblue}20.7k} \\
    Functions & \DSTAR{} & 5.2\% & 14.0\% & 20.4\% & 49.3\% & \textbf{\color{deepblue}0.101} & \textbf{\color{deepblue}5.7k} & 3.4\% & 10.0\% & 14.9\% & 42.9\% & \textbf{\color{deepblue}0.120} & \textbf{\color{deepblue}6.3k} & 2.4\% & 7.4\% & 10.8\% & \textbf{\color{deepblue}34.9\%} & \textbf{\color{deepblue}0.230} & \textbf{\color{deepblue}20.9k} \\
     & \NAISHT{} & 6.1\% & 19.5\% & 25.2\% & 56.0\% & \textbf{\color{deepblue}0.087} & \textbf{\color{deepblue}5.3k} & 4.0\% & 14.9\% & 19.7\% & 49.7\% & \textbf{\color{deepblue}0.106} & \textbf{\color{deepblue}5.9k} & 2.8\% & 11.0\% & 14.7\% & \textbf{\color{deepblue}40.6\%} & \textbf{\color{deepblue}0.222} & \textbf{\color{deepblue}20.7k} \\
     & \GPOT{} & 6.0\% & 19.6\% & 25.5\% & 54.2\% & \textbf{\color{deepblue}0.088} & \textbf{\color{deepblue}5.4k} & 4.0\% & 15.0\% & 19.8\% & 47.6\% & \textbf{\color{deepblue}0.107} & \textbf{\color{deepblue}6.0k} & 2.8\% & 11.1\% & 14.8\% & \textbf{\color{deepblue}38.6\%} & \textbf{\color{deepblue}0.222} & \textbf{\color{deepblue}20.7k} \\[.2em]
\rowcolor{row}
     & \TARANTULA{}\rowstrut{} & 3.4\% & 11.2\% & 16.2\% & 44.0\% & 0.123 & 7.0k & 2.3\% & 7.7\% & 11.6\% & 36.3\% & 0.142 & 7.6k & 1.5\% & 5.8\% & 8.4\% & 28.9\% & 0.244 & 21.9k \\
\rowcolor{row}
     & \OCHIAI{} & 3.5\% & 11.2\% & 16.2\% & 43.8\% & 0.123 & 7.0k & 2.3\% & 7.8\% & 11.7\% & 36.1\% & 0.142 & 7.6k & 1.6\% & 5.8\% & 8.5\% & 28.6\% & 0.244 & 21.9k \\
\rowcolor{row}
    Function Errors & \DSTAR{} & 1.5\% & 6.0\% & 9.2\% & 40.7\% & 0.146 & 7.6k & 0.7\% & 3.5\% & 5.9\% & 33.1\% & 0.163 & 8.1k & 0.5\% & 2.7\% & 4.4\% & 26.2\% & 0.261 & 22.4k \\
\rowcolor{row}
     & \NAISHT{} & 3.4\% & 10.8\% & 15.6\% & 43.6\% & 0.121 & 6.9k & 2.3\% & 7.8\% & 11.5\% & 35.9\% & 0.142 & 7.5k & 1.6\% & 5.7\% & 8.3\% & 28.6\% & 0.246 & 21.9k \\
\rowcolor{row}
     & \GPOT{} & 3.4\% & 10.9\% & 15.9\% & 43.3\% & 0.123 & 7.0k & 2.4\% & 7.7\% & 11.4\% & 35.5\% & 0.142 & 7.6k & 1.6\% & 5.7\% & 8.3\% & 28.0\% & 0.244 & 21.9k \\[.2em]
     & \TARANTULA{}\rowstrut{} & 8.5\% & \textbf{\color{deepblue}25.9\%} & \textbf{\color{deepblue}32.9\%} & \textbf{\color{deepblue}60.0\%} & 0.120 & 10.6k & 5.6\% & \textbf{\color{deepblue}19.2\%} & \textbf{\color{deepblue}24.9\%} & 48.8\% & 0.187 & 13.2k & 4.1\% & \textbf{\color{deepblue}14.8\%} & \textbf{\color{deepblue}18.7\%} & 36.5\% & 0.344 & 37.5k \\
     & \OCHIAI{} & 8.6\% & \textbf{\color{deepblue}25.7\%} & \textbf{\color{deepblue}31.9\%} & \textbf{\color{deepblue}59.0\%} & 0.120 & 10.6k & 5.8\% & \textbf{\color{deepblue}19.2\%} & \textbf{\color{deepblue}24.2\%} & \textbf{\color{deepblue}48.3\%} & 0.187 & 13.2k & 4.3\% & \textbf{\color{deepblue}14.9\%} & \textbf{\color{deepblue}18.5\%} & 36.3\% & 0.344 & 37.5k \\
    Def-Use Pairs & \DSTAR{} & 6.1\% & \textbf{\color{deepblue}14.9\%} & \textbf{\color{deepblue}22.1\%} & \textbf{\color{deepblue}53.4\%} & 0.128 & 10.8k & 4.0\% & \textbf{\color{deepblue}10.0\%} & \textbf{\color{deepblue}16.2\%} & \textbf{\color{deepblue}43.7\%} & 0.194 & 13.3k & 2.9\% & \textbf{\color{deepblue}7.7\%} & \textbf{\color{deepblue}12.4\%} & 32.5\% & 0.349 & 37.6k \\
     & \NAISHT{} & 8.4\% & \textbf{\color{deepblue}24.1\%} & \textbf{\color{deepblue}29.1\%} & \textbf{\color{deepblue}60.5\%} & 0.120 & 10.6k & 5.8\% & \textbf{\color{deepblue}18.7\%} & \textbf{\color{deepblue}22.9\%} & \textbf{\color{deepblue}50.4\%} & 0.187 & 13.1k & 4.2\% & \textbf{\color{deepblue}14.5\%} & \textbf{\color{deepblue}17.9\%} & 38.3\% & 0.343 & 37.5k \\
     & \GPOT{} & 8.3\% & \textbf{\color{deepblue}24.3\%} & \textbf{\color{deepblue}29.5\%} & \textbf{\color{deepblue}58.2\%} & 0.121 & 10.7k & 5.8\% & \textbf{\color{deepblue}18.7\%} & \textbf{\color{deepblue}22.9\%} & \textbf{\color{deepblue}47.8\%} & 0.188 & 13.2k & 4.2\% & \textbf{\color{deepblue}14.5\%} & \textbf{\color{deepblue}17.8\%} & 36.0\% & 0.344 & 37.5k \\[.2em]
\rowcolor{row}
     & \TARANTULA{}\rowstrut{} & 3.6\% & 8.6\% & 8.7\% & 18.6\% & 0.407 & 27.3k & 1.3\% & 5.7\% & 6.1\% & 15.2\% & 0.432 & 32.5k & 0.8\% & 3.7\% & 4.2\% & 12.2\% & 0.467 & 45.0k \\
\rowcolor{row}
     & \OCHIAI{} & 3.6\% & 8.5\% & 9.0\% & 18.6\% & 0.407 & 27.3k & 1.3\% & 5.7\% & 6.4\% & 15.2\% & 0.432 & 32.5k & 0.8\% & 3.7\% & 4.5\% & 12.2\% & 0.466 & 45.0k \\
\rowcolor{row}
    Loops & \DSTAR{} & 2.4\% & 6.6\% & 7.2\% & 18.7\% & 0.407 & 27.3k & 0.9\% & 4.1\% & 4.8\% & 15.1\% & 0.432 & 32.5k & 0.6\% & 2.7\% & 3.3\% & 12.0\% & 0.467 & 45.0k \\
\rowcolor{row}
     & \NAISHT{} & 3.6\% & 8.4\% & 8.6\% & 18.8\% & 0.407 & 27.3k & 1.3\% & 5.4\% & 5.8\% & 15.4\% & 0.432 & 32.5k & 0.8\% & 3.6\% & 4.2\% & 12.3\% & 0.466 & 45.0k \\
\rowcolor{row}
     & \GPOT{} & 3.5\% & 8.4\% & 8.6\% & 18.6\% & 0.407 & 27.3k & 1.2\% & 5.4\% & 5.8\% & 15.2\% & 0.432 & 32.5k & 0.8\% & 3.6\% & 4.2\% & 12.2\% & 0.466 & 45.0k \\[.2em]
     & \TARANTULA{}\rowstrut{} & 6.9\% & 13.6\% & 16.2\% & 28.2\% & 0.349 & 29.3k & 4.0\% & 8.0\% & 9.3\% & 17.9\% & 0.433 & 41.2k & 2.8\% & 5.6\% & 6.3\% & 12.2\% & 0.476 & 45.5k \\
     & \OCHIAI{} & 6.3\% & 13.1\% & 15.9\% & 28.2\% & 0.349 & 29.3k & 3.8\% & 7.8\% & 9.3\% & 17.9\% & 0.433 & 41.2k & 2.7\% & 5.6\% & 6.3\% & 12.2\% & 0.476 & 45.5k \\
    Conditions & \DSTAR{} & 3.8\% & 9.5\% & 12.8\% & 27.0\% & 0.350 & 29.3k & 1.8\% & 5.3\% & 7.2\% & 17.2\% & 0.433 & 41.2k & 1.1\% & 3.5\% & 4.8\% & 11.9\% & 0.476 & 45.5k \\
     & \NAISHT{} & 5.9\% & 11.8\% & 14.5\% & 28.2\% & 0.349 & 29.3k & 3.7\% & 7.3\% & 8.9\% & 18.0\% & 0.433 & 41.2k & 2.7\% & 5.3\% & 6.1\% & 12.3\% & 0.476 & 45.5k \\
     & \GPOT{} & 6.0\% & 11.7\% & 14.4\% & 27.8\% & 0.349 & 29.3k & 3.8\% & 7.2\% & 8.9\% & 17.6\% & 0.433 & 41.2k & 2.7\% & 5.3\% & 6.2\% & 12.1\% & 0.476 & 45.5k \\[.2em]
\rowcolor{row}
     & \TARANTULA{}\rowstrut{} & 5.7\% & 14.3\% & 15.6\% & 34.3\% & 0.254 & 20.2k & 3.3\% & 8.3\% & 9.4\% & 24.4\% & 0.343 & 26.5k & 2.2\% & 5.7\% & 6.1\% & 17.7\% & 0.439 & 42.4k \\
\rowcolor{row}
     & \OCHIAI{} & 5.8\% & 13.7\% & 15.2\% & 33.5\% & 0.254 & 20.2k & 3.7\% & 8.1\% & 9.5\% & 24.2\% & 0.343 & 26.5k & 2.5\% & 5.5\% & 6.3\% & 17.5\% & 0.439 & 42.4k \\
\rowcolor{row}
    Scalar Pairs & \DSTAR{} & 4.0\% & 8.8\% & 9.9\% & 29.3\% & 0.258 & 20.4k & 2.2\% & 4.8\% & 5.7\% & 21.4\% & 0.345 & 26.6k & 1.6\% & 3.4\% & 3.9\% & 15.5\% & 0.440 & 42.4k \\
\rowcolor{row}
     & \NAISHT{} & 5.9\% & 12.7\% & 14.3\% & 33.3\% & 0.255 & 20.2k & 3.7\% & 7.8\% & 9.3\% & 23.8\% & 0.343 & 26.5k & 2.5\% & 5.5\% & 6.4\% & 17.2\% & 0.439 & 42.4k \\
\rowcolor{row}
     & \GPOT{} & 5.9\% & 12.7\% & 14.3\% & 33.2\% & 0.255 & 20.2k & 3.7\% & 7.9\% & 9.3\% & 23.8\% & 0.343 & 26.5k & 2.5\% & 5.5\% & 6.5\% & 17.3\% & 0.439 & 42.4k \\[.2em]
\bottomrule
\end{tabular}

    }
\end{table}

\begin{table}
    \caption{Localization. Continuation of \Cref{tab:results-localization-1}.}%
    \label{tab:results-localization-2}
    \setlength\extrarowheight{-3pt}
    \centering
    \resizebox{\textwidth}{!}{%
    \begin{tabular}{llrrrrrrrrrrrrrrrrrr}
    \toprule
    \multicolumn{1}{c}{\multirow{4}*{Feature}} & \multicolumn{1}{c}{\multirow{4}*{Metric}} & \multicolumn{6}{c}{Best-Case Debugging} & \multicolumn{6}{c}{Average-Case Debugging} & \multicolumn{6}{c}{Worst-Case Debugging} \\\cmidrule(lr){3-8}\cmidrule(lr){9-14}\cmidrule(lr){15-20}
    & & \multicolumn{4}{c}{Top-k} & \multicolumn{1}{c}{\multirow{2}*{\EXAM{}}} & \multicolumn{1}{c}{\multirow{2}*{Effort}}
 & \multicolumn{4}{c}{Top-k} & \multicolumn{1}{c}{\multirow{2}*{\EXAM{}}} & \multicolumn{1}{c}{\multirow{2}*{Effort}}
 & \multicolumn{4}{c}{Top-k} & \multicolumn{1}{c}{\multirow{2}*{\EXAM{}}} & \multicolumn{1}{c}{\multirow{2}*{Effort}}
\\\cmidrule{3-6}\cmidrule{9-12}\cmidrule{15-18}
    & & \multicolumn{1}{c}{1} & \multicolumn{1}{c}{5} & \multicolumn{1}{c}{10} & \multicolumn{1}{c}{200} & &
 & \multicolumn{1}{c}{1} & \multicolumn{1}{c}{5} & \multicolumn{1}{c}{10} & \multicolumn{1}{c}{200} & &
 & \multicolumn{1}{c}{1} & \multicolumn{1}{c}{5} & \multicolumn{1}{c}{10} & \multicolumn{1}{c}{200} & &
\\\midrule
     & \TARANTULA{}\rowstrut{} & 2.7\% & 5.6\% & 7.1\% & 12.5\% & 0.433 & 41.7k & 1.3\% & 2.8\% & 4.3\% & 7.9\% & 0.467 & 45.4k & 1.0\% & 1.9\% & 3.2\% & 6.3\% & 0.482 & 48.7k \\
     & \OCHIAI{} & 2.7\% & 5.3\% & 7.1\% & 12.5\% & 0.433 & 41.7k & 1.5\% & 2.7\% & 4.1\% & 7.9\% & 0.467 & 45.4k & 1.1\% & 1.9\% & 3.1\% & 6.3\% & 0.482 & 48.7k \\
    Variable Values & \DSTAR{} & 1.7\% & 4.3\% & 6.0\% & 12.4\% & 0.433 & 41.7k & 0.9\% & 2.2\% & 3.5\% & 7.9\% & 0.467 & 45.4k & 0.7\% & 1.6\% & 2.7\% & 6.2\% & 0.482 & 48.7k \\
     & \NAISHT{} & 2.5\% & 4.5\% & 6.5\% & 11.9\% & 0.433 & 41.7k & 1.3\% & 2.1\% & 3.7\% & 7.6\% & 0.467 & 45.4k & 1.0\% & 1.5\% & 2.8\% & 6.0\% & 0.482 & 48.7k \\
     & \GPOT{} & 2.5\% & 4.5\% & 6.5\% & 11.9\% & 0.433 & 41.7k & 1.3\% & 2.1\% & 3.7\% & 7.6\% & 0.467 & 45.4k & 1.0\% & 1.5\% & 2.8\% & 6.0\% & 0.482 & 48.7k \\[.2em]
\rowcolor{row}
     & \TARANTULA{}\rowstrut{} & 6.1\% & 8.6\% & 10.2\% & 14.5\% & 0.433 & 45.4k & 4.5\% & 6.1\% & 6.6\% & 10.3\% & 0.461 & 48.3k & 3.6\% & 4.6\% & 4.8\% & 7.1\% & 0.487 & 49.5k \\
\rowcolor{row}
     & \OCHIAI{} & 6.2\% & 8.5\% & 9.9\% & 14.5\% & 0.433 & 45.4k & 4.7\% & 5.9\% & 6.5\% & 10.3\% & 0.461 & 48.3k & 3.7\% & 4.5\% & 4.7\% & 7.1\% & 0.487 & 49.5k \\
\rowcolor{row}
    Return Values & \DSTAR{} & 3.5\% & 6.8\% & 9.3\% & 14.5\% & 0.434 & 45.4k & 2.5\% & 4.7\% & 6.1\% & 10.3\% & 0.461 & 48.3k & 2.0\% & 3.5\% & 4.5\% & 7.1\% & 0.487 & 49.5k \\
\rowcolor{row}
     & \NAISHT{} & 6.2\% & 8.5\% & 9.7\% & 14.5\% & 0.433 & 45.4k & 4.7\% & 5.8\% & 6.5\% & 10.3\% & 0.461 & 48.3k & 3.7\% & 4.5\% & 4.7\% & 7.1\% & 0.487 & 49.5k \\
\rowcolor{row}
     & \GPOT{} & 6.2\% & 8.4\% & 9.6\% & 14.5\% & 0.433 & 45.4k & 4.7\% & 5.9\% & 6.4\% & 10.3\% & 0.461 & 48.3k & 3.7\% & 4.5\% & 4.7\% & 7.1\% & 0.487 & 49.5k \\[.2em]
     & \TARANTULA{}\rowstrut{} & 6.0\% & 10.0\% & 11.6\% & 32.0\% & 0.256 & 20.3k & 3.4\% & 5.5\% & 6.7\% & 22.8\% & 0.342 & 26.5k & 2.3\% & 3.9\% & 4.6\% & 16.7\% & 0.438 & 42.4k \\
     & \OCHIAI{} & 6.1\% & 10.0\% & 11.4\% & 31.9\% & 0.256 & 20.3k & 3.5\% & 5.7\% & 6.6\% & 22.8\% & 0.342 & 26.5k & 2.3\% & 4.0\% & 4.6\% & 16.6\% & 0.438 & 42.4k \\
    Null Values & \DSTAR{} & 3.3\% & 6.3\% & 8.3\% & 29.3\% & 0.258 & 20.4k & 1.3\% & 3.4\% & 4.5\% & 20.9\% & 0.344 & 26.6k & 0.8\% & 2.5\% & 3.2\% & 15.3\% & 0.439 & 42.4k \\
     & \NAISHT{} & 6.1\% & 9.4\% & 10.4\% & 33.2\% & 0.255 & 20.3k & 3.5\% & 5.5\% & 6.3\% & 23.9\% & 0.341 & 26.5k & 2.4\% & 4.0\% & 4.5\% & 17.5\% & 0.438 & 42.4k \\
     & \GPOT{} & 6.1\% & 9.5\% & 10.4\% & 31.7\% & 0.256 & 20.3k & 3.5\% & 5.5\% & 6.3\% & 22.8\% & 0.342 & 26.5k & 2.4\% & 3.9\% & 4.4\% & 16.7\% & 0.438 & 42.4k \\[.2em]
\rowcolor{row}
     & \TARANTULA{}\rowstrut{} & 7.2\% & 12.5\% & 15.0\% & 27.5\% & 0.332 & 28.1k & 3.8\% & 7.9\% & 8.8\% & 17.9\% & 0.415 & 38.6k & 2.4\% & 5.4\% & 5.7\% & 12.4\% & 0.470 & 47.9k \\
\rowcolor{row}
     & \OCHIAI{} & 7.4\% & 12.1\% & 14.7\% & 27.1\% & 0.332 & 28.1k & 4.2\% & 7.7\% & 8.7\% & 17.6\% & 0.415 & 38.6k & 2.6\% & 5.4\% & 5.7\% & 12.3\% & 0.470 & 47.9k \\
\rowcolor{row}
    Lengths & \DSTAR{} & 4.1\% & 6.7\% & 8.9\% & 25.0\% & 0.334 & 28.2k & 1.7\% & 3.7\% & 4.9\% & 16.4\% & 0.416 & 38.6k & 1.1\% & 2.7\% & 3.1\% & 11.6\% & 0.471 & 47.9k \\
\rowcolor{row}
     & \NAISHT{} & 7.7\% & 12.0\% & 13.9\% & 26.8\% & 0.332 & 28.2k & 4.2\% & 8.0\% & 8.5\% & 17.6\% & 0.415 & 38.6k & 2.6\% & 5.6\% & 5.7\% & 12.3\% & 0.470 & 47.9k \\
\rowcolor{row}
     & \GPOT{} & 7.8\% & 11.9\% & 13.9\% & 26.7\% & 0.332 & 28.2k & 4.2\% & 7.9\% & 8.6\% & 17.5\% & 0.415 & 38.6k & 2.6\% & 5.6\% & 5.7\% & 12.3\% & 0.470 & 47.9k \\[.2em]
     & \TARANTULA{}\rowstrut{} & 1.6\% & 3.3\% & 5.5\% & 28.7\% & 0.262 & 20.7k & 0.7\% & 1.7\% & 3.2\% & 20.2\% & 0.344 & 26.8k & 0.4\% & 1.3\% & 2.3\% & 14.4\% & 0.440 & 42.5k \\
     & \OCHIAI{} & 1.6\% & 3.2\% & 5.3\% & 28.6\% & 0.262 & 20.7k & 0.7\% & 1.7\% & 3.1\% & 20.2\% & 0.344 & 26.8k & 0.4\% & 1.3\% & 2.3\% & 14.4\% & 0.440 & 42.5k \\
    Empty Strings & \DSTAR{} & 1.0\% & 3.2\% & 5.3\% & 28.6\% & 0.262 & 20.7k & 0.3\% & 1.7\% & 3.1\% & 20.2\% & 0.345 & 26.8k & 0.2\% & 1.3\% & 2.3\% & 14.5\% & 0.440 & 42.5k \\
     & \NAISHT{} & 1.6\% & 3.2\% & 5.3\% & 28.6\% & 0.262 & 20.7k & 0.7\% & 1.7\% & 3.1\% & 20.2\% & 0.344 & 26.8k & 0.4\% & 1.3\% & 2.3\% & 14.5\% & 0.440 & 42.5k \\
     & \GPOT{} & 1.6\% & 3.2\% & 5.3\% & 28.8\% & 0.262 & 20.7k & 0.7\% & 1.7\% & 3.1\% & 20.1\% & 0.344 & 26.8k & 0.4\% & 1.3\% & 2.3\% & 14.5\% & 0.440 & 42.5k \\[.2em]
\rowcolor{row}
     & \TARANTULA{}\rowstrut{} & 6.7\% & 10.2\% & 11.0\% & 28.1\% & 0.259 & 20.6k & 3.9\% & 6.5\% & 6.9\% & 20.0\% & 0.344 & 26.7k & 2.5\% & 4.7\% & 4.9\% & 14.3\% & 0.439 & 42.5k \\
\rowcolor{row}
     & \OCHIAI{} & 6.5\% & 9.8\% & 10.8\% & 28.1\% & 0.259 & 20.6k & 4.2\% & 6.3\% & 6.8\% & 19.9\% & 0.344 & 26.7k & 2.7\% & 4.6\% & 4.8\% & 14.3\% & 0.439 & 42.5k \\
\rowcolor{row}
    ASCII Strings & \DSTAR{} & 3.6\% & 6.1\% & 8.0\% & 27.3\% & 0.260 & 20.6k & 2.2\% & 3.4\% & 4.7\% & 19.2\% & 0.345 & 26.7k & 1.6\% & 2.7\% & 3.3\% & 13.8\% & 0.440 & 42.5k \\
\rowcolor{row}
     & \NAISHT{} & 6.2\% & 9.6\% & 10.9\% & 27.7\% & 0.259 & 20.6k & 4.1\% & 6.2\% & 7.0\% & 19.8\% & 0.344 & 26.7k & 2.6\% & 4.7\% & 5.0\% & 14.4\% & 0.439 & 42.5k \\
\rowcolor{row}
     & \GPOT{} & 6.1\% & 9.5\% & 10.7\% & 27.9\% & 0.259 & 20.6k & 4.0\% & 6.2\% & 6.9\% & 19.8\% & 0.344 & 26.7k & 2.6\% & 4.7\% & 5.0\% & 14.2\% & 0.439 & 42.5k \\[.2em]
     & \TARANTULA{}\rowstrut{} & 3.3\% & 6.7\% & 7.9\% & 28.2\% & 0.260 & 20.7k & 1.7\% & 4.3\% & 5.2\% & 20.0\% & 0.344 & 26.7k & 1.1\% & 3.2\% & 3.7\% & 14.4\% & 0.439 & 42.5k \\
     & \OCHIAI{} & 3.3\% & 6.6\% & 7.9\% & 28.2\% & 0.260 & 20.7k & 1.8\% & 4.2\% & 5.1\% & 20.0\% & 0.344 & 26.7k & 1.2\% & 3.2\% & 3.6\% & 14.3\% & 0.439 & 42.5k \\
    Digit Strings & \DSTAR{} & 1.6\% & 4.4\% & 6.3\% & 27.9\% & 0.261 & 20.7k & 0.5\% & 2.5\% & 3.8\% & 19.9\% & 0.344 & 26.8k & 0.4\% & 1.8\% & 2.8\% & 14.3\% & 0.439 & 42.5k \\
     & \NAISHT{} & 2.9\% & 6.5\% & 7.7\% & 28.0\% & 0.260 & 20.7k & 1.7\% & 4.1\% & 4.9\% & 20.0\% & 0.344 & 26.7k & 1.1\% & 3.1\% & 3.5\% & 14.4\% & 0.439 & 42.5k \\
     & \GPOT{} & 2.9\% & 6.5\% & 7.7\% & 28.1\% & 0.260 & 20.7k & 1.7\% & 4.2\% & 4.9\% & 20.0\% & 0.344 & 26.7k & 1.1\% & 3.0\% & 3.4\% & 14.4\% & 0.439 & 42.5k \\[.2em]
\rowcolor{row}
     & \TARANTULA{}\rowstrut{} & 6.7\% & 9.5\% & 9.6\% & 28.0\% & 0.260 & 20.6k & 3.9\% & 6.0\% & 6.0\% & 19.9\% & 0.344 & 26.7k & 2.5\% & 4.3\% & 4.6\% & 14.3\% & 0.439 & 42.5k \\
\rowcolor{row}
     & \OCHIAI{} & 6.6\% & 9.3\% & 9.6\% & 27.7\% & 0.260 & 20.6k & 4.0\% & 5.7\% & 6.1\% & 19.8\% & 0.344 & 26.7k & 2.7\% & 4.1\% & 4.6\% & 14.1\% & 0.439 & 42.5k \\
\rowcolor{row}
    Special Strings & \DSTAR{} & 4.2\% & 6.6\% & 7.6\% & 27.1\% & 0.260 & 20.7k & 2.4\% & 3.7\% & 4.5\% & 19.1\% & 0.345 & 26.8k & 1.7\% & 2.8\% & 3.3\% & 13.7\% & 0.440 & 42.5k \\
\rowcolor{row}
     & \NAISHT{} & 6.5\% & 9.1\% & 9.8\% & 27.5\% & 0.259 & 20.6k & 4.0\% & 5.7\% & 6.1\% & 19.7\% & 0.344 & 26.7k & 2.7\% & 4.2\% & 4.5\% & 14.3\% & 0.439 & 42.5k \\
\rowcolor{row}
     & \GPOT{} & 6.5\% & 9.1\% & 9.6\% & 27.8\% & 0.260 & 20.6k & 4.1\% & 5.7\% & 6.0\% & 19.7\% & 0.344 & 26.7k & 2.7\% & 4.1\% & 4.5\% & 14.2\% & 0.439 & 42.5k \\[.2em]
     & \TARANTULA{}\rowstrut{} & 0.6\% & 3.0\% & 5.4\% & 28.6\% & 0.262 & 20.7k & 0.3\% & 1.8\% & 3.2\% & 20.2\% & 0.344 & 26.8k & 0.3\% & 1.3\% & 2.3\% & 14.4\% & 0.439 & 42.5k \\
     & \OCHIAI{} & 0.6\% & 3.0\% & 5.4\% & 28.6\% & 0.262 & 20.7k & 0.4\% & 1.8\% & 3.2\% & 20.2\% & 0.344 & 26.8k & 0.3\% & 1.3\% & 2.3\% & 14.5\% & 0.439 & 42.5k \\
    Empty Bytes & \DSTAR{} & 0.6\% & 3.1\% & 5.4\% & 28.7\% & 0.262 & 20.7k & 0.3\% & 1.8\% & 3.2\% & 20.2\% & 0.344 & 26.8k & 0.3\% & 1.3\% & 2.3\% & 14.4\% & 0.439 & 42.5k \\
     & \NAISHT{} & 0.6\% & 3.0\% & 5.4\% & 28.6\% & 0.262 & 20.7k & 0.4\% & 1.8\% & 3.2\% & 20.1\% & 0.344 & 26.8k & 0.3\% & 1.3\% & 2.3\% & 14.5\% & 0.439 & 42.5k \\
     & \GPOT{} & 0.6\% & 3.0\% & 5.4\% & 28.6\% & 0.262 & 20.7k & 0.3\% & 1.8\% & 3.1\% & 20.1\% & 0.344 & 26.8k & 0.3\% & 1.3\% & 2.3\% & 14.5\% & 0.439 & 42.5k \\[.2em]
\bottomrule
\end{tabular}

    }
\end{table}

Concerning the fault localization and addressing \Cref{rq2}, we can see that \emph{lines} outperform all other feature classes in localizing faulty lines, especially in the top-5 best-case debugging scenario. Lines also perform well in average and worst-case scenarios, only being surpassed by functions for \DSTAR{}.

When considering top-10 and top-200, we see a shift in the performance of the best localization to functions and def-use pairs, where \emph{def-use} pairs slightly outperform functions for the best-case and average-case debugging scenario.
For the \EXAM{} scores, we see that \emph{functions} show the lowest score. 
However, lines, def-use pairs, and function errors also show a comparable low \EXAM{} score, indicating that these features are more likely to suggest the faulty lines early in the suggested lines.
The wasted effort reinforces the results for the \EXAM{} score, where lines, def-use pairs, functions, and function errors show the lowest score.

\begin{result}
    \emph{Lines}, \emph{def-use pairs}, and \emph{functions} are the best features for localizing faults.
\end{result}

However, some features by design are inherently better suited for localizing faults.
For instance, lines and functions covering almost the entire program are likelier to assign reasonable suspiciousness scores to the faulty lines and rank them higher.
Other features can only partially cover the code and may not assign a reasonable suspiciousness score to the actual faulty lines. For instance, loops only cover a small portion of the code and might not be executed in the presence of a failure.
This argument is supported by the high wasted effort of these features, for instance, return values, and the generally poor performance of the particular features, e.g., if a string contains digits or is empty.

\begin{result}
 By design, some features are better suited for localizing faults than others.
\end{result}

\subsubsection{\bfseries\Cref{rq3}: Multi-Feature Localization}%
\label{sub:eval-multi-feature}

\begin{table}
    \caption{Localization of Multiple Features. 
    Results for the localization of faults leveraging \TARANTULA{}, \OCHIAI{}, \DSTAR{}, \NAISHT{}, and \GPOT{} for considering multiple features at once. Each feature is evaluated according to Top-1, Top-5, Top-10, Top-200, \EXAM{} score and wasted effort for three debugging scenarios. The bolted values show that multiple features achieve a better result than any individual one.}%
    \label{tab:results-localization-multiple}
    \centering
    \setlength\extrarowheight{-3pt}
    \resizebox{\textwidth}{!}{%
    \begin{tabular}{llrrrrrrrrrrrrrrrrrr}
    \toprule
    \multicolumn{1}{c}{\multirow{4}*{Feature}} & \multicolumn{1}{c}{\multirow{4}*{Metric}} & \multicolumn{6}{c}{Best-Case Debugging} & \multicolumn{6}{c}{Average-Case Debugging} & \multicolumn{6}{c}{Worst-Case Debugging} \\\cmidrule(lr){3-8}\cmidrule(lr){9-14}\cmidrule(lr){15-20}
    & & \multicolumn{4}{c}{Top-k} & \multicolumn{1}{c}{\multirow{2}*{\EXAM{}}} & \multicolumn{1}{c}{\multirow{2}*{Effort}}
 & \multicolumn{4}{c}{Top-k} & \multicolumn{1}{c}{\multirow{2}*{\EXAM{}}} & \multicolumn{1}{c}{\multirow{2}*{Effort}}
 & \multicolumn{4}{c}{Top-k} & \multicolumn{1}{c}{\multirow{2}*{\EXAM{}}} & \multicolumn{1}{c}{\multirow{2}*{Effort}}
\\\cmidrule{3-6}\cmidrule{9-12}\cmidrule{15-18}
    & & \multicolumn{1}{c}{1} & \multicolumn{1}{c}{5} & \multicolumn{1}{c}{10} & \multicolumn{1}{c}{200} & &
 & \multicolumn{1}{c}{1} & \multicolumn{1}{c}{5} & \multicolumn{1}{c}{10} & \multicolumn{1}{c}{200} & &
 & \multicolumn{1}{c}{1} & \multicolumn{1}{c}{5} & \multicolumn{1}{c}{10} & \multicolumn{1}{c}{200} & &
\\\midrule
     & \TARANTULA{}\rowstrut{} & 8.2\% & 21.2\% & 30.2\% & 55.6\% & \textbf{\color{deepblue}0.056} & \textbf{\color{deepblue}2.2k} & 5.4\% & 15.1\% & 22.7\% & 47.3\% & \textbf{\color{deepblue}0.084} & \textbf{\color{deepblue}5.2k} & 3.8\% & 11.4\% & 16.6\% & 37.5\% & \textbf{\color{deepblue}0.178} & \textbf{\color{deepblue}10.1k} \\
     & \OCHIAI{} & 8.6\% & 21.1\% & 30.3\% & 54.5\% & \textbf{\color{deepblue}0.056} & \textbf{\color{deepblue}2.2k} & 5.8\% & 15.3\% & 23.1\% & 46.4\% & \textbf{\color{deepblue}0.085} & \textbf{\color{deepblue}5.2k} & 4.1\% & 11.7\% & 17.0\% & 36.9\% & \textbf{\color{deepblue}0.178} & \textbf{\color{deepblue}10.1k} \\
    $\text{Multi}_\text{max}$ & \DSTAR{} & 5.7\% & 16.3\% & 20.3\% & 47.2\% & \textbf{\color{deepblue}0.082} & \textbf{\color{deepblue}3.0k} & 3.5\% & 11.6\% & 14.8\% & 40.8\% & \textbf{\color{deepblue}0.110} & \textbf{\color{deepblue}6.0k} & 2.4\% & \textbf{\color{deepblue}8.7\%} & 10.9\% & 32.4\% & \textbf{\color{deepblue}0.197} & \textbf{\color{deepblue}10.6k} \\
     & \NAISHT{} & 8.5\% & 20.2\% & 28.4\% & 53.1\% & \textbf{\color{deepblue}0.060} & \textbf{\color{deepblue}2.3k} & 5.7\% & 15.4\% & 22.2\% & 45.7\% & \textbf{\color{deepblue}0.087} & \textbf{\color{deepblue}5.3k} & 4.1\% & 11.9\% & 16.6\% & 36.5\% & \textbf{\color{deepblue}0.179} & \textbf{\color{deepblue}10.1k} \\
     & \GPOT{} & 8.5\% & 20.2\% & 28.5\% & 53.1\% & \textbf{\color{deepblue}0.060} & \textbf{\color{deepblue}2.3k} & 5.7\% & 15.3\% & 22.2\% & 45.7\% & \textbf{\color{deepblue}0.087} & \textbf{\color{deepblue}5.3k} & 4.1\% & 11.9\% & 16.6\% & 36.4\% & \textbf{\color{deepblue}0.179} & \textbf{\color{deepblue}10.1k} \\[.2em]
\rowcolor{row}
     & \TARANTULA{}\rowstrut{} & 8.2\% & 21.1\% & 30.2\% & 55.5\% & 0.161 & 6.3k & 5.3\% & 15.1\% & 22.7\% & 47.3\% & 0.181 & 9.1k & 3.8\% & 11.4\% & 16.6\% & 37.5\% & 0.249 & 13.3k \\
\rowcolor{row}
     & \OCHIAI{} & 8.6\% & 21.1\% & 30.3\% & 54.6\% & 0.173 & 6.6k & 5.7\% & 15.3\% & 23.1\% & 46.4\% & 0.188 & 9.3k & 4.1\% & 11.7\% & 17.1\% & 37.1\% & 0.252 & 13.4k \\
\rowcolor{row}
    $\text{Multi}_\text{mean}$ & \DSTAR{} & 5.7\% & \textbf{\color{deepblue}16.3\%} & 20.3\% & 47.1\% & 0.173 & 6.6k & 3.5\% & \textbf{\color{deepblue}11.7\%} & 14.7\% & 40.7\% & 0.188 & 9.3k & 2.4\% & 8.7\% & 10.9\% & 32.4\% & 0.252 & 13.4k \\
\rowcolor{row}
     & \NAISHT{} & 8.4\% & 20.2\% & 28.3\% & 53.2\% & 0.173 & 6.6k & 5.7\% & 15.3\% & 22.2\% & 45.7\% & 0.188 & 9.3k & 4.1\% & 11.9\% & 16.6\% & 36.5\% & 0.252 & 13.4k \\
\rowcolor{row}
     & \GPOT{} & 8.5\% & 20.2\% & 28.4\% & 53.1\% & 0.173 & 6.6k & 5.7\% & 15.3\% & 22.2\% & 45.7\% & 0.188 & 9.3k & 4.1\% & 11.9\% & 16.6\% & 36.4\% & 0.252 & 13.4k \\[.2em]
\bottomrule
\end{tabular}

    }
\end{table}

After evaluating the näive multi-feature technique for \Cref{rq3}, we found it generally performs well but does not outperform the best individual feature classes for localizing faults for most top-k metrics and debugging scenarios, which is reflected in \Cref{tab:results-localization-1,tab:results-localization-2}.
However, the approach eliminates the flaws of the individual features.
For instance, when compared with lines, the multi-feature technique, considering the maximum suspiciousness, does not reach the results of lines for the top-5. Still, with increasing k, the multi-feature technique outperforms lines.
The same holds for the approach of leveraging mean suspiciousness.
Vice versa, this observation applies to functions where the multi-feature technique does not reach the results of functions for the top-200, but with decreasing k, it outperforms functions.

Moreover, the multi-feature technique achieves significantly lower \EXAM{} scores and wasted effort than the individual features, indicating that the multi-feature technique is more likely to suggest the faulty lines early.
These findings are consistent across all debugging scenarios, indicating that the multi-feature technique is generally beneficial for localizing faults.
These benefits are compelling, considering a developer must still inspect the suggested lines to find the fault.

\begin{result}
 Leveraging multiple execution features considerably reduces the effort to find the faulty lines.
\end{result}

Finally, we can conclude that the maximum suspiciousness performs better than the mean for the multi-feature technique.

\section{Learning Diagnoses from Execution Features}%
\label{sec:debugging}

Now that we have established the effectiveness of execution features in identifying failure-inducing properties in executions, can we make them actionable for developers?
In software debugging, developers often face the challenge of diagnosing faults based on available test cases.
\EFDD{} (\textbf{E}xecution-\textbf{F}eature-\textbf{D}riven \textbf{D}ebugging) addresses this by systematically transforming raw execution data into actionable insights, helping developers localize faults with greater precision and interpretability.

We focus on scenarios where developers are equipped with a faulty program and a set of labeled test cases—comprising both passing and failing cases—that effectively expose the fault.
The primary objective of \EFDD{} is to utilize these labeled test executions to infer an interpretable diagnosis that explains why specific test cases trigger failures while others succeed.
At its core, \EFDD{} leverages the contrast between passing and failing execution features to extract meaningful diagnoses.
Analyzing execution traces identifies correlations between execution features and observed outcomes, enabling an execution-driven approach to fault diagnosis.

\begin{figure}
    \includegraphics[width=\textwidth]{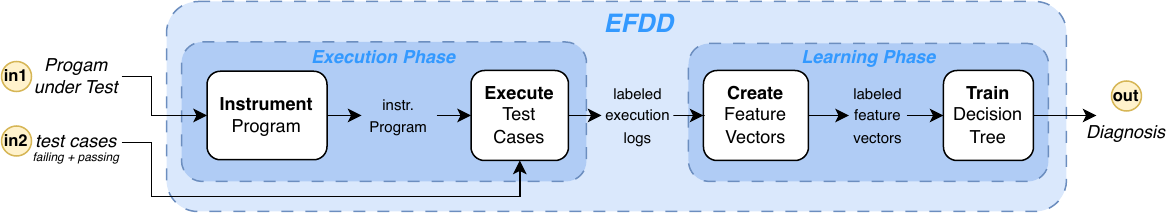}
    \caption{\EFDD{} at a glance. \EFDD{} takes a program and a set of labeled test cases as input, instruments the program, executes the test cases, and captures an execution trace. From this trace, it constructs execution features to train a decision tree. The resulting model offers an interpretable diagnosis for the observed fault.}
    \label{fig:overview-efdd}
\end{figure}

\Cref{fig:overview-efdd} outlines the architecture of \EFDD{}, which is composed of two primary phases: the Execution Phase and the Learning Phase.
The process unfolds through the following sequential steps:

\begin{enumerate*}[label=(\arabic*)]
    \item (\emph{Program Instrumentation}) \EFDD{} begins by instrumenting the program under test. This instrumentation embeds probes within the code to capture execution events, allowing us to understand the program's dynamic behavior.
    \item (\emph{Test Execution}) The instrumented program is then executed using the provided test cases. Both passing and failing tests are run to gather diverse event execution traces. Each trace consists of a chronological sequence of events triggered during the program's execution, offering a fine-grained view of the program's behavior under different inputs.
    \item (\emph{Feature Extraction}) From the collected execution traces, \EFDD{} constructs feature vectors that abstract relevant characteristics of each run. These features capture key execution properties—such as the frequency of certain events, the activation of specific code paths, or data dependencies—that may correlate with faults. The feature engineering process ensures that the resulting vectors are informative and amenable to machine learning.
    \item (\emph{Model Training}) With the labeled feature vectors, \EFDD{} trains a decision tree classifier. This model learns to discriminate between passing and failing executions by identifying patterns in the extracted features. The choice of a decision tree ensures that the resulting diagnosis is transparent and easily understandable by developers, as the tree structure naturally maps to logical conditions that can be traced back to the program's behavior.
    \item (\emph{Diagnosis Generation}) Finally, \EFDD{} analyzes the trained decision tree to produce a diagnosis that pinpoints the program behaviors most strongly associated with failures. The diagnosis highlights critical decision points and feature thresholds that differentiate failing runs from passing ones, providing developers with clear insights into the root cause of the fault.
\end{enumerate*}

For any subsequent input or test, \EFDD{} would streamline this process.
It executes the input, captures the events, formulates the features, and then classifies the run based on the initially trained model.

\subsection{Training a Decision Tree Model}%
\label{sub:classifying}
The concluding step of our approach involves training a machine learning classifier using the feature vectors we have constructed.
Inspired by \ALHAZEN{}~\cite{kampmann2020alhazen}, we opt for decision trees as our classifier of choice for several reasons.
First, our features predominantly consist of discrete values, making them well-suited for decision tree-based classification.
These features enable us to distinguish effectively between failing and passing runs through straightforward combinations of feature values.
Second, decision trees' explainable nature allows us to identify relevant features that contribute to fault detection quickly.
Finally, this model enables classifying an unseen test input's behavior by correlating extracted execution features with program failures.

\begin{wrapfigure}{l}{0.45\textwidth}
    \centering
    \begin{tikzpicture}[font=\sffamily, sibling distance=6em]
    \node {\begin{tabular}{c}\texttt{return y} executed? \\[-.4em] \scriptsize middle.py:6\end{tabular}}
	child {
        node {\PASS{}}
        edge from parent node[left]{no}
    }
	child {
        node {\begin{tabular}{c}y $\ge$ x? \\[-.4em] \scriptsize middle.py:1\end{tabular}}
        child {
            node {\FAIL{}}
            edge from parent node[left]{no}
        }
        child {
            node {\PASS{}}
            edge from parent node[right]{yes}
        }
        edge from parent node[right]{yes}
    };
\end{tikzpicture}
    \caption{The decision tree generated by \EFDD{} for the \texttt{middle()} example. Each node represents a decision, leading to a classification of either \PASS{} (Pass) or \FAIL{} (Fail).}%
    \label{fig:tree}
\end{wrapfigure}
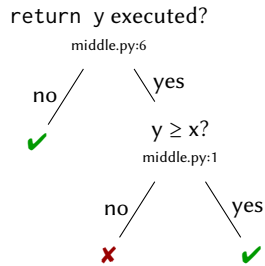

Our preliminary experiments with various classifiers, including other naive classifiers, neural networks, and large language models, consistently revealed that decision trees either outperformed or matched the performance of the best classifiers in all cases, making them ideal candidates for our classification models.
Additionally, imbalanced sample sizes were not an issue, as most experiments were, as in the real world, with fewer failing than passing examples.

We are committed to maintaining impartiality in our classification process and do not favor detecting faults over identifying correct executions.
We consider passing and failing executions equally valuable for learning an adequate diagnosis.
Biasing our classification towards a particular outcome could lead to issues in any downstream applications of our approach.

\subsection{Deriving Diagnoses}%
\label{sub:diagnoses}

A central design goal of \EFDD{} is to ensure the \textbf{interpretability} of its machine learning model, enabling developers to gain clear insights into the underlying causes of software faults.
Using decision trees as the core diagnostic model was a deliberate choice, as they inherently provide a transparent and logical decision-making flow that developers can easily follow.
By framing the diagnosis as a decision tree, \EFDD{} transforms complex execution traces into intuitive, rule-based explanations.
Each node in the tree represents a conditional check derived from the program's execution features, guiding developers through the logical paths that lead to passing or failing outcomes.
This approach localizes faults and offers a deeper understanding of the execution contexts that trigger failures.

We illustrate this interpretability through the \texttt{middle()} example introduced in \Cref{sec:introduction}.
Applying \EFDD{} to this case, using the initial set of labeled test cases, yields the decision tree shown in \Cref{fig:tree}. 

The decision tree reveals two critical features associated with the fault:
\begin{enumerate*}
    \item \textbf{Execution of Line 6:} Whether the statement \texttt{return y} at Line~6 was executed.
    \item \textbf{Comparison of Variables (\texttt{y} $\ge$ \texttt{x}):} Whether the value of \texttt{y} is greater than or equal to \texttt{x} when entering the function.
\end{enumerate*}
The fault occurs under a specific condition: when Line~6 is executed and \texttt{y} is not greater than or equal to \texttt{x}.
This precise insight directs developers to the faulty behavior without combing through the entire execution trace.
From the decision tree, we can infer the following:
\begin{enumerate*}
    \item If Line~6 is \textbf{not} executed, the test case passes—irrespective of other conditions.
    \item If Line~6 \textbf{is} executed and \texttt{y} $\ge$ \texttt{x}, the execution still passes.
    \item However, if Line~6 is executed \textbf{and} \texttt{y} $<$ \texttt{x}, the failure is triggered.
\end{enumerate*}
This diagnosis immediately highlights both the location (\textbf{Line~6}) and the condition leading to the fault (\texttt{y} $<$ \texttt{x}), offering actionable information to the developer.
The comprehensible nature of the decision tree makes it an ideal candidate for diagnosis generation.
Since the model identifies the faulty line and the conditions leading to failure, developers can systematically target these conditions to design a correct fix.

\subsection{Implementation}%
\label{sub:implementation-efdd}

Our implementation of \EFDD{} is built upon \SFLKIT{}~\cite{smytzek2022sflkit}.
\SFLKIT{} provides the means to instrument a \PYTHON{} program under test and collect a trace of execution events.
In our implementation, we utilize \SFLKIT{}'s built-in iteration over the event trace to gather features.
Specifically, while \SFLKIT{} processes the events to extract various predicates and spectra (such as coverage information), we inject our feature collector into this loop.
This feature collector constructs the features based on the extracted data.
Once the feature vectors are constructed, we convert them into data frames suitable for feeding directly into a machine-learning classifier.
We chose \SCIKITLEARN{}'s decision tree learner as the backbone of our diagnosis.
Although \SCIKITLEARN{} offers a variety of classifiers, we restricted ourselves to decision trees for several reasons:
\begin{enumerate*}[label=(\alph*)]
\item The inherent explainability of decision trees, as detailed in \Cref{sub:classifying}.
\item The focus of this work is on the overarching approach, not on classifier comparisons.
\end{enumerate*}

\section{Evaluation}%
\label{sec:evaluation}

To assess the effectiveness of \EFDD{}, we designed our evaluation around the following key research question:

\begin{questions}[topsep=2pt]
    \setcounter{questionsi}{3}
    \item\label[question]{rq4} \textbf{Diagnosis' Quality}. How accurate are the diagnoses generated by \EFDD{} from \Cref{sec:debugging}?
\end{questions}

\subsection{Experimental Setup}%
\label{sec:setup}

The evaluation focuses on measuring \EFDD{}'s capability to generate accurate and interpretable diagnoses by assessing how effectively the generated diagnosis can classify the presence of faults based on program executions.
Specifically, we evaluate the diagnosis as a classifier that distinguishes between failing and passing runs, aligning with established practices in this research domain~\cite{kampmann2020alhazen,eberlein2023avicenna}.
Unlike prior approaches such as \ALHAZEN{} and \AVICENNA{}, which evaluate both the generative and predictive capabilities of their models, our evaluation exclusively focuses on the predictive accuracy of the diagnosis.
This distinction stems from the fact that \EFDD{} does not generate new program inputs—a limitation discussed further in \Cref{sub:related-execution-features}—but instead builds diagnoses based on existing execution data.

To ensure a rigorous and meaningful evaluation, we selected subject programs based on the following criteria:
\begin{description}[leftmargin=5pt,noitemsep,partopsep=0pt,topsep=0pt,parsep=0pt]
    \item[Fault Impact on Output.] Each subject must include a fault that affects the program's state and, consequently, its output, ensuring that the fault is observable through test executions and making diagnosis feasible.
    
    \item[Emphasis on Functional Bugs.] Most subjects focus on functional bugs that cause incorrect outputs rather than bugs that lead to exceptions or crashes. Functional bugs often require deeper analysis, aligning with \EFDD{}'s core strength. While exceptions can signal faults explicitly, functional bugs pose a greater diagnostic challenge.

    \item[Realistic and Isolated Defects.] Subjects are chosen to reflect real-world defects. Each subject contains exactly one bug to isolate the diagnostic process. Although \EFDD{} is currently designed to analyze one fault at a time, it can handle multiple bugs in a single program if tests clearly distinguish between them, allowing separate diagnoses for each.

    \item[Availability of Ground Truth.] Each subject includes a known ground truth, enabling a direct comparison between the diagnosed fault and the actual defect, which is critical for quantitatively assessing the accuracy of \EFDD{}'s diagnoses.

    \item[Labeled Test Cases.] Each subject provides at least one set of pre-labeled test cases, including failing and passing tests. This initial labeled set is a foundation for training and evaluating \EFDD{} and ensures the fault is triggered and avoided across different test executions.
\end{description}

Based on our benchmark requirements, we selected \REFACTORY{}~\cite{hu2019refactory} for our evaluation.
\REFACTORY{} is a comprehensive benchmark of student submissions for five distinct programming tasks.
It provides initial input-output pairs and a correct reference implementation for each task, making it an ideal candidate for evaluating fault localization and diagnosis tools like \EFDD{}.
For each subject in the benchmark, we label the generated inputs as either \textit{passing} or \textit{failing} by comparing the subject's actual output against the expected output from the correct implementation.
A test is labeled passing if the program's output matches the expected output and labeled failing if the program's output deviates from the expected result.
We then use these labeled examples to diagnose each subject based on the initial seed inputs.
However, we apply a filtering step to exclude uninformative subjects that either fail or pass on all the initial seeds.

\begin{wraptable}{l}{.4\textwidth}
	\caption{Quality of the generated diagnoses by \EFDD{}.}%
	\label{tab:results-quality}
    \centering
    %!TEX root = ../bashiri.tex
%!TEX spellcheck = en-US

\begin{tabular}{lrr}    
    \toprule
    \REFACTORY{}& \multicolumn{1}{c}{\BUG{}}    & \multicolumn{1}{c}{\NOBUG{}}  \\\midrule
    \PRECISION{}    & $0.8985$                      & $0.8875$                      \\
    \rowcolor{row}\RECALL{}       & $0.9072$                      & $0.8772$                      \\
    \FONESCORE{}    & $0.9029$                      & $0.8823$                      \\\midrule
    \rowcolor{row}\ACCURACY{}             & \multicolumn{2}{r}{$89.36$\%} \\
    \MACRO{} \PRECISION{}   & \multicolumn{2}{r}{$0.8930$}  \\
    \rowcolor{row}\MACRO{} \RECALL{}      & \multicolumn{2}{r}{$0.8922$}  \\
    \MACRO{} \FONESCORE{}   & \multicolumn{2}{r}{$0.8926$}  \\\midrule
    \rowcolor{row}\# of subjects          & \multicolumn{2}{r}{$1777$}   \\
    avg execution time      & \multicolumn{2}{r}{$3.90$s}  \\\bottomrule
\end{tabular}
\end{wraptable}

To assess the accuracy of the diagnoses generated by \EFDD{}, we generated 400~labeled evaluation input examples for each subject. We evaluated the model's ability to classify passing and failing runs correctly.
We compute the achieved accuracy, precision, recall, and F1 score based on the evaluation input.
To allow an efficient analysis, we evaluate all considered metrics (i.e., accuracy, precision, recall, F1 score, and their macro counterparts) of all subjects not by averaging but by considering all correctly labeled and mislabeled predictions.

\subsection{\Cref{rq4}: Diagnosis Quality}%
\label{sec:results-diagnosis}

For the quality of the diagnosis, we consider the 1777 subjects that pass our requirements from the \REFACTORY{} benchmark.
\Cref{tab:results-quality} comprises all our results over all subjects for each benchmark.

Our approach generated diagnoses that can distinguish passing from failing executions with an overall accuracy of $89.36\%$ and a macro F1 score of $0.89$, showing the predictive power of these diagnoses.
On 884 subjects from \REFACTORY{}, we could infer diagnoses that could distinguish all runs from our evaluation set, which means that in 50\% of the cases, the diagnoses were as sound as they could be in our evaluation setup.
Overall, all our evaluation metrics show stable values without significant outliers, demonstrating again how powerful and balanced, in terms of not favoring non-buggy features over buggy ones since they are more represented in the training sets, the diagnoses generated by \EFDD{} can be.
In addition, an average execution time of \EFDD{} of 3.9 seconds on our evaluation machines is a manageable workload with a highly beneficial outcome.

\begin{result}
    From a small set of given labeled tests, \EFDD{} can generate diagnoses that can differentiate unseen passing and failing program runs with a high predictive power, implying the high accuracy of the diagnoses.
\end{result}

\section{Threats to Validity}%
\label{sec:threats-to-validity}

\subsection{Execution Features Study}%
\label{sub:threats-study}

\subsubsection*{Internal Validity}
One potential threat to internal validity is the accuracy of the event collection process.
Any inaccuracies in capturing execution features could lead to incorrect correlations between features and failures.
To mitigate this, we ensured that the instrumentation and data collection processes were consistent across all experiments.
Another concern is the potential for biases in the test cases the \TESTS4PY{} dataset provides, which we tackled by verifying that the test cases failed and passed as they should in the buggy and fixed versions.

\vspace*{-.5\baselineskip}
\subsubsection*{External Validity}
Selecting projects from the \TESTS4PY{} dataset primarily threatens our study's external validity.
While this dataset offers a diverse set of Python projects, the results may not generalize to projects written in other programming languages or those with different characteristics, such as size or complexity.
Additionally, the dataset's focus on open-source projects may only partially represent the range of proprietary or industrial software failures.

\vspace*{-.5\baselineskip}
\subsubsection*{Construct Validity}
Construct validity concerns arise from the metrics used to evaluate correlation and fault localization.
To mitigate this risk we included multiple well-established coefficients (\TARANTULA{}, \OCHIAI{}, \DSTAR{}, \NAISHT{}, and \GPOT{}) and evaluation metrics (Top-1, Top-5, Top-10, Top-200, \EXAM{} score, and wasted effort).
However, these metrics may only partially align with practical debugging experiences~\cite{parnin2011automated}, but we attempted to address this by providing a comprehensive analysis across multiple metrics and cross-verified our results with these.

\vspace*{-.5\baselineskip}
\subsubsection*{Conclusion Validity}
Conclusion validity could be impacted by statistical methods correlating execution features with failures. These methods should be corrected to avoid incorrect conclusions about the relevance of certain features. We mitigated this threat by employing established statistical fault localization techniques and cross-verifying our results with multiple metrics. Additionally, using a single dataset limits our empirical analysis's diversity, potentially affecting our conclusions' robustness. Future work could involve replicating the study on additional datasets to strengthen the generalizability of our findings.

\subsection{Execution Features Driven Debugging}%
\label{sub:threats-debugging}

\subsubsection*{Internal Validity}
Concerning implementing our diagnosis approach, we cannot verify that it realizes the exact approach presented in \Cref{sec:debugging}.
However, the results of our experiments from \Cref{tab:results-quality} should eliminate the risk of any significant flaws in our implementation.
Moreover, the manual inspection of the diagnostic power of our approach in \Cref{sub:diagnoses} reinforces this claim.
Regarding our experiments for \Cref{rq1} we leveraged our implemented test generation for \REFACTORY{} that may be
\begin{enumerate*}[label=(\alph*)]
    \item incomplete, i.e., there are possible tests that will not be generated,
    \item easy to distinguish by simple features,
    \item or produce inputs that the program was not designed to handle.
\end{enumerate*}
For the test generation for \REFACTORY{}, we tried to stay true to the tests provided by this benchmark and introduce randomness to cover as much of the input space as possible.
Moreover, all generated inputs produced a result when executed on the correct implementation for each question, showing that they are at least accepted.
However, when considering the test generation and our implementation, another threat could be that we learn diagnoses based on the execution features but instead overfit to the generators, i.e., learn only to distinguish passing and failing runs that the generator produces.
We tried to eliminate this risk by ensuring that the generators were as general as possible (considering that they should still trigger the fault) and assuming an adequate number of unseen inputs for our evaluation sets.

\vspace*{-.5\baselineskip}
\subsubsection*{External Validity}
The selection of our subjects might not be sufficient to show the applicability of our approach to an unseen real-world program.
This threat might be heavy, especially considering the subjects in \REFACTORY{} student submissions to relatively small tasks.
Since the critical part of our approach is the underlying \SFLKIT{}~\cite{smytzek2022sflkit} as its base, our approach should be applicable whenever \SFLKIT{} is, which was already verified and tested on the entire \BUGSINPY{}~\cite{widyasari2020bugsinpy} benchmark.
Because of this reason, we would consider that \EFDD{} applies to actual programs, and our evaluation's results are generalizable and can transfer to an unseen fault.

Since our evaluation is done in \PYTHON{}, we cannot mitigate the risk that \EFDD{} does not apply to other programming languages.
However, arguing that our execution features are general and could also be extracted from programs implemented in other programming languages, like \C{} or \JAVA{}, we are convinced that our technique is general.

\vspace*{-.5\baselineskip}
\subsubsection*{Construct Validity}
Another concern is the metrics we have chosen to evaluate the results for \Cref{rq1} that might be insufficient to show our diagnosis quality.
We leveraged several established metrics (precision, recall, F1 score, and accuracy) to counter this threat for all our measurements.

\vspace*{-.5\baselineskip}
\subsubsection*{Conclusion Validity}
The statistical methods used to evaluate the quality of the diagnoses could affect the conclusion's validity.
To mitigate this risk, we cross-verified our results with multiple metrics.

\section{Related Work}%
\label{sec:related-work}

\subsection{Fault Localization}%
\label{sub:related-fault-localization}

Fault localization is a critical area in software debugging.
It aims to pinpoint the exact locations in the code responsible for failures.
Several traditional techniques have been extensively studied, particularly statistical methods that introduce similarity coefficients such as \TARANTULA{}~\cite{jones2005tarantula}, \OCHIAI{}~\cite{abreu2006ochiai}, \DSTAR{}~\cite{wong2012dstar}, \GP{}~\cite{xie2013gp}, \AMPLE{}~\cite{dallmeier2005ample}, \JACCARD{}~\cite{chen2002jaccard}, and many others.
Spectrum-based fault localization (SBFL) techniques, including those mentioned, are especially notable for using execution traces to estimate the likelihood that specific code elements are faulty statistically.
In recent years, these techniques have gained traction in automated debugging, particularly within automatic program repair frameworks, where they significantly contribute to identifying code regions for repair~\cite{qi2013apr,lutellier2020coconut}.

Other research in this area has shifted the focus from using lines as the unit for localization to other coverage-based information.
Zhang et al.~\cite{zhang2009capturing} examined basic blocks, Le et al.~\cite{le2010path} investigated paths, Jiang et al.~\cite{jiang2023variable} leveraged variables and decision trees, and Vancsics et al.~\cite{vancsics2021calls} considered call frequency.
Ribeiro et al.~\cite{ribeiro2019dataflow} used data flow for localizing faults.
The work by Yu et al.~\cite{yu2011models} leverages data and control dependency models to calculate suspiciousness scores.
In contrast, Yan et al.~\cite{yan2023context} employ traditional SBFL but modify the scores afterward based on the context in which the fault propagates.
Other approaches do not consider coverage of the traditional SBFL.
Soremekun et al.~\cite{soremekun2021slicing} present an approach that builds on program slices for localization, while Papadakis et al.~\cite{papadakis2014mutation,papadakis2015mutation} leverage mutation analysis.

The extensive research in this area has led to numerous metrics designed to assess the correlation between code locations and failures~\cite{daniel2013sbfl,landsberg2015sfl,naish2011sbfl}.
Parnin et al.~\cite{parnin2011automated} critically evaluate these techniques and question their practical applicability for developers.
They conclude that while these approaches can reduce the search space, they often overwhelm developers with numerous potential fault locations, making pinpointing the exact faulty lines challenging.
Another comprehensive survey by Soremekun et al.~\cite{soremekun2023evaluating} investigates general assumptions of fault localization, finding that most developers would prefer a diagnosis to locations for debugging.
Other studies, such as those by Abreu et al.~\cite{abreu2009practical}, Pearson et al.~\cite{pearson2017sfl}, Heiden et al.~\cite{heidengkkhfl19}, and Widyasari et al.~\cite{widyasari2022sfl}, have evaluated fault localization techniques across different benchmarks to assess their effectiveness and limitations.
In contrast to our work, they did not consider multiple features in their evaluation but concentrated on lines as the unit of measure for localization.

Beyond traditional fault localization, recent research has expanded to leverage these techniques in novel ways.
For instance, Le et al.~\cite{lelgg16} use a learning-to-rank-based approach that integrates knowledge from mined likely invariants~\cite{ernstpgmptx07} to prioritize functions that are more likely to be the cause of faults.
Additionally, approaches like \ENTBUG{}~\cite{camposafd13} utilize fault localization to drive test generation, incorporating localization metrics into the fitness of a genetic test generator.

Considering how statistical fault localization needs to execute the entire test suite to provide relevant code locations, Jiang et al.~\cite{jiangzctc12} explore how test case prioritization could improve the process.
Yoo et al.~\cite{yoo2013priorization} and Gonzales-Sanchez et al.~\cite{gonzalez-sanchezagg11} build an approach that applies prioritization not only to reduce execution time but also to enhance fault localization results.
Gong et al.~\cite{gongwsm13} directly reduce the test suite to achieve similar improvements.
In contrast, Xuan et al.~\cite{xuanm14b} purify test cases by splitting them into smaller parts, providing more granular insights, and leveraging these purified tests to improve statistical fault localization.
Moreover, test generation can also enhance fault localization results.
For example, Artzi et al.~\cite{artzidtp10b} show that generating directed test cases maximizing the similarity between path constraints of generated tests and those of faulty executions outperforms undirected generation.
While our approach seeks to maximize execution differences, the concept aligns with test generation methodologies.

With the rise of machine learning techniques, researchers have explored using various learning techniques for fault localization~\cite{li2019deepfl,li2021covrepresentation,widyasari2022xai4fl,wang2024mtltransfer,yang2024multilingual,li2022cc}.
Notably, feature-based fault localization approaches by Lei et al.~\cite{lei2022featurefl} abstract program behaviors as features, while Meng et al.~\cite{meng2022transfer} leverage semantic features.
Additionally, research has investigated the impact of large language models on fault localization~\cite{kang2024llm}.

This body of work highlights both the progress made in fault localization and the challenges that remain, particularly in enhancing the practical utility and precision of these techniques for developers.

\subsection{Feature-based Debugging}%
\label{sub:related-execution-features}

Recent research also uses \emph{features} for deriving debugging diagnoses.
\ALHAZEN{}~\cite{kampmann2020alhazen} was one of the first approaches that leveraged features for debugging.
The approach learns a decision tree from a set of \emph{input features} and iteratively refines the tree by generating new inputs that aim to trigger the failure.
From the decision tree, \ALHAZEN{} derives a diagnosis.
Similar to \ALHAZEN{}, \AVICENNA{}~\cite{eberlein2023avicenna} leverages \emph{input features} for debugging.
It follows the same refinement loop but leverages a sophisticated constraint learner to derive a diagnosis and generates new inputs by solving these constraints.

In contrast, our work opens the field for \emph{execution-feature-driven debugging} that can generate diagnoses explaining the actual program behavior that leads to a failure, providing a more comprehensive understanding of the failure while \ALHAZEN{} and \AVICENNA{} focus on the input space that is not as useful for debugging as the execution space.
However, our approach can, at this point, not effortlessly generate new inputs to trigger the failure since we cannot directly map the execution features to the input space, but this is mitigated by the fact that \ALHAZEN{} and \AVICENNA{} require a specification of the input space. In contrast, our approach runs out-of-the-box on any program.
We believe that \ALHAZEN{} or \AVICENNA{} would complement our approach.

\section{Conclusion and Future Work}%
\label{sec:conclusion}

First, we conducted an empirical study that explored alternative execution features that could enhance fault localization in software programs.
By leveraging the \TESTS4PY{} dataset, we analyzed the correlation between these features and the presence of failures using various statistical fault localization techniques, including \TARANTULA{}, \OCHIAI{}, \DSTAR{}, \NAISHT{}, and \GPOT{}.
Our findings indicate that incorporating a diverse set of execution features can improve the accuracy and precision of fault localization.

Second, we introduced a novel technique to generate accurate diagnoses that refer to the features investigated in our study, such as lines executed or variable values, making them easy to read and assess.

Our future work will focus on the following topics:

\begin{description}[leftmargin=5pt,noitemsep,partopsep=0pt,topsep=0pt,parsep=0pt]
     \item[More Programming Languages.] To further validate our findings, we will build on our findings by extending the analysis to a broader range of programming languages.
    \item[Test Generation.] We aim to map execution features to the input space, allowing us to directly generate new inputs to trigger the fault based on the generated diagnosis.
    \item[Automated Repair.] In automated repair, the diagnosis inferred by our approach helps to generate patches that fix the fault.
 For instance, the diagnosis of the \texttt{middle()} example from \Cref{fig:tree} could directly suggest to replace the variable \texttt{y} with \texttt{x} in the \texttt{return} in Line~6 for a program repair.
    \item[More Execution Features.] Despite our capability to produce precise diagnoses, our approach's performance depends on its set of execution features: If a failure does not depend on any of the features collected, it will be hard to produce an accurate diagnosis.
 This limitation can be countered by adding more features, notably \emph{derived} features such as string or arithmetic properties---but if a failure occurs, say, whenever $n$ is a perfect number, and $d$ is a day with a full moon, it will still be hard to detect these features.
    \item[Program Synthesis.] By selecting relevant features, we effectively synthesize a \emph{predicate}. Such predicates can also be synthesized through symbolic means~\cite{gulwani2017synthesis} and possibly yield better results while still being explainable.
    \item[Deep Learning Models.] If one can live without explainability, many machine learning models are available that may all result in applicable diagnoses with even higher accuracy.
 Nevertheless, the decision tree classifier outperformed all other models investigated during our preliminary experiments.
 However, given the vast number of available models and the frequency with which novel models are introduced, it is worth investigating this area further, e.g., by conducting a large-scale study.
    \item[Testing Oracle.] Based on our diagnoses and an adequate mapping approach that allows us to relate features of a program version to an altered version of the same program, we plan on investigating the possibility of generating a testing oracle that can differentiate between passing and failing tests.
    \item[User Study.] To evaluate the usability of our diagnosis approach, we plan to conduct a user study with developers to assess the effectiveness of the diagnoses generated by our approach. 
\end{description}

\section*{Data and Tool Availability}%
\label{sec:tool}

Our evaluation data, all scripts, and our \EFDD{} artifact are all open source.
The current versions of evaluation scripts and \EFDD{} can be downloaded from
\vspace{-.2\baselineskip}
\begin{center}
    \url{https://github.com/smythi93/efdd}%
\end{center}
\vspace{-.2\baselineskip}
The dataset containing the collected events, features, and all intermediate results is available at
\vspace{-.2\baselineskip}
\begin{center}
    \url{https://doi.org/10.5281/zenodo.14909966}%
\end{center}

\begin{acks}
    This research was partially funded by the Deutsche Forschungsgemeinschaft (DFG, German Research Foundation) --- ZE 509/7--2 and GR 3634/4--2 Emperor (261444241) and by the European Union (ERC S3, 101093186). Views and opinions expressed are, however, those of the authors only and do not necessarily reflect those of the European Union or the European Research Council. Neither the European Union nor the granting authority can be held responsible for them. 
\end{acks}

\bibliographystyle{ACM-Reference-Format}
\bibliography{references}


\begin{thebibliography}{63}

%%% the Web address.

\ifx \showCODEN    \undefined \def \showCODEN     #1{\unskip}     \fi
\ifx \showDOI      \undefined \def \showDOI       #1{#1}\fi
\ifx \showISBNx    \undefined \def \showISBNx     #1{\unskip}     \fi
\ifx \showISBNxiii \undefined \def \showISBNxiii  #1{\unskip}     \fi
\ifx \showISSN     \undefined \def \showISSN      #1{\unskip}     \fi
\ifx \showLCCN     \undefined \def \showLCCN      #1{\unskip}     \fi
\ifx \shownote     \undefined \def \shownote      #1{#1}          \fi
\ifx \showarticletitle \undefined \def \showarticletitle #1{#1}   \fi
\ifx \showURL      \undefined \def \showURL       {\relax}        \fi
% invisible to TeX
\providecommand\bibfield[2]{#2}
\providecommand\bibinfo[2]{#2}
\providecommand\natexlab[1]{#1}
\providecommand\showeprint[2][]{arXiv:#2}

\bibitem[\protect\citeauthoryear{Abreu, Zoeteweij, and Gemund}{Abreu
  et~al\mbox{.}}{2006}]%
        {abreu2006ochiai}
\bibfield{author}{\bibinfo{person}{Rui Abreu}, \bibinfo{person}{Peter
  Zoeteweij}, {and} \bibinfo{person}{Arjan J. C.~van Gemund}.}
  \bibinfo{year}{2006}\natexlab{}.
\newblock \showarticletitle{An Evaluation of Similarity Coefficients for
  Software Fault Localization}. In \bibinfo{booktitle}{\emph{Proceedings of the
  12th Pacific Rim International Symposium on Dependable Computing}}
  \emph{(\bibinfo{series}{PRDC '06})}. \bibinfo{publisher}{IEEE Computer
  Society}, \bibinfo{address}{USA}, \bibinfo{pages}{39–46}.
\newblock
\showISBNx{0769527248}
\urldef\tempurl%
\url{https://doi.org/10.1109/PRDC.2006.18}
\showDOI{\tempurl}


\bibitem[\protect\citeauthoryear{Abreu, Zoeteweij, Golsteijn, and van
  Gemund}{Abreu et~al\mbox{.}}{2009}]%
        {abreu2009practical}
\bibfield{author}{\bibinfo{person}{Rui Abreu}, \bibinfo{person}{Peter
  Zoeteweij}, \bibinfo{person}{Rob Golsteijn}, {and} \bibinfo{person}{Arjan
  J.~C. van Gemund}.} \bibinfo{year}{2009}\natexlab{}.
\newblock \showarticletitle{A Practical Evaluation of Spectrum-based Fault
  Localization}.
\newblock \bibinfo{journal}{\emph{J. Syst. Softw.}} \bibinfo{volume}{82},
  \bibinfo{number}{11} (\bibinfo{date}{Nov.} \bibinfo{year}{2009}),
  \bibinfo{pages}{1780--1792}.
\newblock
\showISSN{0164-1212}
\urldef\tempurl%
\url{https://doi.org/10.1016/j.jss.2009.06.035}
\showDOI{\tempurl}


\bibitem[\protect\citeauthoryear{Artzi, Dolby, Tip, and Pistoia}{Artzi
  et~al\mbox{.}}{2010}]%
        {artzidtp10b}
\bibfield{author}{\bibinfo{person}{Shay Artzi}, \bibinfo{person}{Julian Dolby},
  \bibinfo{person}{Frank Tip}, {and} \bibinfo{person}{Marco Pistoia}.}
  \bibinfo{year}{2010}\natexlab{}.
\newblock \showarticletitle{Directed test generation for effective fault
  localization}. In \bibinfo{booktitle}{\emph{Proceedings of the 19th
  International Symposium on Software Testing and Analysis}} (Trento, Italy)
  \emph{(\bibinfo{series}{ISSTA '10})}. \bibinfo{publisher}{Association for
  Computing Machinery}, \bibinfo{address}{New York, NY, USA},
  \bibinfo{pages}{49–60}.
\newblock
\showISBNx{9781605588230}
\urldef\tempurl%
\url{https://doi.org/10.1145/1831708.1831715}
\showDOI{\tempurl}


\bibitem[\protect\citeauthoryear{Campos, Abreu, Fraser, and d'Amorim}{Campos
  et~al\mbox{.}}{2013}]%
        {camposafd13}
\bibfield{author}{\bibinfo{person}{Jos{\'e} Campos}, \bibinfo{person}{Rui
  Abreu}, \bibinfo{person}{Gordon Fraser}, {and} \bibinfo{person}{Marcelo
  d'Amorim}.} \bibinfo{year}{2013}\natexlab{}.
\newblock \showarticletitle{Entropy-based Test Generation for Improved Fault
  Localization}. In \bibinfo{booktitle}{\emph{Proceedings of the 28th IEEE/ACM
  International Conference on Automated Software Engineering}} (Silicon Valley,
  CA, USA). \bibinfo{address}{Piscataway, NJ, USA}, \bibinfo{pages}{257--267}.
\newblock
\showISBNx{978-1-4799-0215-6}
\urldef\tempurl%
\url{https://doi.org/10.1109/ASE.2013.6693085}
\showDOI{\tempurl}


\bibitem[\protect\citeauthoryear{Chen, Kiciman, Fratkin, Fox, and Brewer}{Chen
  et~al\mbox{.}}{2002}]%
        {chen2002jaccard}
\bibfield{author}{\bibinfo{person}{Mike Chen}, \bibinfo{person}{Emre Kiciman},
  \bibinfo{person}{Eugene Fratkin}, \bibinfo{person}{Armando Fox}, {and}
  \bibinfo{person}{Eric Brewer}.} \bibinfo{year}{2002}\natexlab{}.
\newblock \showarticletitle{Pinpoint: problem determination in large, dynamic
  Internet services}. In \bibinfo{booktitle}{\emph{Proceedings of the 2002
  International Conference on Dependable Systems and Networks}}.
  \bibinfo{pages}{595--604}.
\newblock
\showISBNx{0-7695-1101-5}
\urldef\tempurl%
\url{https://doi.org/10.1109/DSN.2002.1029005}
\showDOI{\tempurl}


\bibitem[\protect\citeauthoryear{Dallmeier, Lindig, and Zeller}{Dallmeier
  et~al\mbox{.}}{2005}]%
        {dallmeier2005ample}
\bibfield{author}{\bibinfo{person}{Valentin Dallmeier},
  \bibinfo{person}{Christian Lindig}, {and} \bibinfo{person}{Andreas Zeller}.}
  \bibinfo{year}{2005}\natexlab{}.
\newblock \showarticletitle{Lightweight Bug Localization with {AMPLE}}. In
  \bibinfo{booktitle}{\emph{Proceedings of the Sixth International Symposium on
  Automated Analysis-Driven Debugging}} (Monterey, California, USA)
  \emph{(\bibinfo{series}{AADEBUG'05})}. \bibinfo{publisher}{Association for
  Computing Machinery}, \bibinfo{address}{New York, NY, USA},
  \bibinfo{pages}{99–104}.
\newblock
\showISBNx{1595930507}
\urldef\tempurl%
\url{https://doi.org/10.1145/1085130.1085143}
\showDOI{\tempurl}


\bibitem[\protect\citeauthoryear{Dandan, Tiantian, Xiaohong, and Peijun}{Dandan
  et~al\mbox{.}}{2013}]%
        {gongwsm13}
\bibfield{author}{\bibinfo{person}{Gong Dandan}, \bibinfo{person}{Wang
  Tiantian}, \bibinfo{person}{Su Xiaohong}, {and} \bibinfo{person}{Ma Peijun}.}
  \bibinfo{year}{2013}\natexlab{}.
\newblock \showarticletitle{A Test-suite Reduction Approach to Improving
  Fault-localization Effectiveness}.
\newblock \bibinfo{journal}{\emph{Comput. Lang. Syst. Struct.}}
  \bibinfo{volume}{39}, \bibinfo{number}{3} (\bibinfo{date}{Oct.}
  \bibinfo{year}{2013}), \bibinfo{pages}{95--108}.
\newblock
\showISSN{1477-8424}
\urldef\tempurl%
\url{https://doi.org/10.1016/j.cl.2013.04.001}
\showDOI{\tempurl}


\bibitem[\protect\citeauthoryear{Daniel and Sim}{Daniel and Sim}{2013}]%
        {daniel2013sbfl}
\bibfield{author}{\bibinfo{person}{Patrick Daniel} {and}
  \bibinfo{person}{Kwan~Yong Sim}.} \bibinfo{year}{2013}\natexlab{}.
\newblock \showarticletitle{Spectrum-based Fault Localization: A Pair Scoring
  Approach}.
\newblock \bibinfo{journal}{\emph{Journal of Industrial and Intelligent
  Information}}  \bibinfo{volume}{1} (\bibinfo{year}{2013}),
  \bibinfo{pages}{185--190}.
\newblock
\urldef\tempurl%
\url{https://doi.org/10.12720/jiii.1.4.185-190}
\showDOI{\tempurl}


\bibitem[\protect\citeauthoryear{Eberlein, Smytzek, Steinh{\"{o}}fel, Grunske,
  and Zeller}{Eberlein et~al\mbox{.}}{2023}]%
        {eberlein2023avicenna}
\bibfield{author}{\bibinfo{person}{Martin Eberlein}, \bibinfo{person}{Marius
  Smytzek}, \bibinfo{person}{Dominic Steinh{\"{o}}fel}, \bibinfo{person}{Lars
  Grunske}, {and} \bibinfo{person}{Andreas Zeller}.}
  \bibinfo{year}{2023}\natexlab{}.
\newblock \showarticletitle{Semantic Debugging}. In
  \bibinfo{booktitle}{\emph{Proceedings of the 31st {ACM} Joint European
  Software Engineering Conference and Symposium on the Foundations of Software
  Engineering, {ESEC/FSE} 2023, San Francisco, CA, USA, December 3-9, 2023}},
  \bibfield{editor}{\bibinfo{person}{Satish Chandra}, \bibinfo{person}{Kelly
  Blincoe}, {and} \bibinfo{person}{Paolo Tonella}} (Eds.).
  \bibinfo{publisher}{{ACM}}, \bibinfo{pages}{438--449}.
\newblock
\urldef\tempurl%
\url{https://doi.org/10.1145/3611643.3616296}
\showDOI{\tempurl}


\bibitem[\protect\citeauthoryear{Ernst, Perkins, Guo, McCamant, Pacheco,
  Tschantz, and Xiao}{Ernst et~al\mbox{.}}{2007}]%
        {ernstpgmptx07}
\bibfield{author}{\bibinfo{person}{Michael~D. Ernst}, \bibinfo{person}{Jeff~H.
  Perkins}, \bibinfo{person}{Philip~J. Guo}, \bibinfo{person}{Stephen
  McCamant}, \bibinfo{person}{Carlos Pacheco}, \bibinfo{person}{Matthew~S.
  Tschantz}, {and} \bibinfo{person}{Chen Xiao}.}
  \bibinfo{year}{2007}\natexlab{}.
\newblock \showarticletitle{The {Daikon} system for dynamic detection of likely
  invariants}.
\newblock \bibinfo{journal}{\emph{Sci. Comput. Program.}} \bibinfo{volume}{69},
  \bibinfo{number}{1-3} (\bibinfo{year}{2007}), \bibinfo{pages}{35--45}.
\newblock
\urldef\tempurl%
\url{https://doi.org/10.1016/j.scico.2007.01.015}
\showDOI{\tempurl}


\bibitem[\protect\citeauthoryear{Gonz{\'{a}}lez{-}Sanchez, Abreu, Gross, and
  van Gemund}{Gonz{\'{a}}lez{-}Sanchez et~al\mbox{.}}{2011}]%
        {gonzalez-sanchezagg11}
\bibfield{author}{\bibinfo{person}{Alberto Gonz{\'{a}}lez{-}Sanchez},
  \bibinfo{person}{Rui Abreu}, \bibinfo{person}{Hans{-}Gerhard Gross}, {and}
  \bibinfo{person}{Arjan J.~C. van Gemund}.} \bibinfo{year}{2011}\natexlab{}.
\newblock \showarticletitle{Prioritizing tests for fault localization through
  ambiguity group reduction}. In \bibinfo{booktitle}{\emph{26th {IEEE/ACM}
  International Conference on Automated Software Engineering ({ASE} 2011)}},
  \bibfield{editor}{\bibinfo{person}{Perry Alexander},
  \bibinfo{person}{Corina~S. Pasareanu}, {and} \bibinfo{person}{John~G.
  Hosking}} (Eds.). \bibinfo{publisher}{{IEEE} Computer Society},
  \bibinfo{pages}{83--92}.
\newblock
\urldef\tempurl%
\url{https://doi.org/10.1109/ASE.2011.6100153}
\showDOI{\tempurl}


\bibitem[\protect\citeauthoryear{Gulwani, Polozov, and Singh}{Gulwani
  et~al\mbox{.}}{2017}]%
        {gulwani2017synthesis}
\bibfield{author}{\bibinfo{person}{Sumit Gulwani}, \bibinfo{person}{Oleksandr
  Polozov}, {and} \bibinfo{person}{Rishabh Singh}.}
  \bibinfo{year}{2017}\natexlab{}.
\newblock \showarticletitle{Program Synthesis}.
\newblock \bibinfo{journal}{\emph{Foundations and Trends in Programming
  Languages}} \bibinfo{volume}{4}, \bibinfo{number}{1-2}
  (\bibinfo{year}{2017}), \bibinfo{pages}{1--119}.
\newblock
\showISSN{2325-1107}
\urldef\tempurl%
\url{https://doi.org/10.1561/2500000010}
\showDOI{\tempurl}


\bibitem[\protect\citeauthoryear{Heiden, Grunske, Kehrer, Keller, van Hoorn,
  Filieri, and Lo}{Heiden et~al\mbox{.}}{2019}]%
        {heidengkkhfl19}
\bibfield{author}{\bibinfo{person}{Simon Heiden}, \bibinfo{person}{Lars
  Grunske}, \bibinfo{person}{Timo Kehrer}, \bibinfo{person}{Fabian Keller},
  \bibinfo{person}{Andr{\'{e}} van Hoorn}, \bibinfo{person}{Antonio Filieri},
  {and} \bibinfo{person}{David Lo}.} \bibinfo{year}{2019}\natexlab{}.
\newblock \showarticletitle{An evaluation of pure spectrum-based fault
  localization techniques for large-scale software systems}.
\newblock \bibinfo{journal}{\emph{Softw. Pract. Exp.}} \bibinfo{volume}{49},
  \bibinfo{number}{8} (\bibinfo{year}{2019}), \bibinfo{pages}{1197--1224}.
\newblock
\urldef\tempurl%
\url{https://doi.org/10.1002/spe.2703}
\showDOI{\tempurl}


\bibitem[\protect\citeauthoryear{Hu, Ahmed, Mechtaev, Leong, and
  Roychoudhury}{Hu et~al\mbox{.}}{2019}]%
        {hu2019refactory}
\bibfield{author}{\bibinfo{person}{Yang Hu}, \bibinfo{person}{Umair~Z. Ahmed},
  \bibinfo{person}{Sergey Mechtaev}, \bibinfo{person}{Ben Leong}, {and}
  \bibinfo{person}{Abhik Roychoudhury}.} \bibinfo{year}{2019}\natexlab{}.
\newblock \showarticletitle{Re-Factoring Based Program Repair Applied to
  Programming Assignments}. In \bibinfo{booktitle}{\emph{2019 34th IEEE/ACM
  International Conference on Automated Software Engineering (ASE)}}.
  \bibinfo{pages}{388--398}.
\newblock
\urldef\tempurl%
\url{https://doi.org/10.1109/ASE.2019.00044}
\showDOI{\tempurl}


\bibitem[\protect\citeauthoryear{Jiang, Zhang, Chan, Tse, and Chen}{Jiang
  et~al\mbox{.}}{2012}]%
        {jiangzctc12}
\bibfield{author}{\bibinfo{person}{Bo Jiang}, \bibinfo{person}{Zhenyu Zhang},
  \bibinfo{person}{W.~K. Chan}, \bibinfo{person}{T.~H. Tse}, {and}
  \bibinfo{person}{Tsong~Yueh Chen}.} \bibinfo{year}{2012}\natexlab{}.
\newblock \showarticletitle{How Well Does Test Case Prioritization Integrate
  with Statistical Fault Localization?}
\newblock \bibinfo{journal}{\emph{Inf. Softw. Technol.}} \bibinfo{volume}{54},
  \bibinfo{number}{7} (\bibinfo{date}{July} \bibinfo{year}{2012}),
  \bibinfo{pages}{739--758}.
\newblock
\showISSN{0950-5849}
\urldef\tempurl%
\url{https://doi.org/10.1016/j.infsof.2012.01.006}
\showDOI{\tempurl}


\bibitem[\protect\citeauthoryear{Jiang, Wang, Chen, Lv, and Liu}{Jiang
  et~al\mbox{.}}{2023}]%
        {jiang2023variable}
\bibfield{author}{\bibinfo{person}{Jiajun Jiang}, \bibinfo{person}{Yumeng
  Wang}, \bibinfo{person}{Junjie Chen}, \bibinfo{person}{Delin Lv}, {and}
  \bibinfo{person}{Mengjiao Liu}.} \bibinfo{year}{2023}\natexlab{}.
\newblock \showarticletitle{Variable-based Fault Localization via Enhanced
  Decision Tree}.
\newblock \bibinfo{journal}{\emph{ACM Trans. Softw. Eng. Methodol.}}
  \bibinfo{volume}{33}, \bibinfo{number}{2}, Article \bibinfo{articleno}{41}
  (\bibinfo{date}{dec} \bibinfo{year}{2023}), \bibinfo{numpages}{32}~pages.
\newblock
\showISSN{1049-331X}
\urldef\tempurl%
\url{https://doi.org/10.1145/3624741}
\showDOI{\tempurl}


\bibitem[\protect\citeauthoryear{Jones and Harrold}{Jones and Harrold}{2005}]%
        {jones2005tarantula}
\bibfield{author}{\bibinfo{person}{James~A. Jones} {and}
  \bibinfo{person}{Mary~Jean Harrold}.} \bibinfo{year}{2005}\natexlab{}.
\newblock \showarticletitle{Empirical evaluation of the {Tarantula} automatic
  fault-localization technique}. In \bibinfo{booktitle}{\emph{Proceedings of
  the 20th IEEE/ACM International Conference on Automated Software
  Engineering}} (Long Beach, CA, USA) \emph{(\bibinfo{series}{ASE '05})}.
  \bibinfo{publisher}{Association for Computing Machinery},
  \bibinfo{address}{New York, NY, USA}, \bibinfo{pages}{273–282}.
\newblock
\showISBNx{1581139934}
\urldef\tempurl%
\url{https://doi.org/10.1145/1101908.1101949}
\showDOI{\tempurl}


\bibitem[\protect\citeauthoryear{Jones, Harrold, and Stasko}{Jones
  et~al\mbox{.}}{2002a}]%
        {jones2002tarantula}
\bibfield{author}{\bibinfo{person}{James~A. Jones}, \bibinfo{person}{Mary~Jean
  Harrold}, {and} \bibinfo{person}{John Stasko}.}
  \bibinfo{year}{2002}\natexlab{a}.
\newblock \showarticletitle{Visualization of Test Information to Assist Fault
  Localization}. In \bibinfo{booktitle}{\emph{Proceedings of the 24th
  International Conference on Software Engineering}} (Orlando, Florida).
  \bibinfo{address}{New York, NY, USA}, \bibinfo{pages}{467--477}.
\newblock
\showISBNx{1-58113-472-X}
\urldef\tempurl%
\url{https://doi.org/10.1145/581339.581397}
\showDOI{\tempurl}


\bibitem[\protect\citeauthoryear{Jones, Harrold, and Stasko}{Jones
  et~al\mbox{.}}{2002b}]%
        {jones2002visualization}
\bibfield{author}{\bibinfo{person}{James~A. Jones}, \bibinfo{person}{Mary~Jean
  Harrold}, {and} \bibinfo{person}{John Stasko}.}
  \bibinfo{year}{2002}\natexlab{b}.
\newblock \showarticletitle{Visualization of Test Information to Assist Fault
  Localization}. In \bibinfo{booktitle}{\emph{Proceedings of the 24th
  International Conference on Software Engineering}} (Orlando, Florida).
  \bibinfo{publisher}{ACM}, \bibinfo{address}{New York, NY, USA},
  \bibinfo{pages}{467--477}.
\newblock
\showISBNx{1-58113-472-X}
\urldef\tempurl%
\url{https://doi.org/10.1145/581339.581397}
\showDOI{\tempurl}


\bibitem[\protect\citeauthoryear{Just, Parnin, Drosos, and Ernst}{Just
  et~al\mbox{.}}{2018}]%
        {just2018sfl}
\bibfield{author}{\bibinfo{person}{Ren\'{e} Just}, \bibinfo{person}{Chris
  Parnin}, \bibinfo{person}{Ian Drosos}, {and} \bibinfo{person}{Michael~D.
  Ernst}.} \bibinfo{year}{2018}\natexlab{}.
\newblock \showarticletitle{Comparing developer-provided to user-provided tests
  for fault localization and automated program repair}. In
  \bibinfo{booktitle}{\emph{Proceedings of the 27th ACM SIGSOFT International
  Symposium on Software Testing and Analysis}} (Amsterdam, Netherlands)
  \emph{(\bibinfo{series}{ISSTA 2018})}. \bibinfo{publisher}{Association for
  Computing Machinery}, \bibinfo{address}{New York, NY, USA},
  \bibinfo{pages}{287–297}.
\newblock
\showISBNx{9781450356992}
\urldef\tempurl%
\url{https://doi.org/10.1145/3213846.3213870}
\showDOI{\tempurl}


\bibitem[\protect\citeauthoryear{Kampmann, Havrikov, Soremekun, and
  Zeller}{Kampmann et~al\mbox{.}}{2020}]%
        {kampmann2020alhazen}
\bibfield{author}{\bibinfo{person}{Alexander Kampmann},
  \bibinfo{person}{Nikolas Havrikov}, \bibinfo{person}{Ezekiel~O. Soremekun},
  {and} \bibinfo{person}{Andreas Zeller}.} \bibinfo{year}{2020}\natexlab{}.
\newblock \showarticletitle{When does my program do this? learning
  circumstances of software behavior}. In \bibinfo{booktitle}{\emph{Proceedings
  of the 28th ACM Joint Meeting on European Software Engineering Conference and
  Symposium on the Foundations of Software Engineering}} (Virtual Event, USA)
  \emph{(\bibinfo{series}{ESEC/FSE 2020})}. \bibinfo{publisher}{Association for
  Computing Machinery}, \bibinfo{address}{New York, NY, USA},
  \bibinfo{pages}{1228–1239}.
\newblock
\showISBNx{9781450370431}
\urldef\tempurl%
\url{https://doi.org/10.1145/3368089.3409687}
\showDOI{\tempurl}


\bibitem[\protect\citeauthoryear{Kang, An, and Yoo}{Kang et~al\mbox{.}}{2024}]%
        {kang2024llm}
\bibfield{author}{\bibinfo{person}{Sungmin Kang}, \bibinfo{person}{Gabin An},
  {and} \bibinfo{person}{Shin Yoo}.} \bibinfo{year}{2024}\natexlab{}.
\newblock \showarticletitle{A Quantitative and Qualitative Evaluation of
  LLM-Based Explainable Fault Localization}.
\newblock \bibinfo{journal}{\emph{Proc. ACM Softw. Eng.}} \bibinfo{volume}{1},
  \bibinfo{number}{FSE}, Article \bibinfo{articleno}{64} (\bibinfo{date}{jul}
  \bibinfo{year}{2024}), \bibinfo{numpages}{23}~pages.
\newblock
\urldef\tempurl%
\url{https://doi.org/10.1145/3660771}
\showDOI{\tempurl}


\bibitem[\protect\citeauthoryear{Landsberg, Chockler, Kroening, and
  Lewis}{Landsberg et~al\mbox{.}}{2015}]%
        {landsberg2015sfl}
\bibfield{author}{\bibinfo{person}{David Landsberg}, \bibinfo{person}{Hana
  Chockler}, \bibinfo{person}{Daniel Kroening}, {and} \bibinfo{person}{Matt
  Lewis}.} \bibinfo{year}{2015}\natexlab{}.
\newblock \showarticletitle{Evaluation of Measures for Statistical Fault
  Localisation and an Optimising Scheme}. In
  \bibinfo{booktitle}{\emph{Fundamental Approaches to Software Engineering}},
  \bibfield{editor}{\bibinfo{person}{Alexander Egyed} {and}
  \bibinfo{person}{Ina Schaefer}} (Eds.). \bibinfo{publisher}{Springer Berlin
  Heidelberg}, \bibinfo{address}{Berlin, Heidelberg},
  \bibinfo{pages}{115--129}.
\newblock
\showISBNx{978-3-662-46675-9}
\urldef\tempurl%
\url{https://doi.org/10.1007/978-3-662-46675-9_8}
\showDOI{\tempurl}


\bibitem[\protect\citeauthoryear{Le, Lo, {Le Goues}, and Grunske}{Le
  et~al\mbox{.}}{2016}]%
        {lelgg16}
\bibfield{author}{\bibinfo{person}{Tien{-}Duy~B. Le}, \bibinfo{person}{David
  Lo}, \bibinfo{person}{Claire {Le Goues}}, {and} \bibinfo{person}{Lars
  Grunske}.} \bibinfo{year}{2016}\natexlab{}.
\newblock \showarticletitle{A learning-to-rank based fault localization
  approach using likely invariants}. In \bibinfo{booktitle}{\emph{Proceedings
  of the 25th International Symposium on Software Testing and Analysis, {ISSTA}
  2016}}. \bibinfo{pages}{177--188}.
\newblock
\urldef\tempurl%
\url{https://doi.org/10.1145/2931037.2931049}
\showDOI{\tempurl}


\bibitem[\protect\citeauthoryear{Le, Lo, and Li}{Le et~al\mbox{.}}{2015}]%
        {le2015topk}
\bibfield{author}{\bibinfo{person}{Tien-Duy~B. Le}, \bibinfo{person}{David Lo},
  {and} \bibinfo{person}{Ming Li}.} \bibinfo{year}{2015}\natexlab{}.
\newblock \showarticletitle{Constrained feature selection for localizing
  faults}. In \bibinfo{booktitle}{\emph{Proceedings of the 2015 IEEE
  International Conference on Software Maintenance and Evolution (ICSME)}}
  \emph{(\bibinfo{series}{ICSME '15})}. \bibinfo{publisher}{IEEE Computer
  Society}, \bibinfo{address}{USA}, \bibinfo{pages}{501–505}.
\newblock
\showISBNx{9781467375320}
\urldef\tempurl%
\url{https://doi.org/10.1109/ICSM.2015.7332502}
\showDOI{\tempurl}


\bibitem[\protect\citeauthoryear{Le and Soffa}{Le and Soffa}{2010}]%
        {le2010path}
\bibfield{author}{\bibinfo{person}{Wei Le} {and} \bibinfo{person}{Mary~Lou
  Soffa}.} \bibinfo{year}{2010}\natexlab{}.
\newblock \showarticletitle{Path-based fault correlations}. In
  \bibinfo{booktitle}{\emph{Proceedings of the Eighteenth ACM SIGSOFT
  International Symposium on Foundations of Software Engineering}} (Santa Fe,
  New Mexico, USA) \emph{(\bibinfo{series}{FSE '10})}.
  \bibinfo{publisher}{Association for Computing Machinery},
  \bibinfo{address}{New York, NY, USA}, \bibinfo{pages}{307–316}.
\newblock
\showISBNx{9781605587912}
\urldef\tempurl%
\url{https://doi.org/10.1145/1882291.1882336}
\showDOI{\tempurl}


\bibitem[\protect\citeauthoryear{Lei, Xie, Zhang, Yan, Xu, and Sun}{Lei
  et~al\mbox{.}}{2022}]%
        {lei2022featurefl}
\bibfield{author}{\bibinfo{person}{Yan Lei}, \bibinfo{person}{Huan Xie},
  \bibinfo{person}{Tao Zhang}, \bibinfo{person}{Meng Yan},
  \bibinfo{person}{Zhou Xu}, {and} \bibinfo{person}{Chengnian Sun}.}
  \bibinfo{year}{2022}\natexlab{}.
\newblock \showarticletitle{Feature-FL: Feature-Based Fault Localization}.
\newblock \bibinfo{journal}{\emph{IEEE Transactions on Reliability}}
  \bibinfo{volume}{71}, \bibinfo{number}{1} (\bibinfo{year}{2022}),
  \bibinfo{pages}{264--283}.
\newblock
\urldef\tempurl%
\url{https://doi.org/10.1109/TR.2022.3140453}
\showDOI{\tempurl}


\bibitem[\protect\citeauthoryear{Li, Li, Zhang, and Zhang}{Li
  et~al\mbox{.}}{2019}]%
        {li2019deepfl}
\bibfield{author}{\bibinfo{person}{Xia Li}, \bibinfo{person}{Wei Li},
  \bibinfo{person}{Yuqun Zhang}, {and} \bibinfo{person}{Lingming Zhang}.}
  \bibinfo{year}{2019}\natexlab{}.
\newblock \showarticletitle{DeepFL: integrating multiple fault diagnosis
  dimensions for deep fault localization}. In
  \bibinfo{booktitle}{\emph{Proceedings of the 28th ACM SIGSOFT International
  Symposium on Software Testing and Analysis}} (Beijing, China)
  \emph{(\bibinfo{series}{ISSTA 2019})}. \bibinfo{publisher}{Association for
  Computing Machinery}, \bibinfo{address}{New York, NY, USA},
  \bibinfo{pages}{169–180}.
\newblock
\showISBNx{9781450362245}
\urldef\tempurl%
\url{https://doi.org/10.1145/3293882.3330574}
\showDOI{\tempurl}


\bibitem[\protect\citeauthoryear{Li, Wang, and Nguyen}{Li
  et~al\mbox{.}}{2021}]%
        {li2021covrepresentation}
\bibfield{author}{\bibinfo{person}{Yi Li}, \bibinfo{person}{Shaohua Wang},
  {and} \bibinfo{person}{Tien~N. Nguyen}.} \bibinfo{year}{2021}\natexlab{}.
\newblock \showarticletitle{Fault Localization with Code Coverage
  Representation Learning}. In \bibinfo{booktitle}{\emph{Proceedings of the
  43rd International Conference on Software Engineering}} (Madrid, Spain)
  \emph{(\bibinfo{series}{ICSE '21})}. \bibinfo{publisher}{IEEE Press},
  \bibinfo{pages}{661–673}.
\newblock
\showISBNx{9781450390859}
\urldef\tempurl%
\url{https://doi.org/10.1109/ICSE43902.2021.00067}
\showDOI{\tempurl}


\bibitem[\protect\citeauthoryear{Li, Wang, and Nguyen}{Li
  et~al\mbox{.}}{2022}]%
        {li2022cc}
\bibfield{author}{\bibinfo{person}{Yi Li}, \bibinfo{person}{Shaohua Wang},
  {and} \bibinfo{person}{Tien~N. Nguyen}.} \bibinfo{year}{2022}\natexlab{}.
\newblock \showarticletitle{Fault localization to detect co-change fixing
  locations}. In \bibinfo{booktitle}{\emph{Proceedings of the 30th ACM Joint
  European Software Engineering Conference and Symposium on the Foundations of
  Software Engineering}} (Singapore, Singapore)
  \emph{(\bibinfo{series}{ESEC/FSE 2022})}. \bibinfo{publisher}{Association for
  Computing Machinery}, \bibinfo{address}{New York, NY, USA},
  \bibinfo{pages}{659–671}.
\newblock
\showISBNx{9781450394130}
\urldef\tempurl%
\url{https://doi.org/10.1145/3540250.3549137}
\showDOI{\tempurl}


\bibitem[\protect\citeauthoryear{Liblit, Naik, Zheng, Aiken, and Jordan}{Liblit
  et~al\mbox{.}}{2005}]%
        {liblit2005sd}
\bibfield{author}{\bibinfo{person}{Ben Liblit}, \bibinfo{person}{Mayur Naik},
  \bibinfo{person}{Alice~X. Zheng}, \bibinfo{person}{Alex Aiken}, {and}
  \bibinfo{person}{Michael~I. Jordan}.} \bibinfo{year}{2005}\natexlab{}.
\newblock \showarticletitle{Scalable Statistical Bug Isolation}.
\newblock \bibinfo{journal}{\emph{SIGPLAN Not.}} \bibinfo{volume}{40},
  \bibinfo{number}{6} (\bibinfo{date}{jun} \bibinfo{year}{2005}),
  \bibinfo{pages}{15–26}.
\newblock
\showISSN{0362-1340}
\urldef\tempurl%
\url{https://doi.org/10.1145/1064978.1065014}
\showDOI{\tempurl}


\bibitem[\protect\citeauthoryear{Long and Rinard}{Long and Rinard}{2016}]%
        {long2016search}
\bibfield{author}{\bibinfo{person}{Fan Long} {and} \bibinfo{person}{Martin
  Rinard}.} \bibinfo{year}{2016}\natexlab{}.
\newblock \showarticletitle{An analysis of the search spaces for generate and
  validate patch generation systems}. In \bibinfo{booktitle}{\emph{Proceedings
  of the 38th International Conference on Software Engineering}} (Austin,
  Texas) \emph{(\bibinfo{series}{ICSE '16})}. \bibinfo{publisher}{Association
  for Computing Machinery}, \bibinfo{address}{New York, NY, USA},
  \bibinfo{pages}{702–713}.
\newblock
\showISBNx{9781450339001}
\urldef\tempurl%
\url{https://doi.org/10.1145/2884781.2884872}
\showDOI{\tempurl}


\bibitem[\protect\citeauthoryear{Lutellier, Pham, Pang, Li, Wei, and
  Tan}{Lutellier et~al\mbox{.}}{2020}]%
        {lutellier2020coconut}
\bibfield{author}{\bibinfo{person}{Thibaud Lutellier},
  \bibinfo{person}{Hung~Viet Pham}, \bibinfo{person}{Lawrence Pang},
  \bibinfo{person}{Yitong Li}, \bibinfo{person}{Moshi Wei}, {and}
  \bibinfo{person}{Lin Tan}.} \bibinfo{year}{2020}\natexlab{}.
\newblock \showarticletitle{CoCoNuT: combining context-aware neural translation
  models using ensemble for program repair}. In
  \bibinfo{booktitle}{\emph{Proceedings of the 29th ACM SIGSOFT International
  Symposium on Software Testing and Analysis}} (Virtual Event, USA)
  \emph{(\bibinfo{series}{ISSTA 2020})}. \bibinfo{publisher}{Association for
  Computing Machinery}, \bibinfo{address}{New York, NY, USA},
  \bibinfo{pages}{101–114}.
\newblock
\showISBNx{9781450380089}
\urldef\tempurl%
\url{https://doi.org/10.1145/3395363.3397369}
\showDOI{\tempurl}


\bibitem[\protect\citeauthoryear{Meng, Wang, Zhang, Sun, and Liu}{Meng
  et~al\mbox{.}}{2022}]%
        {meng2022transfer}
\bibfield{author}{\bibinfo{person}{Xiangxin Meng}, \bibinfo{person}{Xu Wang},
  \bibinfo{person}{Hongyu Zhang}, \bibinfo{person}{Hailong Sun}, {and}
  \bibinfo{person}{Xudong Liu}.} \bibinfo{year}{2022}\natexlab{}.
\newblock \showarticletitle{Improving fault localization and program repair
  with deep semantic features and transferred knowledge}. In
  \bibinfo{booktitle}{\emph{Proceedings of the 44th International Conference on
  Software Engineering}} (Pittsburgh, Pennsylvania)
  \emph{(\bibinfo{series}{ICSE '22})}. \bibinfo{publisher}{Association for
  Computing Machinery}, \bibinfo{address}{New York, NY, USA},
  \bibinfo{pages}{1169–1180}.
\newblock
\showISBNx{9781450392211}
\urldef\tempurl%
\url{https://doi.org/10.1145/3510003.3510147}
\showDOI{\tempurl}


\bibitem[\protect\citeauthoryear{Naish, Lee, and Ramamohanarao}{Naish
  et~al\mbox{.}}{2011}]%
        {naish2011sbfl}
\bibfield{author}{\bibinfo{person}{Lee Naish}, \bibinfo{person}{Hua~Jie Lee},
  {and} \bibinfo{person}{Kotagiri Ramamohanarao}.}
  \bibinfo{year}{2011}\natexlab{}.
\newblock \showarticletitle{A Model for Spectra-Based Software Diagnosis}.
\newblock \bibinfo{journal}{\emph{ACM Trans. Softw. Eng. Methodol.}}
  \bibinfo{volume}{20}, \bibinfo{number}{3}, Article \bibinfo{articleno}{11}
  (\bibinfo{date}{aug} \bibinfo{year}{2011}), \bibinfo{numpages}{32}~pages.
\newblock
\showISSN{1049-331X}
\urldef\tempurl%
\url{https://doi.org/10.1145/2000791.2000795}
\showDOI{\tempurl}


\bibitem[\protect\citeauthoryear{Papadakis and Le~Traon}{Papadakis and
  Le~Traon}{2014}]%
        {papadakis2014mutation}
\bibfield{author}{\bibinfo{person}{Mike Papadakis} {and} \bibinfo{person}{Yves
  Le~Traon}.} \bibinfo{year}{2014}\natexlab{}.
\newblock \showarticletitle{Effective fault localization via mutation analysis:
  a selective mutation approach}. In \bibinfo{booktitle}{\emph{Proceedings of
  the 29th Annual ACM Symposium on Applied Computing}} (Gyeongju, Republic of
  Korea) \emph{(\bibinfo{series}{SAC '14})}. \bibinfo{publisher}{Association
  for Computing Machinery}, \bibinfo{address}{New York, NY, USA},
  \bibinfo{pages}{1293–1300}.
\newblock
\showISBNx{9781450324694}
\urldef\tempurl%
\url{https://doi.org/10.1145/2554850.2554978}
\showDOI{\tempurl}


\bibitem[\protect\citeauthoryear{Papadakis and Le~Traon}{Papadakis and
  Le~Traon}{2015}]%
        {papadakis2015mutation}
\bibfield{author}{\bibinfo{person}{Mike Papadakis} {and} \bibinfo{person}{Yves
  Le~Traon}.} \bibinfo{year}{2015}\natexlab{}.
\newblock \showarticletitle{Metallaxis-FL: mutation-based fault localization}.
\newblock \bibinfo{journal}{\emph{Softw. Test. Verif. Reliab.}}
  \bibinfo{volume}{25}, \bibinfo{number}{5–7} (\bibinfo{date}{aug}
  \bibinfo{year}{2015}), \bibinfo{pages}{605–628}.
\newblock
\showISSN{0960-0833}
\urldef\tempurl%
\url{https://doi.org/10.1002/stvr.1509}
\showDOI{\tempurl}


\bibitem[\protect\citeauthoryear{Parnin and Orso}{Parnin and Orso}{2011}]%
        {parnin2011automated}
\bibfield{author}{\bibinfo{person}{Chris Parnin} {and}
  \bibinfo{person}{Alessandro Orso}.} \bibinfo{year}{2011}\natexlab{}.
\newblock \showarticletitle{Are Automated Debugging Techniques Actually Helping
  Programmers?}. In \bibinfo{booktitle}{\emph{Proceedings of the 2011
  International Symposium on Software Testing and Analysis}} (Toronto, Ontario,
  Canada) \emph{(\bibinfo{series}{ISSTA '11})}. \bibinfo{publisher}{Association
  for Computing Machinery}, \bibinfo{address}{New York, NY, USA},
  \bibinfo{pages}{199–209}.
\newblock
\showISBNx{9781450305624}
\urldef\tempurl%
\url{https://doi.org/10.1145/2001420.2001445}
\showDOI{\tempurl}


\bibitem[\protect\citeauthoryear{Pearson, Campos, Just, Fraser, Abreu, Ernst,
  Pang, and Keller}{Pearson et~al\mbox{.}}{2017}]%
        {pearson2017sfl}
\bibfield{author}{\bibinfo{person}{Spencer Pearson}, \bibinfo{person}{Jos\'{e}
  Campos}, \bibinfo{person}{Ren\'{e} Just}, \bibinfo{person}{Gordon Fraser},
  \bibinfo{person}{Rui Abreu}, \bibinfo{person}{Michael~D. Ernst},
  \bibinfo{person}{Deric Pang}, {and} \bibinfo{person}{Benjamin Keller}.}
  \bibinfo{year}{2017}\natexlab{}.
\newblock \showarticletitle{Evaluating and improving fault localization}. In
  \bibinfo{booktitle}{\emph{Proceedings of the 39th International Conference on
  Software Engineering}} (Buenos Aires, Argentina) \emph{(\bibinfo{series}{ICSE
  '17})}. \bibinfo{publisher}{IEEE Press}, \bibinfo{pages}{609–620}.
\newblock
\showISBNx{9781538638682}
\urldef\tempurl%
\url{https://doi.org/10.1109/ICSE.2017.62}
\showDOI{\tempurl}


\bibitem[\protect\citeauthoryear{Qi, Mao, Lei, and Wang}{Qi
  et~al\mbox{.}}{2013}]%
        {qi2013apr}
\bibfield{author}{\bibinfo{person}{Yuhua Qi}, \bibinfo{person}{Xiaoguang Mao},
  \bibinfo{person}{Yan Lei}, {and} \bibinfo{person}{Chengsong Wang}.}
  \bibinfo{year}{2013}\natexlab{}.
\newblock \showarticletitle{Using Automated Program Repair for Evaluating the
  Effectiveness of Fault Localization Techniques}. In
  \bibinfo{booktitle}{\emph{Proceedings of the 2013 International Symposium on
  Software Testing and Analysis}} (Lugano, Switzerland)
  \emph{(\bibinfo{series}{ISSTA 2013})}. \bibinfo{publisher}{Association for
  Computing Machinery}, \bibinfo{address}{New York, NY, USA},
  \bibinfo{pages}{191–201}.
\newblock
\showISBNx{9781450321594}
\urldef\tempurl%
\url{https://doi.org/10.1145/2483760.2483785}
\showDOI{\tempurl}


\bibitem[\protect\citeauthoryear{Ribeiro, Roberto~de Araujo, Chaim, Souza, and
  Kon}{Ribeiro et~al\mbox{.}}{2019}]%
        {ribeiro2019dataflow}
\bibfield{author}{\bibinfo{person}{Henrique~L. Ribeiro}, \bibinfo{person}{P.~A.
  Roberto~de Araujo}, \bibinfo{person}{Marcos~L. Chaim}, \bibinfo{person}{Higor
  A.~de Souza}, {and} \bibinfo{person}{Fabio Kon}.}
  \bibinfo{year}{2019}\natexlab{}.
\newblock \showarticletitle{Evaluating data-flow coverage in spectrum-based
  fault localization}. In \bibinfo{booktitle}{\emph{2019 ACM/IEEE International
  Symposium on Empirical Software Engineering and Measurement (ESEM)}}.
  \bibinfo{pages}{1--11}.
\newblock
\urldef\tempurl%
\url{https://doi.org/10.1109/ESEM.2019.8870182}
\showDOI{\tempurl}


\bibitem[\protect\citeauthoryear{Santelices, Jones, Yu, and Harrold}{Santelices
  et~al\mbox{.}}{2009}]%
        {santelices2009defuse}
\bibfield{author}{\bibinfo{person}{Raul Santelices}, \bibinfo{person}{James~A.
  Jones}, \bibinfo{person}{Yanbing Yu}, {and} \bibinfo{person}{Mary~Jean
  Harrold}.} \bibinfo{year}{2009}\natexlab{}.
\newblock \showarticletitle{Lightweight Fault-Localization Using Multiple
  Coverage Types}. In \bibinfo{booktitle}{\emph{Proceedings of the 31st
  International Conference on Software Engineering}}
  \emph{(\bibinfo{series}{ICSE '09})}. \bibinfo{publisher}{IEEE Computer
  Society}, \bibinfo{address}{USA}, \bibinfo{pages}{56–66}.
\newblock
\showISBNx{9781424434534}
\urldef\tempurl%
\url{https://doi.org/10.1109/ICSE.2009.5070508}
\showDOI{\tempurl}


\bibitem[\protect\citeauthoryear{Smytzek, Eberlein, Ser\c{c}e, Grunske, and
  Zeller}{Smytzek et~al\mbox{.}}{2024}]%
        {smytzek2024tests4py}
\bibfield{author}{\bibinfo{person}{Marius Smytzek}, \bibinfo{person}{Martin
  Eberlein}, \bibinfo{person}{Batuhan Ser\c{c}e}, \bibinfo{person}{Lars
  Grunske}, {and} \bibinfo{person}{Andreas Zeller}.}
  \bibinfo{year}{2024}\natexlab{}.
\newblock \showarticletitle{Tests4Py: A Benchmark for System Testing}. In
  \bibinfo{booktitle}{\emph{Companion Proceedings of the 32nd ACM International
  Conference on the Foundations of Software Engineering}} (Porto de Galinhas,
  Brazil) \emph{(\bibinfo{series}{FSE 2024})}. \bibinfo{publisher}{Association
  for Computing Machinery}, \bibinfo{address}{New York, NY, USA},
  \bibinfo{pages}{557–561}.
\newblock
\showISBNx{9798400706585}
\urldef\tempurl%
\url{https://doi.org/10.1145/3663529.3663798}
\showDOI{\tempurl}


\bibitem[\protect\citeauthoryear{Smytzek and Zeller}{Smytzek and
  Zeller}{2022}]%
        {smytzek2022sflkit}
\bibfield{author}{\bibinfo{person}{Marius Smytzek} {and}
  \bibinfo{person}{Andreas Zeller}.} \bibinfo{year}{2022}\natexlab{}.
\newblock \showarticletitle{{SFLKit}: {A} workbench for statistical fault
  localization}. In \bibinfo{booktitle}{\emph{Proceedings of the 30th ACM Joint
  European Software Engineering Conference and Symposium on the Foundations of
  Software Engineering}} (Singapore, Singapore)
  \emph{(\bibinfo{series}{ESEC/FSE 2022})}. \bibinfo{publisher}{Association for
  Computing Machinery}, \bibinfo{address}{New York, NY, USA},
  \bibinfo{pages}{1701–1705}.
\newblock
\showISBNx{9781450394130}
\urldef\tempurl%
\url{https://doi.org/10.1145/3540250.3558915}
\showDOI{\tempurl}


\bibitem[\protect\citeauthoryear{Soremekun, Kirschner, B\"{o}hme, and
  Papadakis}{Soremekun et~al\mbox{.}}{2023}]%
        {soremekun2023evaluating}
\bibfield{author}{\bibinfo{person}{Ezekiel Soremekun}, \bibinfo{person}{Lukas
  Kirschner}, \bibinfo{person}{Marcel B\"{o}hme}, {and} \bibinfo{person}{Mike
  Papadakis}.} \bibinfo{year}{2023}\natexlab{}.
\newblock \showarticletitle{Evaluating the Impact of Experimental Assumptions
  in Automated Fault Localization}. In \bibinfo{booktitle}{\emph{Proceedings of
  the 45th International Conference on Software Engineering}} (Melbourne,
  Victoria, Australia) \emph{(\bibinfo{series}{ICSE '23})}.
  \bibinfo{publisher}{IEEE Press}, \bibinfo{pages}{159–171}.
\newblock
\showISBNx{9781665457019}
\urldef\tempurl%
\url{https://doi.org/10.1109/ICSE48619.2023.00025}
\showDOI{\tempurl}


\bibitem[\protect\citeauthoryear{Soremekun, Kirschner, B\"{o}hme, and
  Zeller}{Soremekun et~al\mbox{.}}{2021}]%
        {soremekun2021slicing}
\bibfield{author}{\bibinfo{person}{Ezekiel Soremekun}, \bibinfo{person}{Lukas
  Kirschner}, \bibinfo{person}{Marcel B\"{o}hme}, {and}
  \bibinfo{person}{Andreas Zeller}.} \bibinfo{year}{2021}\natexlab{}.
\newblock \showarticletitle{Locating faults with program slicing: an empirical
  analysis}.
\newblock \bibinfo{journal}{\emph{Empirical Softw. Engg.}}
  \bibinfo{volume}{26}, \bibinfo{number}{3} (\bibinfo{date}{may}
  \bibinfo{year}{2021}), \bibinfo{numpages}{45}~pages.
\newblock
\showISSN{1382-3256}
\urldef\tempurl%
\url{https://doi.org/10.1007/s10664-020-09931-7}
\showDOI{\tempurl}


\bibitem[\protect\citeauthoryear{Steimann, Frenkel, and Abreu}{Steimann
  et~al\mbox{.}}{2013}]%
        {steimann2013threats}
\bibfield{author}{\bibinfo{person}{Friedrich Steimann}, \bibinfo{person}{Marcus
  Frenkel}, {and} \bibinfo{person}{Rui Abreu}.}
  \bibinfo{year}{2013}\natexlab{}.
\newblock \showarticletitle{Threats to the validity and value of empirical
  assessments of the accuracy of coverage-based fault locators}. In
  \bibinfo{booktitle}{\emph{Proceedings of the 2013 International Symposium on
  Software Testing and Analysis}} (Lugano, Switzerland)
  \emph{(\bibinfo{series}{ISSTA 2013})}. \bibinfo{publisher}{Association for
  Computing Machinery}, \bibinfo{address}{New York, NY, USA},
  \bibinfo{pages}{314–324}.
\newblock
\showISBNx{9781450321594}
\urldef\tempurl%
\url{https://doi.org/10.1145/2483760.2483767}
\showDOI{\tempurl}


\bibitem[\protect\citeauthoryear{Vancsics, Horv\'{a}th, Szatm\'{a}ri, and
  Besz\'{e}des}{Vancsics et~al\mbox{.}}{2021}]%
        {vancsics2021calls}
\bibfield{author}{\bibinfo{person}{B\'{e}la Vancsics}, \bibinfo{person}{Ferenc
  Horv\'{a}th}, \bibinfo{person}{Attila Szatm\'{a}ri}, {and}
  \bibinfo{person}{\'{A}rp\'{a}d Besz\'{e}des}.}
  \bibinfo{year}{2021}\natexlab{}.
\newblock \showarticletitle{Call Frequency-Based Fault Localization}. In
  \bibinfo{booktitle}{\emph{2021 IEEE International Conference on Software
  Analysis, Evolution and Reengineering (SANER)}}. \bibinfo{pages}{365--376}.
\newblock
\urldef\tempurl%
\url{https://doi.org/10.1109/SANER50967.2021.00041}
\showDOI{\tempurl}


\bibitem[\protect\citeauthoryear{Wang, Yu, Meng, Cao, Zhang, Sun, Liu, and
  Hu}{Wang et~al\mbox{.}}{2024}]%
        {wang2024mtltransfer}
\bibfield{author}{\bibinfo{person}{Xu Wang}, \bibinfo{person}{Hongwei Yu},
  \bibinfo{person}{Xiangxin Meng}, \bibinfo{person}{Hongliang Cao},
  \bibinfo{person}{Hongyu Zhang}, \bibinfo{person}{Hailong Sun},
  \bibinfo{person}{Xudong Liu}, {and} \bibinfo{person}{Chunming Hu}.}
  \bibinfo{year}{2024}\natexlab{}.
\newblock \showarticletitle{MTL-TRANSFER: Leveraging Multi-task Learning and
  Transferred Knowledge for Improving Fault Localization and Program Repair}.
\newblock \bibinfo{journal}{\emph{ACM Trans. Softw. Eng. Methodol.}}
  \bibinfo{volume}{33}, \bibinfo{number}{6}, Article \bibinfo{articleno}{148}
  (\bibinfo{date}{jun} \bibinfo{year}{2024}), \bibinfo{numpages}{31}~pages.
\newblock
\showISSN{1049-331X}
\urldef\tempurl%
\url{https://doi.org/10.1145/3654441}
\showDOI{\tempurl}


\bibitem[\protect\citeauthoryear{Widyasari, Prana, Haryono, Tian, Zachiary, and
  Lo}{Widyasari et~al\mbox{.}}{2022a}]%
        {widyasari2022xai4fl}
\bibfield{author}{\bibinfo{person}{Ratnadira Widyasari}, \bibinfo{person}{Gede
  Artha~Azriadi Prana}, \bibinfo{person}{Stefanus~A. Haryono},
  \bibinfo{person}{Yuan Tian}, \bibinfo{person}{Hafil~Noer Zachiary}, {and}
  \bibinfo{person}{David Lo}.} \bibinfo{year}{2022}\natexlab{a}.
\newblock \showarticletitle{XAI4FL: enhancing spectrum-based fault localization
  with explainable artificial intelligence}. In
  \bibinfo{booktitle}{\emph{Proceedings of the 30th IEEE/ACM International
  Conference on Program Comprehension}} (Virtual Event)
  \emph{(\bibinfo{series}{ICPC '22})}. \bibinfo{publisher}{Association for
  Computing Machinery}, \bibinfo{address}{New York, NY, USA},
  \bibinfo{pages}{499–510}.
\newblock
\showISBNx{9781450392983}
\urldef\tempurl%
\url{https://doi.org/10.1145/3524610.3527902}
\showDOI{\tempurl}


\bibitem[\protect\citeauthoryear{Widyasari, Prana, Haryono, Wang, and
  Lo}{Widyasari et~al\mbox{.}}{2022b}]%
        {widyasari2022sfl}
\bibfield{author}{\bibinfo{person}{Ratnadira Widyasari}, \bibinfo{person}{Gede
  Artha~Azriadi Prana}, \bibinfo{person}{Stefanus~Agus Haryono},
  \bibinfo{person}{Shaowei Wang}, {and} \bibinfo{person}{David Lo}.}
  \bibinfo{year}{2022}\natexlab{b}.
\newblock \showarticletitle{Real world projects, real faults: evaluating
  spectrum based fault localization techniques on Python projects}.
\newblock \bibinfo{journal}{\emph{Empirical Softw. Engg.}}
  \bibinfo{volume}{27}, \bibinfo{number}{6} (\bibinfo{date}{nov}
  \bibinfo{year}{2022}), \bibinfo{numpages}{50}~pages.
\newblock
\showISSN{1382-3256}
\urldef\tempurl%
\url{https://doi.org/10.1007/s10664-022-10189-4}
\showDOI{\tempurl}


\bibitem[\protect\citeauthoryear{Widyasari, Sim, Lok, Qi, Phan, Tay, Tan, Wee,
  Tan, Yieh, Goh, Thung, Kang, Hoang, Lo, and Ouh}{Widyasari
  et~al\mbox{.}}{2020}]%
        {widyasari2020bugsinpy}
\bibfield{author}{\bibinfo{person}{Ratnadira Widyasari},
  \bibinfo{person}{Sheng~Qin Sim}, \bibinfo{person}{Camellia Lok},
  \bibinfo{person}{Haodi Qi}, \bibinfo{person}{Jack Phan},
  \bibinfo{person}{Qijin Tay}, \bibinfo{person}{Constance Tan},
  \bibinfo{person}{Fiona Wee}, \bibinfo{person}{Jodie~Ethelda Tan},
  \bibinfo{person}{Yuheng Yieh}, \bibinfo{person}{Brian Goh},
  \bibinfo{person}{Ferdian Thung}, \bibinfo{person}{Hong~Jin Kang},
  \bibinfo{person}{Thong Hoang}, \bibinfo{person}{David Lo}, {and}
  \bibinfo{person}{Eng~Lieh Ouh}.} \bibinfo{year}{2020}\natexlab{}.
\newblock \showarticletitle{{BugsInPy}: {A} database of existing bugs in
  {Python} programs to enable controlled testing and debugging studies}. In
  \bibinfo{booktitle}{\emph{Proceedings of the 28th ACM Joint Meeting on
  European Software Engineering Conference and Symposium on the Foundations of
  Software Engineering}} (Virtual Event, USA) \emph{(\bibinfo{series}{ESEC/FSE
  2020})}. \bibinfo{publisher}{Association for Computing Machinery},
  \bibinfo{address}{New York, NY, USA}, \bibinfo{pages}{1556–1560}.
\newblock
\showISBNx{9781450370431}
\urldef\tempurl%
\url{https://doi.org/10.1145/3368089.3417943}
\showDOI{\tempurl}


\bibitem[\protect\citeauthoryear{Wong, Wei, Qi, and Zhao}{Wong
  et~al\mbox{.}}{2008}]%
        {wong2008crosstab}
\bibfield{author}{\bibinfo{person}{Eric Wong}, \bibinfo{person}{Tingting Wei},
  \bibinfo{person}{Yu Qi}, {and} \bibinfo{person}{Lei Zhao}.}
  \bibinfo{year}{2008}\natexlab{}.
\newblock \showarticletitle{A Crosstab-based Statistical Method for Effective
  Fault Localization}. In \bibinfo{booktitle}{\emph{Proceedings of the 2008
  International Conference on Software Testing, Verification, and Validation}}
  \emph{(\bibinfo{series}{ICST '08})}. \bibinfo{publisher}{IEEE Computer
  Society}, \bibinfo{address}{USA}, \bibinfo{pages}{42–51}.
\newblock
\showISBNx{9780769531274}
\urldef\tempurl%
\url{https://doi.org/10.1109/ICST.2008.65}
\showDOI{\tempurl}


\bibitem[\protect\citeauthoryear{Wong, Debroy, Li, and Gao}{Wong
  et~al\mbox{.}}{2012}]%
        {wong2012dstar}
\bibfield{author}{\bibinfo{person}{W.~Eric Wong}, \bibinfo{person}{Vidroha
  Debroy}, \bibinfo{person}{Yihao Li}, {and} \bibinfo{person}{Ruizhi Gao}.}
  \bibinfo{year}{2012}\natexlab{}.
\newblock \showarticletitle{Software Fault Localization Using {DStar} {(D*)}}.
  In \bibinfo{booktitle}{\emph{2012 IEEE Sixth International Conference on
  Software Security and Reliability}}. \bibinfo{pages}{21--30}.
\newblock
\urldef\tempurl%
\url{https://doi.org/10.1109/SERE.2012.12}
\showDOI{\tempurl}


\bibitem[\protect\citeauthoryear{Wong, Qi, Zhao, and Cai}{Wong
  et~al\mbox{.}}{2007}]%
        {wong2007wong}
\bibfield{author}{\bibinfo{person}{W.~Eric Wong}, \bibinfo{person}{Yu Qi},
  \bibinfo{person}{Lei Zhao}, {and} \bibinfo{person}{Kai-Yuan Cai}.}
  \bibinfo{year}{2007}\natexlab{}.
\newblock \showarticletitle{Effective Fault Localization using Code Coverage}.
  In \bibinfo{booktitle}{\emph{31st Annual International Computer Software and
  Applications Conference (COMPSAC 2007)}}, Vol.~\bibinfo{volume}{1}.
  \bibinfo{pages}{449--456}.
\newblock
\urldef\tempurl%
\url{https://doi.org/10.1109/COMPSAC.2007.109}
\showDOI{\tempurl}


\bibitem[\protect\citeauthoryear{Xie, Kuo, Chen, Yoo, and Harman}{Xie
  et~al\mbox{.}}{2013}]%
        {xie2013gp}
\bibfield{author}{\bibinfo{person}{Xiaoyuan Xie}, \bibinfo{person}{Fei-Ching
  Kuo}, \bibinfo{person}{Tsong~Yueh Chen}, \bibinfo{person}{Shin Yoo}, {and}
  \bibinfo{person}{Mark Harman}.} \bibinfo{year}{2013}\natexlab{}.
\newblock \showarticletitle{Provably Optimal and Human-Competitive Results in
  SBSE for Spectrum Based Fault Localisation}. In
  \bibinfo{booktitle}{\emph{Proceedings of the 5th International Symposium on
  Search Based Software Engineering - Volume 8084}} (St. Petersburg, Russia)
  \emph{(\bibinfo{series}{SSBSE 2013})}. \bibinfo{publisher}{Springer-Verlag},
  \bibinfo{address}{Berlin, Heidelberg}, \bibinfo{pages}{224–238}.
\newblock
\showISBNx{9783642397417}
\urldef\tempurl%
\url{https://doi.org/10.1007/978-3-642-39742-4_17}
\showDOI{\tempurl}


\bibitem[\protect\citeauthoryear{Xuan and Monperrus}{Xuan and
  Monperrus}{2014a}]%
        {xuan2014test}
\bibfield{author}{\bibinfo{person}{Jifeng Xuan} {and} \bibinfo{person}{Martin
  Monperrus}.} \bibinfo{year}{2014}\natexlab{a}.
\newblock \showarticletitle{Test case purification for improving fault
  localization}. In \bibinfo{booktitle}{\emph{Proceedings of the 22nd ACM
  SIGSOFT International Symposium on Foundations of Software Engineering}}
  (Hong Kong, China) \emph{(\bibinfo{series}{FSE 2014})}.
  \bibinfo{publisher}{Association for Computing Machinery},
  \bibinfo{address}{New York, NY, USA}, \bibinfo{pages}{52–63}.
\newblock
\showISBNx{9781450330565}
\urldef\tempurl%
\url{https://doi.org/10.1145/2635868.2635906}
\showDOI{\tempurl}


\bibitem[\protect\citeauthoryear{Xuan and Monperrus}{Xuan and
  Monperrus}{2014b}]%
        {xuanm14b}
\bibfield{author}{\bibinfo{person}{Jifeng Xuan} {and} \bibinfo{person}{Martin
  Monperrus}.} \bibinfo{year}{2014}\natexlab{b}.
\newblock \showarticletitle{Test case purification for improving fault
  localization}. In \bibinfo{booktitle}{\emph{Proceedings of the 22nd {ACM}
  {SIGSOFT} International Symposium on Foundations of Software Engineering,
  (FSE-22), Hong Kong, China, November 16 - 22, 2014}},
  \bibfield{editor}{\bibinfo{person}{Shing{-}Chi Cheung},
  \bibinfo{person}{Alessandro Orso}, {and} \bibinfo{person}{Margaret{-}Anne~D.
  Storey}} (Eds.). \bibinfo{publisher}{{ACM}}, \bibinfo{pages}{52--63}.
\newblock
\urldef\tempurl%
\url{https://doi.org/10.1145/2635868.2635906}
\showDOI{\tempurl}


\bibitem[\protect\citeauthoryear{Yan, Jiang, Zhang, Zhang, and Zhang}{Yan
  et~al\mbox{.}}{2023}]%
        {yan2023context}
\bibfield{author}{\bibinfo{person}{Yue Yan}, \bibinfo{person}{Shujuan Jiang},
  \bibinfo{person}{Yanmei Zhang}, \bibinfo{person}{Shenggang Zhang}, {and}
  \bibinfo{person}{Cheng Zhang}.} \bibinfo{year}{2023}\natexlab{}.
\newblock \showarticletitle{A fault localization approach based on fault
  propagation context}.
\newblock \bibinfo{journal}{\emph{Inf. Softw. Technol.}} \bibinfo{volume}{160},
  \bibinfo{number}{C} (\bibinfo{date}{aug} \bibinfo{year}{2023}),
  \bibinfo{numpages}{11}~pages.
\newblock
\showISSN{0950-5849}
\urldef\tempurl%
\url{https://doi.org/10.1016/j.infsof.2023.107245}
\showDOI{\tempurl}


\bibitem[\protect\citeauthoryear{Yang, Nong, Zhang, Luo, and Cai}{Yang
  et~al\mbox{.}}{2024}]%
        {yang2024multilingual}
\bibfield{author}{\bibinfo{person}{Haoran Yang}, \bibinfo{person}{Yu Nong},
  \bibinfo{person}{Tao Zhang}, \bibinfo{person}{Xiapu Luo}, {and}
  \bibinfo{person}{Haipeng Cai}.} \bibinfo{year}{2024}\natexlab{}.
\newblock \showarticletitle{Learning to Detect and Localize Multilingual Bugs}.
\newblock \bibinfo{journal}{\emph{Proc. ACM Softw. Eng.}} \bibinfo{volume}{1},
  \bibinfo{number}{FSE}, Article \bibinfo{articleno}{97} (\bibinfo{date}{jul}
  \bibinfo{year}{2024}), \bibinfo{numpages}{24}~pages.
\newblock
\urldef\tempurl%
\url{https://doi.org/10.1145/3660804}
\showDOI{\tempurl}


\bibitem[\protect\citeauthoryear{Yoo, Harman, and Clark}{Yoo
  et~al\mbox{.}}{2013}]%
        {yoo2013priorization}
\bibfield{author}{\bibinfo{person}{Shin Yoo}, \bibinfo{person}{Mark Harman},
  {and} \bibinfo{person}{David Clark}.} \bibinfo{year}{2013}\natexlab{}.
\newblock \showarticletitle{Fault Localization Prioritization: Comparing
  Information-Theoretic and Coverage-Based Approaches}.
\newblock \bibinfo{journal}{\emph{ACM Trans. Softw. Eng. Methodol.}}
  \bibinfo{volume}{22}, \bibinfo{number}{3}, Article \bibinfo{articleno}{19}
  (\bibinfo{date}{jul} \bibinfo{year}{2013}), \bibinfo{numpages}{29}~pages.
\newblock
\showISSN{1049-331X}
\urldef\tempurl%
\url{https://doi.org/10.1145/2491509.2491513}
\showDOI{\tempurl}


\bibitem[\protect\citeauthoryear{Yu, Lin, Gao, Zhang, and Zhang}{Yu
  et~al\mbox{.}}{2011}]%
        {yu2011models}
\bibfield{author}{\bibinfo{person}{Kai Yu}, \bibinfo{person}{Mengxiang Lin},
  \bibinfo{person}{Qing Gao}, \bibinfo{person}{Hui Zhang}, {and}
  \bibinfo{person}{Xiangyu Zhang}.} \bibinfo{year}{2011}\natexlab{}.
\newblock \showarticletitle{Locating faults using multiple spectra-specific
  models}. In \bibinfo{booktitle}{\emph{Proceedings of the 2011 ACM Symposium
  on Applied Computing}} (TaiChung, Taiwan) \emph{(\bibinfo{series}{SAC '11})}.
  \bibinfo{publisher}{Association for Computing Machinery},
  \bibinfo{address}{New York, NY, USA}, \bibinfo{pages}{1404–1410}.
\newblock
\showISBNx{9781450301138}
\urldef\tempurl%
\url{https://doi.org/10.1145/1982185.1982490}
\showDOI{\tempurl}


\bibitem[\protect\citeauthoryear{Zhang, Chan, Tse, Jiang, and Wang}{Zhang
  et~al\mbox{.}}{2009}]%
        {zhang2009capturing}
\bibfield{author}{\bibinfo{person}{Zhenyu Zhang}, \bibinfo{person}{W.~K. Chan},
  \bibinfo{person}{T.~H. Tse}, \bibinfo{person}{Bo Jiang}, {and}
  \bibinfo{person}{Xinming Wang}.} \bibinfo{year}{2009}\natexlab{}.
\newblock \showarticletitle{Capturing propagation of infected program states}.
  In \bibinfo{booktitle}{\emph{Proceedings of the 7th Joint Meeting of the
  European Software Engineering Conference and the ACM SIGSOFT Symposium on The
  Foundations of Software Engineering}} (Amsterdam, The Netherlands)
  \emph{(\bibinfo{series}{ESEC/FSE '09})}. \bibinfo{publisher}{Association for
  Computing Machinery}, \bibinfo{address}{New York, NY, USA},
  \bibinfo{pages}{43–52}.
\newblock
\showISBNx{9781605580012}
\urldef\tempurl%
\url{https://doi.org/10.1145/1595696.1595705}
\showDOI{\tempurl}


\end{thebibliography}

\end{document}